\documentclass[aps,pra,preprintnumbers,twoside,twocolumn,
superscriptaddress,floatfix,10pt]{revtex4-2}

\pdfoutput=1

\usepackage[utf8]{inputenc} 
\usepackage[]{amsmath} 
\usepackage[]{bbold}
\usepackage[]{bm}
\usepackage{graphicx} 
\usepackage{siunitx}
\usepackage{xcolor} 
\usepackage[version=3]{mhchem} 
\usepackage{upgreek} 
\usepackage{xr-hyper}
\usepackage[colorlinks=true,citecolor=blue,urlcolor=red,linkcolor=red,filecolor=cyan]{hyperref} 
\usepackage{cleveref} 

\makeatletter
\newcommand*{\addFileDependency}[1]{
  \typeout{(#1)}
  \@addtofilelist{#1}
  \IfFileExists{#1}{}{\typeout{No file #1.}}
}
\makeatother

\newcommand*{\myexternaldocument}[1]{%
    \externaldocument{#1}%
    \addFileDependency{#1.tex}%
    \addFileDependency{#1.aux}%
}

\usepackage{natbib} 
\usepackage[mathscr]{eucal}
\usepackage[normalem]{ulem}
\setcitestyle{super}
\hypersetup{pdfauthor={Name}}

\renewcommand{\vec}[1]{\bm{#1}}
\newcommand{\ten}[1]{\bm{#1}}

\newcommand{\maxwellstress}{\ten{\Sigma}^{\ce{M}}}
\newcommand{\grad}{\mbox{grad}}
\renewcommand{\div}{\mbox{div}}

\newcommand{\suml}{\sum\limits}

\newcommand{\eye}{\ten{\mathbb{1}}}
\newcommand{\tr}{\mbox{tr}}
\newcommand{\matder}[1]{\dot{ #1 }}
\makeatletter
\newcommand{\polder}[2][0ex]{%
  \mathrel{\mathop{#2}\limits^{
    \vbox to#1{\kern-2\ex@
    \hbox{$\scriptstyle \prime$}\vss}}} }
\makeatother

\newcommand{\dervar}{\bigg|}
\newcommand{\varstaset}{\boldsymbol{\mathbb{V}}}		
\newcommand{\efieldcom}{E}						
\newcommand{\efield}{\vec{\efieldcom}}			
\newcommand{\efieldinv}{\vec{\mathcal{E}}}		
\newcommand{\dfieldcom}{D}						
\newcommand{\dfield}{\vec{\dfieldcom}}			
\newcommand{\hfield}{\vec{H}}					
\newcommand{\hfieldinv}{\vec{\mathcal{H}}}		
\newcommand{\bfield}{\vec{B}}					

\let\volt\relax

\let\meter\relax
\let\milli\relax

\newcommand{\cpeo}{c_{\ce{p}}}
\newcommand{\cli}{c_{\ce{Li}}}
\newcommand{\ctfsi}{c_{\ce{TFSI}}}
\newcommand{\p}{\ce{p}}
\newcommand{\li}{\ce{Li}}
\newcommand{\tfsi}{\ce{TFSI}}
\newcommand{\litfsi}{\ce{LiTFSI}}

\renewcommand{\cpeo}{c_{\p}}
\renewcommand{\cli}{c_{\li}}
\renewcommand{\ctfsi}{c_{\tfsi}}

\myexternaldocument{SuppInfo}

\begin{document}

\author{Daniel O. Möhrle}
\affiliation{German Aerospace Center,
  Wilhelm-Runge-Straße 10, 89081 Ulm, Germany}
\affiliation{Helmholtz Institute Ulm, Helmholtzstraße 11, 89081
  Ulm, Germany}

\author{Max Schammer}
\affiliation{German Aerospace Center,
  Wilhelm-Runge-Straße 10, 89081 Ulm, Germany}
\affiliation{Helmholtz Institute Ulm, Helmholtzstraße 11, 89081
  Ulm, Germany}
  
\author{Katharina Becker-Steinberger}
\affiliation{German Aerospace Center,
  Wilhelm-Runge-Straße 10, 89081 Ulm, Germany}
\affiliation{Helmholtz Institute Ulm, Helmholtzstraße 11, 89081
  Ulm, Germany}

\author{Birger Horstmann}
\email{birger.horstmann@dlr.de}
\affiliation{German Aerospace Center, Wilhelm-Runge-Straße 10, 89081 Ulm, Germany}
\affiliation{Helmholtz Institute Ulm,
  Helmholtzstraße 11, 89081 Ulm, Germany}
\affiliation{Universität Ulm, Albert-Einstein-Allee 47, 89081 Ulm,
  Germany}

\author{Arnulf Latz}
\affiliation{German Aerospace Center,
  Wilhelm-Runge-Straße 10, 89081 Ulm, Germany}
\affiliation{Helmholtz Institute Ulm, Helmholtzstraße 11, 89081
  Ulm, Germany}
\affiliation{Universität Ulm, Albert-Einstein-Allee
  47, 89081 Ulm, Germany}

\title{Electro-Chemo-Mechanical Model for Polymer Electrolytes}

\allowdisplaybreaks

\begin{abstract}
    Polymer electrolytes (PEs) are promising candidates for use in next-generation high-voltage batteries, as they possess advantageous elastic and electrochemical properties.
    However, PEs still suffer from low ionic conductivity and need to be operated at higher temperatures.
    Furthermore, the wide variety of different types of PEs and the complexity of the internal interactions constitute challenging tasks for progressing towards a systematic understanding of PEs.
    Here, we present a continuum transport theory which enables a straight-forward and thermodynamically consistent method to couple different aspects of PEs relevant for battery performance.
    Our approach combines mechanics and electrochemistry in non-equilibrium thermodynamics, and is based on modeling the free energy, which comprises all relevant bulk properties.
    In our model, the dynamics of the polymer-based electrolyte are formulated relative to the highly elastic structure of the polymer.
    For validation, we discuss a benchmark polymer electrolyte. Based on our theoretical description, we perform numerical simulations and compare the results with data from the literature. In addition, we apply our theoretical framework to a novel type of single-ion conducting PE and derive a detailed understanding of the internal dynamics.
\end{abstract}
	
\maketitle



Batteries play a significant role as energy storage devices in the transition to a renewable energy system. \cite{Chu2012}
This comes with an increasing demand for low-cost, environmentally friendly batteries with high energy and power densities, especially in the electric automotive sector. As result, there exists a tremendous research stimulus for improved battery materials.\cite{Bresser2018} Hence, substantial effort has been put into improving established materials and developing novel materials for all cell components. Among them, the electrolyte plays a significant role for the performance of a battery,\cite{Armand2008}
as it provides the transport pathway of the ions from electrode to electrode. 

Currently, most commercially available batteries use liquid electrolytes (LE). However, these electrolytes are limited due to several factors.
One factor is the typically low transference number of LEs, which results in concentration polarization.
This increases the electrolyte overpotential, limits the charging rate,  while it also creates a more uneven lithium deposition and promotes dendrite growth.\cite{Whittingham2004,Kim2015,Weiss2021}
Another factor is limited electrochemical stability, which makes them not suited for high-voltage cell concepts.\cite{Meng2021}

One approach to overcome these obstacles and to increase the performance and safety of batteries is to change from liquid to solid electrolytes (SEs).\cite{Kim2015, Janek2016} SEs have many advantageous properties. Among them is the property that they can suppress the growth of dendrites, thereby enabling the use of Lithium metal anodes for batteries having high energy densities.\cite{Armand2008, Krauskopf2020, Li2014, Cheng2015} SEs can be split up into two groups, inorganic crystalline SEs,\cite{Zheng2018} and (organic) polymer electrolytes.\cite{Mindemark2018}

Inorganic SEs posses various advantageous properties. They usually inhibit good thermal stability and high ionic conductivities in their bulk phases. Because the transference number of inorganic SEs is often close to unity, they are competitive to conventional liquid electrolytes.\cite{Weiss2021}
Nevertheless, inorganic SEs also exhibit some undesirable properties as electrolytes.\cite{Huo2019,Weiss2020} Among them are grain boundaries inside the material acting as barriers for the ion transport, thus reducing the effective conductivity.\cite{Neumann2021,Jetybayeva2021} Also, imperfect mechanical contacts or brittle mechanical properties constitute another challenge for the commercialization of SEs.\cite{Liu2020}

Polymer electrolytes comprise a large class of materials consisting of long polymeric chains with high ion concentrations.\cite{doi:10.1021/acs.macromol.9b02742} Because the degree of crystalline structure  varies significantly between these materials, they share liquid-like properties (\textit{e.g.} for gel polymer electrolytes) with solid-like properties (\textit{e.g.} for solid polymer electrolytes). This diversity implies a wide range of polymer chemistry, which allows for a high degree of tunability, most importantly of their  elastic properties. Thus, polymers  can be tailor-cut to satisfy desired characteristics for task-specific applications, \textit{e.g.} large thermal and electrochemical stabilities or low material and processing costs. In particular, through their elastic properties, they can also inhibit dendrite growth,\cite{Frenck2019} which makes them promising materials for Li-based batteries. However, in contrast to inorganic SEs, (organic) polymer electrolytes generally show lower conductivities and need to be operated at elevated temperatures.\cite{Weiss2020}

The widely studied polyethylene glycole (PEO) constitutes a benchmark material among the wide class of polymer electrolytes. PEOs are low cost and easy to process and were among the first polymers studied for electrolyte applications. They exhibit promising transport properties and have a good stability against reduction, including the contact with Lithium metal electrodes.\cite{Wright1975,Bresser2019}
However, PEOs have a low ionic conductivity at room temperature and limited stability against oxidation.
This has led to the development of several distinct polymer electrolytes, following different strategies to overcome these shortcomings.\cite{Cheng2015,Nguyen2018,Bresser2019,Nematdoust2020,Butzelaar2021}
Various approaches have been proposed in the literature to increase the conductivity of PEOs. One approach is based on increasing the molar ratio between the Li salt and the polymer, which showed higher ionic conductivities accompanied by high transference numbers at ambient temperatures. \cite{Angell1993}
Another approach is the creation of composite electrolytes, which consist of combinations of two or more different materials.\cite{Huo2019,Nematdoust2020}
As such, composite electrolytes consisting of polymer electrolytes and inorganic SEs combine the advantageous elastic properties of the polymer, especially the good adhesion to Lithium surfaces, with the high ion-conductivity of the inorganic solid material. This combination promises the suppression of dendrite growth and a more homogeneous Li-ion flux at the interface.\cite{Zhou2016, Huo2019, Weiss2020}

However, there are still major challenges for our understanding of these systems. Here, theoretical methods can deliver beneficial insights and evaluate performance characteristics. 
Atomistic methods like density functional theory or molecular dynamics are able to illuminate important aspects of PEs, from transport mechanisms \cite{Maitra2007,Diddens2010} to electrochemical properties.\cite{Mackanic2018,Liivat2011,Ebadi2017,Thum2021,Johansson2015}
However, due to their numerical complexity, they are confined to studies of small systems. For larger systems, continuum models are more capable. In particular,  several (semi)-empirical approaches have been employed for the description of complete battery cells, {\it e.g.} Monte-Carlo simulations or resistor network approaches. \cite{Siekierski2007,Katzenmeier2022}
Also, continuum models which were originally developed for liquid electrolytes have been modified to the description of polymer electrolytes.\cite{Natsiavas2016, Bucci2016, Grazioli2019} However, because polymer electrolytes exhibit some quite unique features, such approaches prove difficult.\cite{Johansson2015,Dickinson2020}
In recent years, a coupled electro-chemo-mechanical model was proposed for polyelectrolyte gels.\cite{Narayan2021}
This model describes the swelling and deswelling of these gels in baths of varying pH and ionic strengths.
In contrast, our approach focuses on  polymer electrolytes used in batteries. Here, migration is a crucial transport process occurring in the polymer when subjected to electric fields.

In this work, we propose a continuum model for polymer electrolytes using non-equilibrium thermodynamics.
Our approach is based on the work of Latz, Horstmann, Schammer and coworkers, who developed  transport models for concentrated electrolytes, ionic liquids and inorganic solid electrolytes.\cite{Latz2011,Latz2015,Braun2015,Schammer2021,VonKolzenberg2021,Kilchert2022} Their approach is based on a rigorous physical basis, and takes account for universal balancing laws, \textit{e.g.} for momentum, charge and energy. As consequence of this fully coupled description, the resulting transport theory ensures a non-negative entropy production ("thermodynamic completeness"), in accordance with the second law of thermodynamics. The focal quantity in this description is the  (Helmholtz) free energy, which incorporates all material specific properties. Hence, in order to apply this description of liquid electrolytes to polymer electrolytes, it mostly suffices to modify the free energy. 

One important difference between polymer electrolytes and liquid electrolytes which must be taken into account in the free energy is that there is typically an excess amount of polymer species present in the electrolyte mixture (both with respect to mass and volume). In combination with the property that convection becomes important in highly correlated electrolytes,\cite{Schammer2021} this suggests using an internal description for our transport model, \textit{i.e.} for the frame of reference, which is either based on the motion of the polymer species or on the motion of the volume-averaged convection velocity.\cite{Kilchert2022} It is important to note that the choice of reference frame plays an important role when discussing transport parameters, too.\cite{doi:10.1021/acs.jpclett.2c02398} However, both descriptions have the advantage that they can be parameterized using results from MD simulations,\cite{Diddens2010} or results based on eNMR experiments.\cite{Kilchert2022} Another important material specific property of polymer based electrolytes which must be incorporated into the free energy is their advantageous mechanical behaviour, which makes it highly attractive for using them in lithium metal batteries or in composite electrolytes. Here, we derive a consistent, kinematical description for the electrolyte transport, which comprises all mechanical couplings.

We structure this document as follows.
In the section Transport Theory, we derive a transport theory for elastic materials and taylor-cut it to the case of polymer electrolytes.
In the section Validation: Simulation of a Li Cell With Polymer Based Electrolyte, we show the results obtained from numerical simulations of a standard PEO polymer electrolyte, validate our transport theory by comparison with in-situ experments, and discuss the influence of the parameters appearing in our electrolyte description on the cell performance.
In the section Application: Single-Ion Conducting Block Copolymer, we apply our model to a novel single ion  conducting polymer electrolyte.
In the section Discussion, we discuss the positioning of our polymer theory within the current status of the literature.

\section*{Transport Theory}
\label{sec:transporttheory}

Depending upon the perspective, highly viscoelastic polymer electrolytes can be either classified as liquid or solid electrolytes.\cite{rudin1998elements} For example, upon the exertion of mechanical stress, they behave like elastic materials on short time scales. In contrast, on longer time scales, they behave more like liquids, as they exhibit viscous flow.\cite{Ligia2009}

Recently, we derived thermodynamically consistent transport theories for both types of electrolyte-materials, \textit{i.e.} highly correlated liquid electrolytes,\cite{Latz2015, Schammer2021,doi:10.1021/acs.jpcb.2c00215} and solid electrolytes.\cite{Braun2015,VonKolzenberg2021,KBSarxiv}
Both frameworks are based on the methodology of rational thermodynamics (RT), and couple non-equilibrium thermodynamics with electromagnetic theory and mechanics. In RT, the focal quantity is the Helmholtz free energy density, which incorporates all material-specific electrolyte properties, and which determines the description of the system via constitutive equations. In this work, we utilize the broad generality of our transport theory for multi-component liquid electrolytes presented in Ref.~\citenum{Schammer2021}, and extend it to the description of polymers via modification of the model for the free energy. 

We split this theory chapter into three main sections.
First, in the section General Transport Theory of Elastic Electrolytes we derive a framework for multi-component electrolytes, where one species is designated by bulk-excess of mass and volume and determines the elastic behaviour.
Second, in the section Polymer Electrolyte Model, we tailor-cut this universal description to polymer electrolytes and close the set of equations of motion. 

Throughout the text, we use a notation which is similar to Ref.~\citenum{Schammer2021}, \textit{i.e.} Greek indices relate to electrolyte species, Latin indices relate to spatial components, and capital Latin indices relate to elements of a set of variables. We highlight the prominent role of the po\-ly\-mer-species in the N-component electrolyte mixture, and assign the polymer to the first species using the index "$\ce{p}$".
\newline

\paragraph*{\textit{\textbf{General transport theory of elastic electrolytes.}}}
\label{sec:general_tt}
In this section we derive our transport theory for elastic electrolytes. 

Our derivation is based on the previous work on highly correlated liquid electrolytes presented in Ref.~\citenum{Schammer2021}, and follows the same logical structure. However, there are two major conceptual differences between our model for viscous and for elastic electrolytes. First, instead of using a description based on the bulk momentum by the center-of-mass convection velocity, we here use the species-related frame of reference defined by the polymer velocity. Second, we replace the rate of strain tensor appearing in the viscous model by the strain tensor, which incorporates the elastic properties. Apart from these conceptual differences, the derivation of our polymer description is similar as outlined in Ref.~\citenum{Schammer2021}. In particular, we formulate universal balance equations for mass, energy and momentum, and derive the corresponding entropy inequality.
Then, we state our model for the free energy density and determine the constitutive equations.
Finally, we state the closure-relations for the fluxes using an Onsager approach.

We begin our derivation by accounting for the bulk-excess of mass and volume-fraction of the dominant polymer species.
As consequence, it can be advantageous for the description of polymers to use the velocity of the polymer species $\vec{v}_{\ce{p}}$ as convection velocity.
However, for liquid electrolytes, the convection velocity is often defined by  the center-of-mass motion, $\vec{v} = \sum_{\alpha=1}^N \uprho_{\alpha} \vec{v}_{\alpha}/\uprho$.
We emphasize that both descriptions are related by suitable transformation rules,\cite{Kilchert2022} and the evolution of a physical quantity $\Psi=\int_V\uprho \cdot\psi\ce{d}V$ can be described in both ways. In particular, in the mass-based description, we have
\begin{gather}
	\label{eqn:material_derivative}
	\matder{\psi} = \partial_t \psi + \vec{v} \cdot \grad{\left( \psi \right)},
	\intertext{or, in the polymer-based description, }
	\label{eqn:polymer_derivative}
	\polder{\psi}\, =  \partial_t \psi + \vec{v}_{\ce{p}} \cdot \grad\left( \psi \right).
\end{gather}
Here, $\partial_t=\partial/\partial t$ is the change with time at fixed laboratory coordinates.

Because of the bulk-excess of the polymer species, we assume that the mechanical properties of the multi-component electrolyte are mainly determined by the elastic deformation / swelling of the polymer-matrix.\cite{Mueller2001} For the description of the mechanical properties of the polymer, we use an elastic model based on finite strain theory.\cite{Holzapfel2000} According to this material-based description, the deformation of the system is described relative to a reference configuration $\vec{x}_{\ce{p}}^0$ of the polymer matrix, where the volume occupied by the polymer in this reference configuration is $V_{\ce{p}}^0 = \upnu_{\ce{p}}^0 \mathcal{N}_{\ce{p}}$.
Here, $\upnu_{\alpha}^0$ and  $\mathcal{N}_{\alpha}$ are the reference partial molar volume and the molar number of species $\alpha$, respectively. In this description,\cite{Holzapfel2000} the deformation gradient tensor ("polymer strain") reads
\begin{equation}
    \label{eqn:polymer_deformation_gradient_definition}
	\ten{F} = \frac{\text{d}\vec{x}_{\ce{p}}}{\text{d}\vec{x}_{\ce{p}}^0}.
\end{equation}
Here, the position $\vec{x}_{\ce{p}}$ describes the current configuration with respect to fixed laboratory coordinates, and the volume occupied by the polymer in this configuration is $V_{\ce{p}} = \upnu_{\ce{p}} \mathcal{N}_{\ce{p}}$.
However, in the isotropic liquid one is more interested in the determinant of the deformation gradient, 
\begin{equation}
\label{eq:definition_J}
J = \det\left( \ten{F} \right),
\end{equation}
as it defines a transformation between the polymer-frame and the fixed laboratory frame, and measures the volume-expansion  of the system via $V = J V^0$. Here, $V^0 = V_{\ce{p}}^0$ is the polymer volume of the strain-free polymer-matrix in the reference configuration ($\ten{F} = \eye$).
Furthermore, the polymer derivative connects the polymer strain-tensor with the polymer rate-of-strain tensor\cite{Holzapfel2000}
\begin{equation}
	\label{eqn:deformation_tensor_velocity}
	\grad\left( \vec{v}_{\ce{p}} \right) = \polder{\ten{F}} \cdot \ten{F}^{-1}.
\end{equation}

The whole volume of the complete multi-component electrolyte consists of the volume of the polymer, and the volume of minor non-polymeric species, $V = J \upnu_{\ce{p}}^0 \mathcal{N}_{\ce{p}} =\upnu_{\ce{p}} \mathcal{N}_{\ce{p}} + \sum_{\alpha = 2}^{N} \upnu_{\alpha} \mathcal{N}_{\alpha}$.
Thus, beneath  polymer-swelling, we must account for molar volumes of non-polymeric species in the evolution of the whole electrolyte-volume too.
Altogether, we find for the Euler equation for the volume,\cite{Schammer2021}
\begin{equation}
	\label{eqn:polymer_deformation_concentration}
	1 = J \upnu_{\ce{p}}^0 c_{\ce{p}} = \suml_{\alpha=1}^N \nu_{\alpha} c_{\alpha},
\end{equation}
where $c_{\alpha} = \mathcal{N}_{\alpha}/V$ are the specific molar concentrations.
Note that, in the literature, the deformation gradient is sometimes factorized such that the individual terms account for specific properties, \textit{e.g.} polymer swelling.\cite{VonKolzenberg2021,Narayan2021}
Here, the swelling of the polymer matrix due to the presence of other species leads to (isotropic) mechanical stresses. However, our mechanical model can be modified easily to account for additional elastic or plastic effects.\cite{Holzapfel2000}
Altogether, this constitutes our mechanical description of the material properties, which shall be incorporated into our existing framework. The next steps in our derivation are closely aligned to Ref.~\citenum{Schammer2021}.

The assumption of mass conservation, $\partial_t\uprho = -\div(\uprho\vec{v})$, leads to continuity equations for the species con\-cen\-tra\-tions.\cite{Schammer2021} These are expressed by frame-dependent fluxes.  For example, 
in the center-of-mass description, there exist N fluxes $\vec{N_\alpha}=c_\alpha (\vec{v}_\alpha -\vec{v}) $. Furthermore, in the polymer-based description, we use polymer-fluxes  $\vec{N}_{\alpha}^{\text{p}} = c_{\alpha} \left( \vec{v}_{\alpha} - \vec{v}_{\text{p}} \right)$ such that 
\begin{equation}
 \label{eqn:continuity_equation}
	\partial_t c_{\alpha} 
	=  - \div{(\vec{N}_{\alpha}^{\ce{p}})} - \div{(c_{\alpha} \vec{v}_{\ce{p}})} + r_\alpha.
\end{equation}
Here, we model reactions as source-terms in our transport equations, where $r_\alpha$ denote reaction rates of species $\alpha$.
By construction, $\vec{N}_{\ce{p}}^{\ce{p}}=0$ such that only N-1 polymer-frame fluxes are independent. The property that only N-1 fluxes are independent is true in  any internal frame of reference.\cite{Schammer2021,Kilchert2022} For example, in the mass-based description, there exists a flux constraint $\sum_{\alpha=1}^{\ce{N}}M_\alpha\vec{N}_\alpha=0$. Assuming charge continuity, $\varrho^{\ce{F}}=F\sum_{\alpha=1}^{\ce{N}} z_\alpha c_\alpha$, \cref{eqn:continuity_equation} implies
\begin{equation}
\label{eqn:charge_conservation}
\partial_t \varrho^{\text{F}} = - \div\left( \vec{\mathcal{J}}^{\text{p}} \right) - \div\left( \varrho^{\text{F}} \vec{v}_{\text{p}} \right) + F\sum_{\alpha=1}^{\ce{N}}z_\alpha r_\alpha,
\end{equation}
with the current density $\vec{\mathcal{J}}^{\ce{p}} = F\sum_{\alpha=2}^{\ce{N}} z_\alpha \vec{N}_{\alpha}^{\ce{p}}$.

In the following, we couple the balance equation for momentum and for energy, which are both formulated with respect to the center-of-mass convection velocity.
We express the balance of total momentum $\vec{G} =\int \rho \vec{g} \ce{ d}V$, where $\vec{G}$ comprises kinematic, mechanical and electromagnetic contributions, via the standard approach
\begin{equation}
	\label{eqn:balance_momentum}
	\rho \matder{\vec{g}} = \uprho \vec{b} + \div\left( \ten{\uptau} \right),
\end{equation}
with long ranged body-forces $\uprho \vec{b}$, and a stress tensor $\ten{\tau}$.\cite{Schammer2021}

We formulate the balance of total energy as the sum of kinematic contributions $\Pi =  \uprho \vec{v} \cdot \vec{b} + \div ( \ten{\uptau}^{\text{T}} \cdot \vec{v} )$, and heating contributions $Q = \uprho h - \div \left( \vec{q} \right) - \div ( \vec{\mathcal{S}} )$, \textit{viz.}
\begin{equation}
\rho\matder{\epsilon}=\Pi+Q.
\end{equation}
Here, $\uprho h$ is the body heating, $\vec{q}$ is the heat-flux, and $\vec{\mathcal{S}} = \efieldinv \times \hfieldinv$ is the Poynting vector expressed via Galilei-invariant electric and magnetic fields $\efieldinv =  \efield + \vec{v} \times \bfield$ and $	\hfieldinv =  \hfield - \vec{v} \times \dfield$. \cite{Schammer2021,Kovetz2000}
From this, we obtain balance of internal energy by eliminating the kinematic parts $\vec{v}(\uprho\matder{\vec{v}} + \matder{\vec{\dfield} \times \vec{\bfield}})$ from the total energy,\cite{Medina2014} $\uprho \matder{u} = 	\rho h 
+ \vec{v} \cdot ( \uprho \matder{\vec{g}} - \uprho \matder{\vec{v}} - \matder{\left( \vec{\dfield} \times \vec{\bfield} \right)} )
+ \ten{\uptau} : \grad( \vec{v} ) 
- \div( \vec{q} + \vec{\mathcal{S}} )$.

However, focal modeling quantity in our framework is the Helmholtz free energy $F_{\ce{H}} = \int \ce{d}V \uprho \upvarphi_{\ce{H}}$, where  $\upvarphi_{\ce{H}} = u - Ts$, \textit{i.e.} the Legendre-transformed quantity with respect to the internal energy. We constitute our material model for viscoelastic electrolytes via the variable set $\varstaset_0 = \left\{ T, \ten{F}, c_{\alpha}, \dfield, \bfield \right\}$ for the free energy in the mass-based description,\cite{Schammer2021, DeGroot1984}  where the canonic expansion\cite{Holzapfel2000,henjes1993pressure}
\begin{multline}
\label{eqn:free_energy_total_differential}
	\text{d}F_\text{H}(\varstaset_0) =\int \text{d}V
	\Big( - \uprho s \cdot \text{d}T + \ten{\upsigma} : \left(  \text{d}\ten{F} \cdot \ten{F}^{-1} \right)
	\\
	+ \sum_{\alpha=1}^N \upmu_{\alpha}	\cdot \text{d}c_{\alpha}
	+ \efieldinv \cdot\text{d}\dfield + \hfieldinv \cdot \text{d}\bfield \Big) 
\end{multline}
identifies the conjugate variables and determines constitutive equations,\cite{Schammer2021}
\begin{gather}
\label{eq:constitutive_equation_entropy}
	s = - \frac{\partial  \upvarphi_\text{H} }{\partial T} \dervar_{\varstaset_0\setminus T},
	\\	\label{eq:constitutive_equation_stress}
	\ten{\upsigma} =  \frac{\partial \left( \rho \varphi_\text{H} \right)}{\partial \ten{F}} \dervar_{\varstaset_0\setminus \ten{F}} \cdot \ten{F}^{\text{T}},
	\\
\label{eq:constitutive_equation_chepot}
	\upmu_{\alpha} =  \frac{\partial \left( \uprho \upvarphi_\text{H} \right)}{\partial c_{\alpha}} \dervar_{\varstaset_0\setminus c_\alpha},
	\\
\label{eq:constitutive_equation_efield}
	\efieldinv =  \frac{\partial \left( \uprho \upvarphi_\text{H} \right)}{\partial \dfield} \dervar_{\varstaset_0\setminus \dfield},
	\\
\label{eq:constitutive_equation_hfield}
	\hfieldinv =  \frac{\partial \left( \uprho \upvarphi_\text{H} \right)}{\partial \bfield} \dervar_{\varstaset_0\setminus \bfield}.
\end{gather}
Here, $\upmu_{\alpha}$ are the chemical potentials of the species. We defined the deformation tensor using the polymer species coordinates ({\it c.f.} \cref{eqn:polymer_deformation_gradient_definition}), so the stress tensor $\ten{\upsigma}$ is conjugate to the polymer rate of strain tensor (see  also \cref{eqn:deformation_tensor_velocity}).
We name $\ten{\upsigma}$ the polymer stress tensor, since it describes the elastic stress in the polymer matrix.

Finally, we describe evolution of entropy via the standard Clausius-Duhem inequality, $\uprho\matder{s} +\div \vec{j}_{\ce{s}} - \uprho h/T\geq 0$, where $\vec{j}_{\text{s}} = ( \vec{q} - \sum_{\alpha=1}^N \upmu_{\alpha} \vec{N}_{\alpha} + \vec{d}_{\text{p}} \cdot \ten{\upsigma} )/T$ is the entropy flux density (see \cref*{subsec:SIentropy_flux}).
Here, the velocity difference $\vec{d}_{\p} = \vec{v}_{\p} - \vec{v}$ defines the relative motion between the mass-based and the polymer-based frame of reference. 

As measure for the deviation from equilibrium, we define the rate of entropy production $R_{\text{s}}$,\cite{Schammer2021} such that $T\uprho \matder{s} = R_{\text{s}} +  \uprho h - T\div{ (\vec{j}_{\ce{s}})}$.  Note that consistency with the second law of thermodynamics requires that this quantity is non-negative, $R_{\ce{s}}\geq 0$. As we show in \cref*{subsec:SIentropy_flux}, this thermodynamic constraint can be used to resolve the mutual couplings between the balance laws, and yields an expression for the Minkowski-momentum $\matder{\vec{g}} = \matder{\vec{v}} + \matder{( \dfield \times \bfield)}$ and for the generalized stress tensor
\begin{equation}
    \label{eq:const_eq_for_tau}
    \ten{\uptau} 
=  \ten{\sigma} + \maxwellstress
 + \left( \uprho \upvarphi_\text{H} {-} \sum_{\alpha=1}^{\ce{N}} \upmu_{\alpha} c_{\alpha} {-} \frac{\efieldinv {\cdot} \dfield {+} \hfieldinv {\cdot} \bfield}{2}  \right) \eye,
\end{equation}
where $\maxwellstress=\efieldinv \otimes \dfield + \hfieldinv \otimes \bfield - ( \efieldinv {\cdot} \dfield/2 + \hfieldinv {\cdot} \bfield/2)\eye$ is the Maxwell-stress tensor.

From now on, we neglect magnetic fields such that  the electric field is determined by the electrostatic potential $\Phi$ via $\efieldinv\to\vec{E} =-\grad (\Phi)$. 

After evaluation of the constitutive equations, the residual part of the entropy production rate is given by products of the thermodynamic fluxes and forces,
\begin{gather}
    \label{eqn:entropy_production_electrochempot}
    R_{\text{s}}
    = 
    - \vec{N}_{\ce{p}} \cdot \vec{\mathcal{F}}_{\text{mech}}
    - \vec{j}_{\ce{s}} \cdot \vec{\mathcal{F}}_{\ce{T}} 
    - \sum_{\alpha = 1}^\text{N} \vec{N}_\alpha \cdot \vec{\mathcal{F}}_\alpha^{\text{el}} .
\intertext{Here, we defined a mechanical force}
    \label{eq:def_forces_mech}
\vec{\mathcal{F}}_{\ce{mech}} 
= -  \left[ \div( \ten{\sigma} ) + \left( \grad( \ten{F} ) \cdot \ten{F}^{-1} \right) : \ten{\sigma} \right]/c_{\ce{p}},
\intertext{and N+1 electrochemical and thermal forces}
    \label{eq:def_forces_TD}
    \vec{\mathcal{F}}_{\alpha}^{\ce{el}}
    = \grad( \upmu_\alpha^{\ce{el}} ),
    \quad \textnormal{and} \quad 
        \vec{\mathcal{F}}_{\ce{T}}  
    {=}
        \grad( T ),
\end{gather}
 with electrochemical potentials $\upmu_\alpha^{\ce{el}}=\upmu_\alpha+Fz_\alpha\Phi$.
 
We make use of the thermodynamical requirement that \cref{eqn:entropy_production_electrochempot} be non-negative, and determine the N-1 independent fluxes $\vec{N}_{\alpha}^{\text{p}}$ using an Onsager approach with respect to the  polymer-frame. For this purpose, we first transform the N-1 mass-based species fluxes $\vec{N}_{\alpha}$  to the polymer-frame (see \cref*{subsec:SIentropy_flux}),  
\begin{gather}
    \label{eqn:flux_frame_transformation}
	\vec{N}_{\alpha}^{\text{p}} 
	= \sum_{\beta=2}^{\ce{N}} A_{\alpha \beta} \vec{N}_{\beta},
\end{gather}
such that $\vec{\mathcal{J}}^{\text{p}} = \sum_{\alpha=2}^{\ce{N}}Fz_\alpha\vec{N}_{\alpha}^{\text{p}}$. The matrix $\vec{A}$ is a  $(\ce{N}{-}1){\times}(\ce{N}{-}1)$-dimensional representation of the frame-transformation from the center-of-mass description to the polymer-based description. It is defined via $A_{\alpha \beta} = ( \delta_{\alpha \beta} + c_{\alpha} M_{\beta}/\uprho_{\text{p}} )$ (see \cref*{subsec:SIentropy_flux}). Furthermore, the entropy flux transforms via $    \vec{j}_{\text{s}}^{\text{p}} 
	= \vec{j}_{\text{s}} +   \uprho s /\uprho_{\text{p}}\cdot \sum_{\alpha, \beta=2}^{\ce{N}}M_{\alpha} A_{\alpha \beta}^{-1} \vec{N}_{\beta}^{\text{p}}$, and the non-thermal forces via
\begin{multline}
	\label{eqn:driving_forces_definition}
\vec{\mathcal{F}}_{\beta}^{\text{el,p}} {=} \sum_{\alpha {=} 2}^{\ce{N}}
	\left(	\vec{\mathcal{F}}_{\alpha}^{\ce{el}}
        {-}M_{\alpha}/M_{\ce{p}}\cdot
 	\vec{\mathcal{F}}_{\ce{p}}^{\ce{el}}
  	\right)A_{\alpha \beta}^{-1} 
	\\
	- M_\beta c_{\ce{p}}/ \uprho\cdot  \vec{\mathcal{F}}_{\ce{mech}}
	- M_\beta s \vec{\mathcal{F}}_{\ce{T}},
\end{multline}
where $\vec{\mathcal{F}}_{\text{T}}^{\ce{p}} = 
\vec{\mathcal{F}}_{\text{T}}$ is frame invariant. 
 Thus,  \cref{eqn:entropy_production_electrochempot} reads  
\begin{equation}
\label{eqn:entropy_production_polymer_frame}
	R_{\text{s}} =  - \sum_{\alpha=2}^{\ce{N}} \vec{N}_{\alpha}^{\text{p}} \cdot \vec{\mathcal{F}}_{\alpha}^{\text{el,p}} - \vec{j}_{\text{s}}^{\text{p}} \cdot \vec{\mathcal{F}}_{\text{T}}^{\text{p}} .
\end{equation}
Via an Onsager-Ansatz, we ensure non-negativity of $R_{\text{s}}$,
\begin{equation}
    \label{eqn:onsager_matrixform}
    \left(
    \begin{matrix}
    \vec{N}^{\ce{p}}_2
    \\
    \vdots
    \\
    \vec{N}^{\ce{p}}_{\ce{N}}
    \\
    \vec{j}_{\text{s}}^{\text{p}}
    \end{matrix}
    \right)
    = -
    \left(
    \begin{matrix}
    \mathcal{L}_{22}^{\text{p}}
    & \ldots & 
    \mathcal{L}_{2\ce{N}}^{\text{p}}
    & \mathcal{L}_{2T}^{\text{p}}
    \\
    \vdots & \ddots & \vdots & \vdots
    \\
    \mathcal{L}_{\ce{N}2}^{\text{p}}
    & \ldots & 
    \mathcal{L}_{\ce{N}\ce{N}}^{\text{p}}
    & \mathcal{L}_{\ce{N}T}^{\text{p}}
    \\
     \mathcal{L}_{2T}^{\text{p}} & \ldots 
    & \mathcal{L}_{\ce{N}T}^{\text{p}}
    & \mathcal{L}_{TT}^{\text{p}}
    \end{matrix}
    \right)    
    \cdot     \left(
    \begin{matrix}
    \vec{\mathcal{F}}_{\text{2}}^{\text{el,p}}
    \\
    \vdots
    \\
    \vec{\mathcal{F}}_{\text{N}}^{\text{el,p}}
    \\
    \vec{\mathcal{F}}_{\text{T}}^{\text{p}}
    \end{matrix}
    \right).
\end{equation}
Here, $\vec{\mathcal{L}}^{\text{p}}$ is the positive semi-definite Onsager matrix of dimension $\ce{N}{\times}\ce{N}$. Since we neglect magnetic fields, the Onsager-matrix is symmetric 
,\cite{DeGroot1984} and only N(N+1)/2 Onsager coefficients are independent. 

Because the independent Onsager coefficients determine the complete set of independent transport parameters,\cite{Schammer2021} a total of N(N+1)/2 independent transport parameters exists. We use the rationale described in Ref.~\citenum{Schammer2021} and expand the fluxes via 
\begin{gather}
    \label{eqn:current_tpform}
    \vec{\mathcal{J}}^{\ce{p}}
    = - \kappa^{\ce{p}} \cdot \grad\left( \varphi \right)
    - \suml_{\alpha=3}^{\ce{N}} \frac{t_\alpha^{\ce{p}} \kappa^{\ce{p}}}{Fz_\alpha} \cdot \tilde{\vec{\mathcal{F}}}_\alpha^{\ce{p}},
\\
    \label{eqn:fluxes_tpform}
    \vec{N}_\alpha^{\ce{p}}
    = \frac{t_\alpha^{\ce{p}}}{Fz_\alpha} \cdot \vec{\mathcal{J}}^{\ce{p}}
    - \suml_{\beta=3}^{\ce{N}} \mathcal{D}_{\alpha \beta}^{\ce{p}} \cdot \tilde{\vec{\mathcal{F}}}_\beta^{\ce{p}},
\end{gather}
where $\grad\left(\varphi\right) = \grad\left(\Phi\right) + \vec{\mathcal{F}}_2^p/Fz_2$ is the chemo-eletrical potential,\cite{Latz2015}  and $\tilde{\vec{\mathcal{F}}}_\alpha^{\ce{p}} = \vec{\mathcal{F}}_\alpha^{\ce{p}} - z_\alpha/z_2\cdot  \vec{\mathcal{F}}_2^{\ce{p}}$. The transport parameters appearing above are the ionic conductivity $\kappa^{\ce{p}}$, N-1 transference numbers $t_\alpha^{\ce{p}}$ and  diffusion coefficients $\mathcal{D}_{\alpha\beta}^{\ce{p}}$, defined by \cite{Latz2015,Schammer2021,Kilchert2022}
\begin{gather}
    \kappa^{\ce{p}}
    = \suml_{\alpha,\beta=2}^{\ce{N}} F^2 z_\alpha \mathcal{L}_{\alpha\beta}^{\ce{p}} z_\beta,
    \\
    t_\alpha^{\ce{p}}
    = \frac{F^2 z_\alpha}{\kappa^{\ce{p}}} \suml_{\beta=2}^{\ce{N}} z_\beta \mathcal{L}_{\beta\alpha}^{\ce{p}},
    \\
    \mathcal{D}_{\alpha\beta}^{\ce{p}}
    = \mathcal{L}_{\alpha\beta}^{\ce{p}}
    - \frac{t_\alpha^{\ce{p}} t_\beta^{\ce{p}} \kappa^{\ce{p}}}{F^2 z_\alpha z_\beta}.
\end{gather}
However, the transport parameters are subject to the constraints $\sum_{\alpha=2}^{\ce{N}}t_\alpha^{\text{p}}=1$, and $    \sum_{\alpha=2}^{\ce{N}} z_\alpha D_{\alpha\beta}^{\ce{p}}=0$, for $\beta {\geq}\num{ 2}$. The Onsager coefficients, and, by extension, the transport parameters, encode the kinetic transport contributions. These macroscopic quantities comprise  microscopic effects, such as energy barriers, ion-hopping along polymer chains or polymer segmental motion.\cite{Diddens2010, Bresser2019}
In contrast, the driving forces comprise thermodynamic transport contributions. These encode the tendency of the macroscopic system to reach a stable configuration (minimal free energy) of thermodynamic equilibrium.
In the section Polymer Electrolyte Model, they are derived in a consistent way from the (Helmholtz) free energy.
In contrast, in the commonly used concentrated solution theory (CST), these driving forces are obtained experimentally.
The difference between these two approaches is examined in more detail in the SI, see \cref{sec:SI_thermodynamic_transport_contributions}.
\newline

\paragraph*{\textit{\textbf{Polymer electrolyte model.}}}
\label{sec:tt_polymers}

In this section we specify our yet universal framework to polymer electrolytes. First, in the section Free Energy we state our model for the free energy density.  Next, in the section Volume Constraint, Polymer Convection and Variable Reduction For Incompressible Polymers we discuss incompressible electrolytes and determine the set of independent material variables.
Finally, in the section Independent Equations of Motion, we state the equations of motion.
\newline

\paragraph*{Free Energy.}
\label{sec:free_energy_model}

Here, we state our isothermal model for the free energy density, which closes our constitutive modelling (see \cref{eq:constitutive_equation_entropy,eq:constitutive_equation_chepot,eq:constitutive_equation_efield,eq:constitutive_equation_hfield,eq:constitutive_equation_stress}). Because we  assume constant temperature throughout the system, our model for the free energy does not comprise thermal contributions.\cite{Schammer2021} 

We model the free energy density in the mass-based description in analogy to the model  for viscous electrolytes outlined in  Ref.\citenum{Schammer2021},
\begin{align}
\nonumber
	\uprho \upvarphi_{\ce{H}} {=} &
	\sum_{\alpha=1}^{\ce{N}} c_{\alpha} \upmu_{\alpha}^{0}
	{+} \frac{1}{2} \frac{\dfield^2}{\varepsilon_0 \varepsilon_r} + \frac{\mathscr{K}}{2} \left( 1 - \sum_{\alpha=1}^{\ce{N}} \upnu_{\alpha}^0 c_{\alpha} \right)^2
	\\
	\nonumber
	& {+} R T \sum_{\alpha=1}^{\ce{N}} c_{\alpha} \ln\left( \upnu_{\alpha}^0 c_{\alpha} \right) 
  + \frac{RT}{2} \sum_{\substack{\alpha,\beta{=}1\\ \alpha\neq\beta}}^{\ce{N}} \upxi_{\alpha \beta} \upnu_{\alpha}^0 c_{\alpha} \upnu_{\beta}^0 c_{\beta} 
  \\
	& {+} G_{\ce{p}} {\cdot} \frac{  \tr( J^{{-}2/3} \ten{F}^T {\cdot} \ten{F} {-} \eye) }{2J}
	  {+} K_{\ce{p}} {\cdot} \frac{J^2 {-} 1 {-} 2 \ln J }{4J}  .\label{eq:helmholtz_free_energy_total}
\end{align}

The first term appearing in \cref{eq:helmholtz_free_energy_total} measures the constant  reference chemical potentials of the pure constituents (species).  

The second and third terms in \cref{eq:helmholtz_free_energy_total} are standard contributions, which appear also in models for viscous electrolytes.\cite{Schammer2021}. The second term in \cref{eq:helmholtz_free_energy_total} measures the electrostatic energy comprised in polarizable media, where we assume a linear relation $\dfield = \varepsilon_0 \varepsilon_{\ce{r}} \vec{E}$ with constant dielectric parameter $\varepsilon_{\ce{r}}$. The third term in \cref{eq:helmholtz_free_energy_total} accounts for volumetric energy contributions, and is defined relative to a reference configuration where the partial molar volumes take the values $\upnu_\alpha^0$,\cite{Schammer2021} and where $\mathscr{K}$ is the bulk modulus (see also the section Volume Constraint, Polymer Convection And Variable Reduction For incompressible Polymers).

The fourth and fifth terms are entropic contributions. The fourth term measures the energy of the mixing entropy and accounts for packing of the highly compressed polymer-states using volume-fractions $\upnu^0_\alpha c_\alpha$ (instead of mole-fractions $c_\alpha/c$).\cite{doi:10.1021/acs.jpcb.2c00215} 
The fifth term additionally accounts for non-ideal local species-interactions via  Flory-Huggins interaction parameters $\upxi_{\alpha\beta}$.\cite{Flory1965}
Note that the Flory-Huggins interaction parameters appearing here differ slightly from parameters $\upchi_{\alpha\beta}$ appearing in the literature,\cite{Flory1965} where both are related via $\upchi_{\alpha\beta}{=}\upxi_{\alpha\beta}\upnu_\alpha$. For a more detailed discussion of this, see the section Free Energy Parameters.

The last two terms in \cref{eq:helmholtz_free_energy_total} measure the energy of the elastic polymer matrix. Here, we use an Ogden-model for compressible rubber-like materials,\cite{Ogden1972} where $\frac{1}{2} G_{\text{p}} {\cdot} \tr( J^{-2/3} \ten{F}^T {\cdot} \ten{F} - \eye )$  accounts for isochoric deformations, \textit{i.e.} volume-preserving shearing, with a shear modulus $G_{\ce{p}}$. In addition, $\frac{1}{4} K_{\text{p}} ( J^2 - 1 - 2 \log( J ) )$ accounts for isotropic volume-expansion, with a polymer bulk-modulus  $K_{\ce{p}}$. 
Note that these two contributions are formulated in the material (Lagrange) frame. However, because our theory is formulated with respect to the external, resting frame of the laboratory (Eulerian frame), an additional factor $1/J$ is required in the formulation of the free energy density.

The chemical potentials follow from \cref{eq:constitutive_equation_chepot},
\begin{multline}
    \label{eq:chempot_eval}
    \upmu_\alpha
    = \upmu_\alpha^0  - \mathscr{K}\upnu^0_\alpha \Bigl(
        1 {-} \sum_{\beta=1}^{\ce{N}} \upnu_\beta^0 c_\beta
    \Bigr)
    \\
    + RT\Bigl(
        1 {+}\ln(\upnu_\alpha^0 c_\alpha)
        {+} \upnu_\alpha^0 \sum_{\beta\neq \alpha}^{\ce{N}}
        \upxi_{\alpha\beta} \upnu_\beta^0 c_\beta
    \Bigr).
\end{multline}
The polymer stresses in $\ten{\sigma}$ follow from \cref{eq:constitutive_equation_stress} and determine the total stress $\ten{\uptau}$ (see \cref{eq:const_eq_for_tau} and \cref*{subsec:SI_evaluation_driving_forces})
\begin{multline}	\label{eqn:generalized_stress_volpar}
	\ten{\uptau} =  \ten{\uptau}^{\text{mech}} + \maxwellstress
	 + \Biggl( 
	 \frac{\mathscr{K}}{2} \Big(
	    1 - \Bigl[
	        \sum_{\alpha=1}^{\ce{N}} c_\alpha \upnu_\alpha^0
	    \Bigr]{}^2 \Bigr)-RTc
	    \\
	 - RT\frac{1}{2} \sum_{\substack{\alpha,\beta{=}1\\ \beta\neq \alpha}}^{\ce{N}} \upxi_{\alpha\beta}\upnu_\alpha^0 c_\alpha \upnu_\beta^0 c_\beta
	  \Biggr) \eye,
\end{multline}
where 
\begin{equation}
		\label{eq:mechanical_stress}
    \ten{\uptau}^{\ce{mech}}
    =
    \frac{G_{\ce{p}} }{J^{5/3}}\left( 
    \ten{F} {\cdot} \ten{F}^T
    { - } \frac{  \tr(  \ten{F}^T {\cdot} \ten{F})}{3} \eye
    \right)
    + K_{\ce{p}}\frac{J^2 {-} 1 }{2 J} \eye
\end{equation}
comprises all mechanical contributions. The pressure is determined by the stress tensor $\ten{\uptau}$ via
\begin{equation}
	\label{eq:pressure_general}
	p =  - \tr\left( \ten{\uptau} \right) / 3,
\end{equation}
and thus comprises thermodynamic, mechanical and electrostatic contributions (see also \cref*{subsec:SI_mechanochemical_coupling}).
\newline

\paragraph*{Volume Constraint, Polymer Convection And Variable Reduction For Incompressible Polymers.}
\label{sec:incompressible_elyte}

Next, we state the equation for the polymer convection velocity and discuss microscopic compressibility of our electrolyte-model. For a more detailed derivation, we refer to Ref.~\citenum{Schammer2021} and to  the SI (see \cref*{subsec:SI_convection_incompressibility}).

From now on, we assume the incompressible limit $\mathscr{K}{\to}\infty$, where the partial molar volumes do not depend on the pressure (\textit{i.e.} $\upnu_\alpha^0{=}\upnu_\alpha$).
In this limit, the convection velocity is determined by a differential equation which involves only spatial derivatives,\cite{Schammer2021}
\begin{equation}
	\label{eq:polymer_convection}
	\div(\vec{v}_{\text{p}}) =  - \sum_{\alpha=2}^{\ce{N}} \upnu_{\alpha} \cdot \div(\vec{N}_{\alpha}^{\text{p}}) + \sum_{\alpha=1}^{\ce{N}}\upnu_\alpha r_\alpha.
\end{equation}

Charge continuity, $\varrho^{\ce{F}}{=}\sum_{\alpha=1}^{\ce{N}}Fz_\alpha c_\alpha$, the Euler equation for volume, $\sum_{\alpha=1}^{\ce{N}}c_\alpha\upnu_\alpha{=}1$ and $\det\left(\ten{F}\right) \upnu_{\ce{p}} c_{\ce{p}}{=}1$ (see \cref{eqn:polymer_deformation_concentration}), constitute three constraints for the species concentrations. These constraints imply that only N-3 independent species concentrations $c_4,\ldots , c_{\ce{N}}$ exist.
This reduces the set of independent material variables for the free energy density. By convention, for $\upvarphi_{\ce{H}}(\varstaset_1)$, we choose 
\begin{equation}
\label{eq:independent_materiallaw}
	\varstaset_1 = \left\{ \ten{F},  \varrho^{\text{F}}, c_{4},\ldots , c_{\ce{N}}, \Phi \right\}.
\end{equation} 

We use charge continuity, \cref{eqn:flux_frame_transformation}, and the property $\vec{N}_{\ce{p}}^{\text{p}}=0$ to reduce the set of independent fluxes to $(\vec{\mathcal{J}}^{\text{p}},\vec{N}_{3}^{\text{p}},{\ldots},\vec{N}_{\ce{N}}^{\text{p}} )$. Hence,   \cref{eq:polymer_convection} becomes,
\begin{multline}
\label{eq:polymer_convection_reducedform}
	\div(\vec{v}_{\text{p}}) {=} {-} \frac{\upnu_{2}}{F z_{2}}  \div(\vec{\mathcal{J}}^{\text{p}}) {-} \sum_{\alpha=3}^{\ce{N}} \left( \upnu_{\alpha} {-} \frac{z_{\alpha}}{z_{2}} \upnu_{2} \right) \div(\vec{N}_{\alpha}^{\text{p}})
	\\
    + \sum_{\alpha=1}^{\ce{N}} \upnu_\alpha r_\alpha.
\end{multline}
Note that $\div(\vec{\mathcal{J}}^{\ce{p}}){=}F\sum_{\alpha=1}^{\ce{N}} z_\alpha r_\alpha$ for electroneutral electrolytes.\cite{Latz2011}
\newline

\paragraph*{Independent Equations of Motion.}
\label{sec:ind_eq_of_mot}

Finally, we state the equations of motion for our polymer description.

The stress tensor and the chemical potentials (see \cref{eqn:generalized_stress_volpar,eq:mechanical_stress,eq:chempot_eval}) are yet not fully closed, as they still depend on unknown pressure-like forces involving the bulk modulus $\mathscr{K}$. We resolve this issue based on the kinematical approach described in Ref.~\citenum{Schammer2021}, and assume that the elastic  polymer-matrix absorbs momentum diffusion via highly effective dissipation, \textit{i.e.}  $\div\left( \ten{\uptau} \right)=0$. This assumption of  mechanical equilibrium  implies a coupling between the stresses and the chemical potentials which closes our description. In the SI (see \cref*{subsec:SI_mechanochemical_coupling}), we evaluate this coupling to incorporate elastic contributions into the chemical potentials, \textit{i.e.} the transport equations. Altogether, the set of transport equations reads, 
\begin{gather}
	\label{eq:summary_eom_F}
	\partial_t \ten{F} =  \grad\left( \vec{v}_{\text{p}} \right) \cdot \ten{F} - \vec{v}_{\text{p}} \cdot \grad\left( \ten{F} \right),
	\\
	\label{eq:summary_eom_charge}
	\partial_t \varrho^{\text{F}} =  - \div(\vec{\mathcal{J}}^{\text{p}})
	 - \div\left(\varrho^{\text{F}} \vec{v}_{\text{p}} \right)
  + F \sum_{\alpha=1}^{\ce{N}} z_\alpha r_\alpha,
	\\
	\label{eq:summary_eom_conc}
 \partial_t c_{\alpha} =  - \div(\vec{N}_{\alpha}^{\text{p}})
	- \div\left(c_\alpha \vec{v}_{\text{p}}\right)
 + r_\alpha, \quad \textnormal{for} \ \alpha \geq 4
	\\
	\label{eq:summary_eom_poisson}
	- \varepsilon \div(\grad( \Phi )) = \varrho^{\text{F}},
\end{gather}
where the fluxes are given by \cref{eqn:current_tpform,eqn:fluxes_tpform,eq:polymer_convection_reducedform}.

\section*{Validation: Simulation of a Li Cell With Polymer Based Electrolyte}
\label{sec:simulation}

In this section, we validate our transport theory for polymer electrolytes. 

For this purpose, we focus on the work described by Steinrück and co-workers in the publication Ref.~\citenum{Steinrueck2020}, and investigate the polymer based electrolyte mixture composed of PEO and LiTFSI salt in a symmetric Li-metal cell.
We compare our numerical results with the experimental data and with numerical results (based on an alternative theory) stated by Steinrück and coworkers.
Their approach combines experimental methods with theoretical methods based on continuum modeling and molecular dynamics (MD).
For their continuum model, the authors use concentrated solution theory (CST), which differs conceptually from our approach.
In our transport theory, the thermodynamic transport contributions are given by the driving forces, which are derived from the free energy, while in CST they are consolidated in a thermodynamic factor, which is fitted to reproduce experimental results.\cite{Pesko2018}

\begin{figure}[!htb]
    \centering
    \includegraphics[width=0.8\columnwidth]{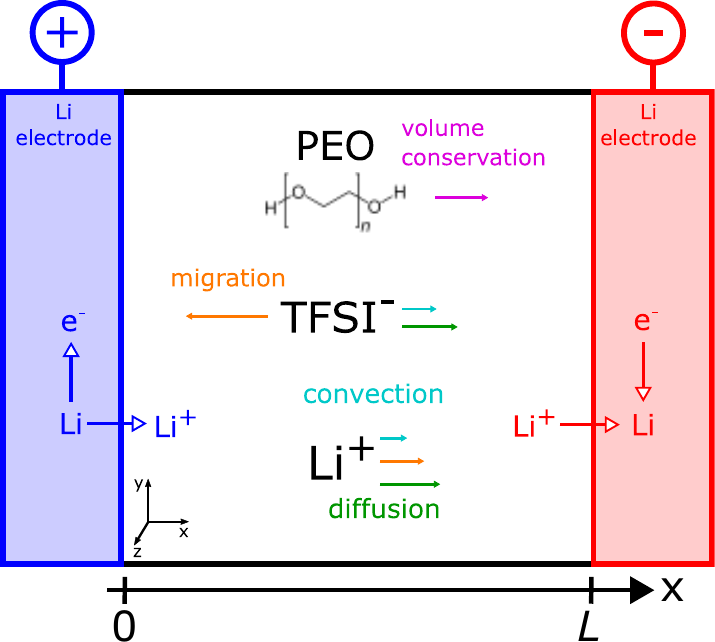}
    \caption{Illustration of the symmetric cell setup with two Li-metal electrodes and  PEO/LiTFSI polymer electrolyte. The length of the cell is $L{=}\SI{3}{\milli\meter}$. The migration of anions creates concentration gradients, which in turn induce diffusion and convection fluxes.}
    \label{fig:peo_setup}
\end{figure}

The system polyethylene glycole (PEO) with LiTFSI salt is often used as "benchmark" polymer for evaluation and comparison with alternate polymer electrolytes, and is among the most extensively investigated polymer electrolytes, both with respect to experimental methods and with respect to theoretical models.\cite{Wright1975, Wen2003, Diddens2010, Steinrueck2020}
Due to its widespread use, polymer electrolytes with PEO as host material are  parametrized very well, which makes them attractive as reference for continuum-scale simulations.\cite{Steinrueck2020}
In addition, PEO has a relatively simple molecular structure, which facilitates atomic scale investigations.\cite{Diddens2010, Thum2021}.

We assume a symmetric cell set-up consisting of two Li-metal electrodes with the polymer electrolyte in between (see \cref{fig:peo_setup} for an illustration). At the more positive electrode (left electrode in \cref{fig:peo_setup}), the surface reaction 
\begin{gather}
    \label{eq:surface_rct_Li_prdctn}
    \ce{Li -> Li^+ + e^-}
    \intertext{and at the negative electrode (right electrode in \cref{fig:peo_setup})}
    \label{eq:surface_rct_Li_destruction}
    \ce{Li^+ + e^- -> Li}
\end{gather}
In this work, we neglect degradation processes occurring at the electrodes (\textit{e.g.}, SEI formation). 

First, in the section One Dimensional Model Equations, we state the one-dimensional equations.
Second, in the section Potentiostatic Discharge Simulation, we present the results obtained from a potentiostatic discharge simulation, and validate our results by a comparison with literature.\cite{Steinrueck2020}
\newline

\paragraph*{\textit{\textbf{One dimensional model equations.}}}
\label{sec:model_equations}

We assume complete dissociation of the salt ions, which yields a number of three electrolyte species.
Furthermore, we reduce our description to one spatial dimension on the \textmu m-scale, where we can safely assume electroneutrality.\cite{Latz2015}
Thus, the set of independent material variables $\varstaset_{\ce{PEO}}$ consists only of the two variables $\ten{F}$ and $\Phi$, see \cref{eq:independent_materiallaw}. 
Since  $\ten{F}=\left(\begin{smallmatrix}J&0&0\\ 0&1&0\\0&0&1\end{smallmatrix}\right)$, 
 the deformation gradient is completely determined by the volume ratio (see SI)
\begin{equation}
\label{eq:const_eq_J}
	 J 
  =\frac{1 }{\cpeo\upnu_{\p}}=\frac{1 }{1-  \upnu_{\litfsi} \cli},
\end{equation}
with $\upnu_{\litfsi}{=}\upnu_{\li}{+}\upnu_{\tfsi}$. Hence, we replace $\ten{F}$ by $\cli$ as independent variable such that $\varstaset_{\ce{PEO}} {=} \{ \cli,\Phi \}$.
The set of transport equations reads (see \cref{eq:summary_eom_F,eq:summary_eom_charge,eq:summary_eom_conc,eq:summary_eom_poisson})
\begin{gather}
    \label{eqn:equations_of_motion_concentration}
    \partial_t \cli = - \div{(\vec{N}_{\ce{Li}}^{\text{p}})} - \div{(\cli \Vec{v}^{\p})} + r_{\ce{Li}},
	\\
	\label{eqn:equations_of_motion_charge}
	0 = - \div{(\vec{\mathcal{J}}^{\text{p}})} + Fr_{\ce{Li}},
 \intertext{with the polymer velocity given by (see \cref{eq:polymer_convection_reducedform})}
    \label{eqn:1d_eom_vel}
	\div( v_{\p} ) = \frac{\upnu_{\tfsi}}{F} \div\left(\vec{\mathcal{J}}^{\p}\right) - \upnu_{\litfsi} \cdot \div( N_{\li}^{\p} ) + \upnu_{\li}r_{\li}. 
\end{gather}
We model the reactions at the Li electrode surfaces via source-terms based on a Butler-Volmer-Ansatz (see \cref*{subsec:SI_parameters}). The fluxes appearing above are determined by \cref{eqn:current_tpform,eqn:fluxes_tpform}, where the driving force $\tilde{\vec{\mathcal{F}}}_{\li}^{\ce{p}}$ is a function of $\cli$ alone.
However, by using \cref{eqn:equations_of_motion_charge} in \cref{eqn:1d_eom_vel}, it follows that the flux of the Li-ions becomes $\div\left(\vec{N}_{\li}^{\p}\right)=r_{\li} - \div\left(\vec{v}^{\p}\right)/\upnu_{\litfsi}$.
Altogether, \cref{eqn:equations_of_motion_concentration} thus reads
\begin{equation}
    \label{eq:peo_eom_cli_volfracform}
    \partial_t \cli 
    = \frac{1}{\upnu_{\litfsi}} \div\left(\upnu_{\p}c_{\p}\vec{v}^{\p}\right),
\end{equation}
where $\cpeo\upnu_{\p}=1/J$, see \cref{eq:const_eq_J}. Thus, the evolution of the Li-concentration is completely determined by the volume-flux of the polymer species. 

Alternatively, we find 
\begin{multline}
    \label{eq:peo_eom_cli_final}
    \partial_t \cli 
    =
     \upnu_{\ce{p}}c_{\ce{p}}\big[(1-t_{\ce{Li}}^{\ce{p}})r_{\ce{Li}} 
    - \vec{\mathcal{J}}^{\ce{p}}/F\cdot \nabla t_{\ce{Li}}^{\ce{p}}
    \\
    + \nabla \cdot (\mathcal{D}_{\ce{Li}}^{\ce{p}} \tilde{\vec{\mathcal{F}}}^{\ce{p}}_{\ce{Li}})\big] -(\vec{v}^{\ce{p}} \cdot \nabla)\cli.
\end{multline}
This is equivalent to the form $\partial_t\cli=(1-t_{\li}^{\p})r_{\li}\upnu_{\p}c_{\p}- (\hat{\mathbb{D}}_{\li}\cdot \nabla)\cli$, where $\hat{\mathbb{D}}_{\li}(\upnu_\alpha,r_{\li},t_{\li}^{\p},\mathcal{D}_{\li}^{\p})$ is an operator-valued diffusion parameter. Hence, the stationary state constitutes an equilibrium of the surface reactions with the polymer deformation and diffusive Li-transport, $(1-t_{\li}^{\p})r_{\li}\upnu_{\p}c_{\p}= (\hat{\mathbb{D}}_{\li}\cdot \nabla)\cli$. Because, in general, $t_{\li}^{\p}{\neq}\num{1}$ and $\cpeo\upnu_{\p}{\neq}\num{0}$, the occurrence of concentration polarization is to be expected in the stationary state.

In our approach based on the modeling of the free energy $\uprho \upvarphi_{\ce{H}}$, we account for the specific electrolyte characteristics (e.g. mechanics, species interactions) via explicit contributions to $\uprho \upvarphi_{\ce{H}}$ (see \cref{eq:helmholtz_free_energy_total}). However, in the literature, the resulting "excess" contributions to the chemical potentials $\upmu_\alpha=\upmu_\alpha^0+RT\ln\left(f_\alpha c_\alpha\right)$ are often collected cumulatively in one single phenomenological parameter via the molar activity coefficients $f_\alpha$.\cite{Newman2004}
The corresponding expression for the molar activity coefficients resulting from our model reads (see  \cref{eq:chempot_eval,eqn:SI_gradchepot})
\begin{multline}
    f_\alpha = \upnu_\alpha \cdot
    \exp\Bigg[ \suml_{\beta=1}^{\ce{N}}\left(\upnu_\beta-\upnu_\alpha\right)c_\beta
    \\
    + \frac{1}{2}\sum_{\substack{\beta,\gamma{=}1\\ \beta\neq\gamma}}^{\ce{N}}\upxi_{\beta\gamma}\upnu_\beta\upnu_\gamma\left(\delta_{\alpha\beta}c_\gamma+\delta_{\alpha\gamma}c_\beta-\upnu_\alpha c_\beta c_\gamma\right)
    \\
    +\frac{2}{3}\upnu_\alpha\frac{G_{\ce{p}}}{RT}\left(J^{\frac{1}{3}}-J^{-\frac{5}{3}}\right)
    +\frac{1}{2}\upnu_\alpha\frac{K_{\ce{p}}}{RT}\left(J-J^{-1}\right)\Bigg].
\end{multline}
Activity coefficients can be obtained from experiment,\cite{Pesko2017,Zhang2023} or atomistic simulations.\cite{Lazaridis1993,Ingenmey2019,Fang2021}
An agnostic approach which casts all excess contributions into one single empirical parameter can be beneficial for practical purposes.
However, our explicit modeling which relies on multiple empirical parameters (e.g. the elastic moduli of the polymer, or  the Flory-Huggins interaction parameters) has the advantage that each excess contribution can be investigated separately and may yield insights into the influence of material specific properties on the overall electrolyte performance.
Furthermore, each parameter can be rigorously obtained from first principles, i.e. atomistic modeling.
Altogether, our approach constitutes a rational approach to the investigation of the cell performance by understanding the behaviour of the polymer electrolyte (see also the section Free Energy Parameters).
We assume constant bulk and shear moduli of the electrolyte ($K_{\ce{p}} {=} \SI{8}{\mega\pascal}$ and $G_{\ce{p}} {=} \SI{3}{\mega\pascal}$),\cite{Jee2013, Ushakova2020, Lee2022}. 
Because we model all interactions between the electrolyte species we have a total of three Flory-Huggins parameters, which we set to $\chi_{\li\tfsi} {=} \num{-6}$, $\chi_{\li\p} {=}\num{-20}$ and $\chi_{\tfsi\p} {=} \num{-15}$.\cite{Nikolic2013}
Due to our different definition of the Flory-Huggins parameters $\upchi_{\alpha\beta}{=}\upxi_{\alpha\beta}\upnu_\alpha$, these values for $\upchi_{\alpha\beta}$ correspond to $\upxi_{\li\tfsi}{\approx}\SI{-5.41e6}{\mol\per\cubic\metre}$, $\upxi_{\li\p}{\approx}\SI{-1.81e7}{\mol\per\cubic\metre}$, and $\upxi_{\tfsi\p}{\approx}\SI{-1.12e5}{\mol\per\cubic\metre}$.
As can be seen by these values, the interaction of the lithium cations with the polymer species is two orders of magnitude larger than that of anions with the polymer.
This corresponds to the strong interaction of lithium with PEO via the crown ether structure.\cite{Ding2023}

In the section Free Energy Parameters, we discuss this parameter choice in more detail and investigate the influence of the bulk and shear moduli and the Flory-Huggins parameters on the cell performance.
Only three independent transport parameters exist in this mixture.
These are the ionic conductivity, one transference number and one diffusion coefficient.
For polymer electrolytes, the exact nature of the microscopic transport mechanisms has important consequences for the overall performance.\cite{Maitra2008}
However, we parametrize them based on experimental results presented in Ref.~\citenum{Steinrueck2020}.
Therefore, we assume that all relevant transport processes are included.
The transfer of the transport parameters from CST to our theory is described in more detail in \cref*{subsec:SI_comparison_cst}, and a complete overview of our parametrization of the system is given in \cref*{subsec:SI_parameters}. In \cref*{sec:SI_simulation} we specify our numerical methods. 
\newline

\paragraph*{\textit{\textbf{Potentiostatic discharge simulation.}}}
\label{sec:simulations}

In this section we present the numerical results of our potentiostatic simulations for the as-modelled Li-cell. First, in the section Numerical Results, we discuss our simulation results, and, based on these observations, we give a detailed analysis of the electrolyte dynamics during discharging the cell.
Second, in the section Validation, we compare our numerical results with experimental results and with numerical results from Ref.~\citenum{Steinrueck2020}.
Third, in the section Free Energy Parameters, we discuss the influence of the parameters appearing in the free energy on the resulting current density and concentration distribution.
\newline

\paragraph*{Numerical Results.}
\label{sec:numerical_results}

\begin{figure}[!htb]
    \centering
    \includegraphics[width=\columnwidth]{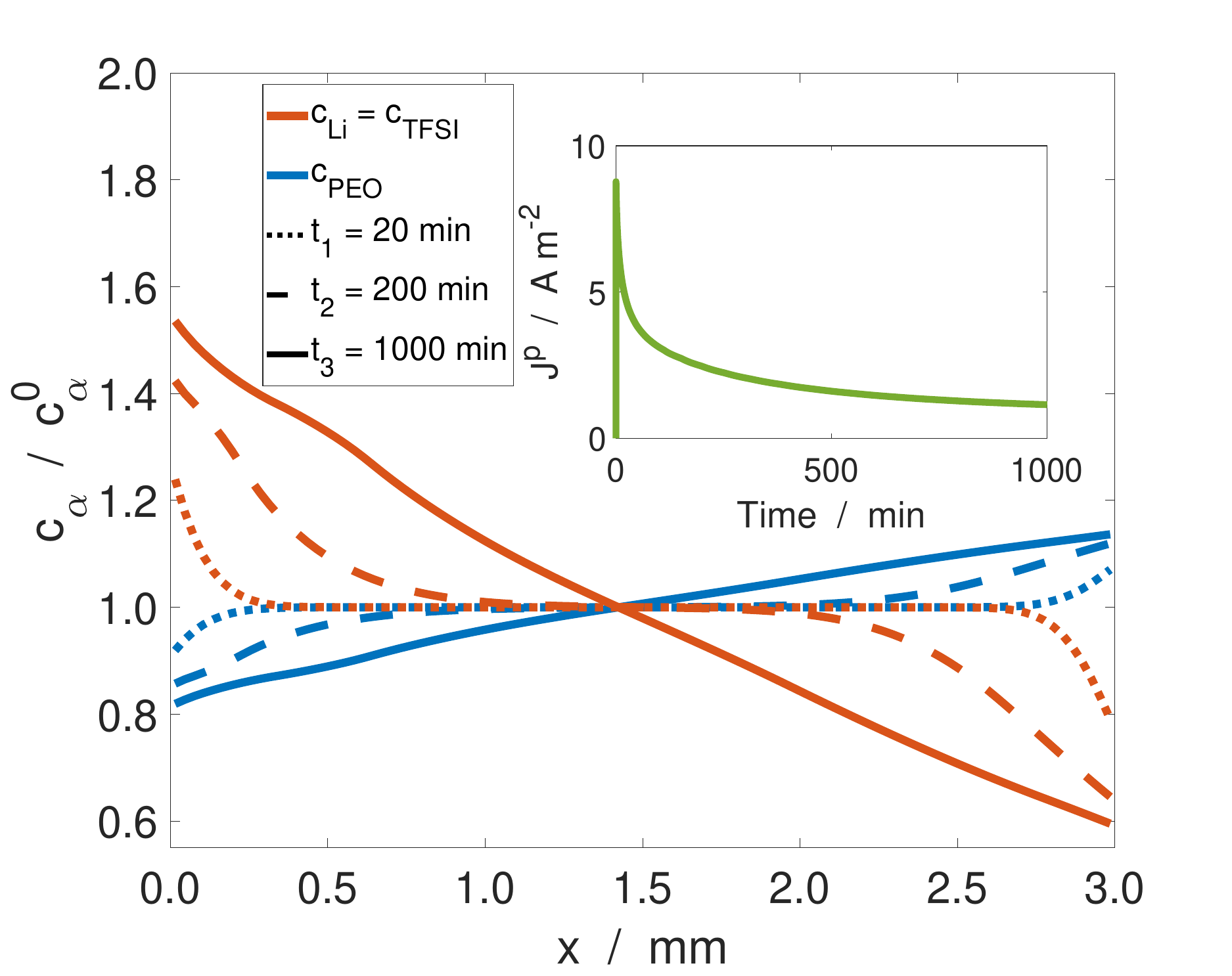}
    \caption{Species concentrations for all three electrolyte species at different time steps, normalized by their initial concentration $c_\alpha^0=c_\alpha(t{=}0)$. Here, the color indicates the species, and the line type indicates the time steps $t_i$. Note that $\cli=\ctfsi$ due to our assumption of electroneutrality.  The green line in the inset shows the current density (green) as function of discharge time.}
    \label{fig:peo_results_DOM_concentrations}
\end{figure}

In our potentiostatic discharge simulation, we apply a constant potential difference of \SI{0.3}{\volt} between the positive electrode (left electrode in \cref{fig:peo_setup}) and the negative electrode (right electrode in \cref{fig:peo_setup}). 

\Cref{fig:peo_results_DOM_concentrations} shows the concentration profiles of the electrolyte species, and the inset  shows the profile of the discharge current $\vec{\mathcal{J}}^{\text{p}}$.
In the beginning,  a dynamical phase can be observed (up to \SI{80}{\minute}), during which the current  decays exponentially from its initial peak at  $\vec{\mathcal{J}}^{\text{p}}\approx\SI{8}{\ampere\meter\squared}$.
This dynamical phase is followed by a relaxation phase, during which the electrolyte dynamics,   as indicated by $\vec{\mathcal{J}}^p$, 
 slows down with time. Hence, the system is approaching a stationary state.
The concentration profiles $c_\alpha(x,t_i)$ of the three electrolyte species are shown  at three different characteristic times $t_i$, and are normalized by their initial value $c_\alpha^0{=}c_\alpha( t{=}0)$.
Note that due to the constraint of electroneutrality, the profiles of the ionic species are exactly the same for all times, \textit{i.e.} $\cli{=}\ctfsi$ (red curves).
The first time $t_1{=}\SI{20}{\minute}$ represents the dynamical phase of enhanced electrolyte dynamics. Apparently, during this phase, concentration gradients develop near the electrodes, where the concentrations of the ions increase at the positive (left) electrode and decrease at the negative (right) electrode. 
Over time, this concentration polarization extends further into the electrolyte, see the profile at $t_2{=}\SI{200}{\minute}$, which  corresponds to the relaxation phase of the electrolyte.
Eventually, at $t_3{=}\SI{1000}{\minute}$ (end of discharge), the zones of  accumulation and depletion of the ions extend throughout the complete bulk, and form almost a constant concentration gradient between the electrodes.
The polymer species displays a similar, but inverse behaviour, with a concentration polarization in the opposite direction.

\begin{figure}[!htb]
    \centering
    \includegraphics[width=\columnwidth]{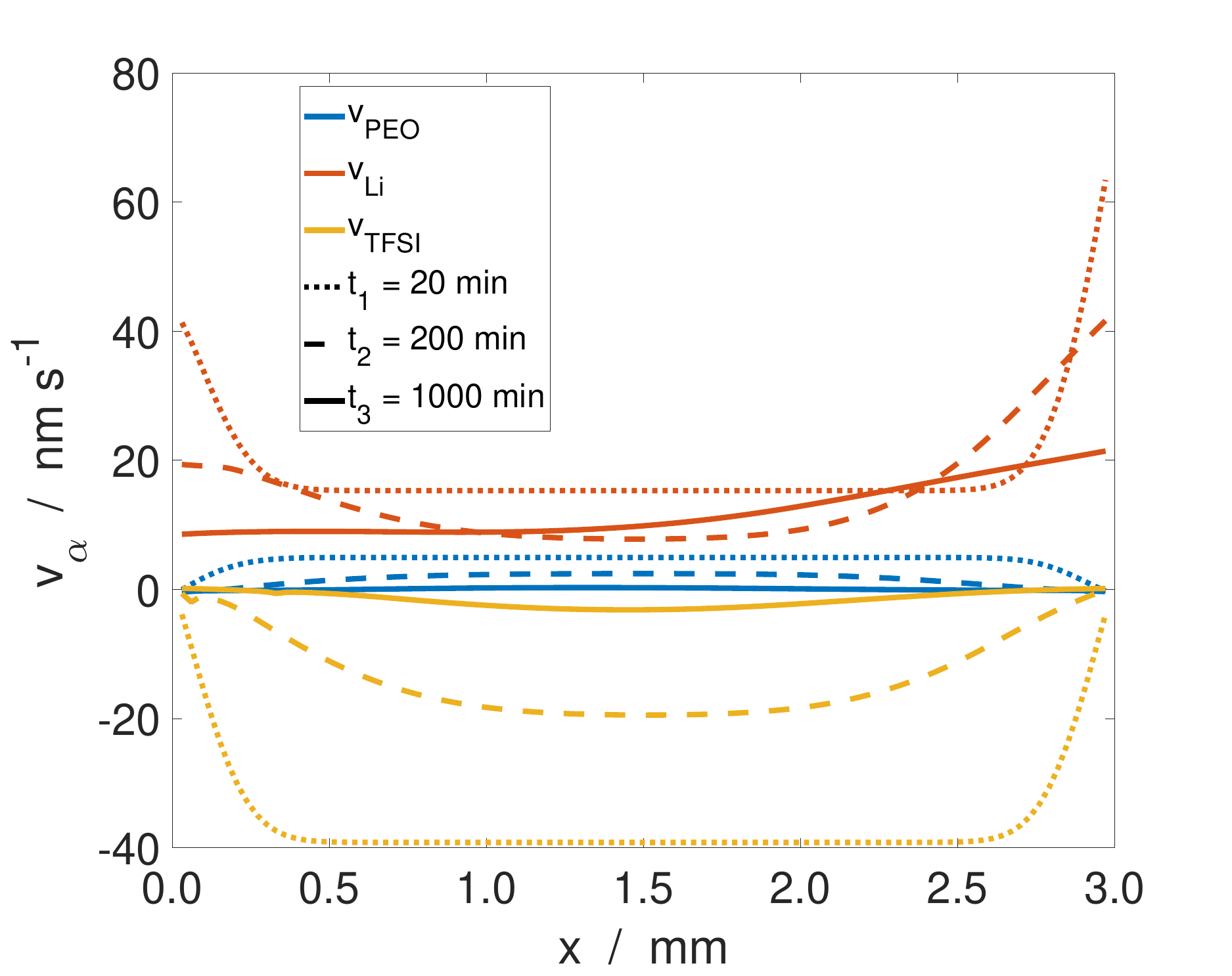}
    \caption{Species velocities at different discharge times. The different line types indicate the time step $t_i$, and the colors indicate the different species. }
    \label{fig:peo_results_DOM_velocities}
\end{figure}

The corresponding species velocities are shown in \cref{fig:peo_results_DOM_velocities}.
During the complete discharge time, the anions move towards the more positive electrode at the left (yellow curves), whereas the polymer (blue curves) and the Li-ions (red curves) move towards the more negative electrode at the right.
However, for all three species, the magnitudes of the species velocities are enhanced during the dynamical phase (dotted lines), and become more relaxed with increasing discharge time (dashed and solid curves).

\begin{figure}
    \centering
    \includegraphics[width=\columnwidth]{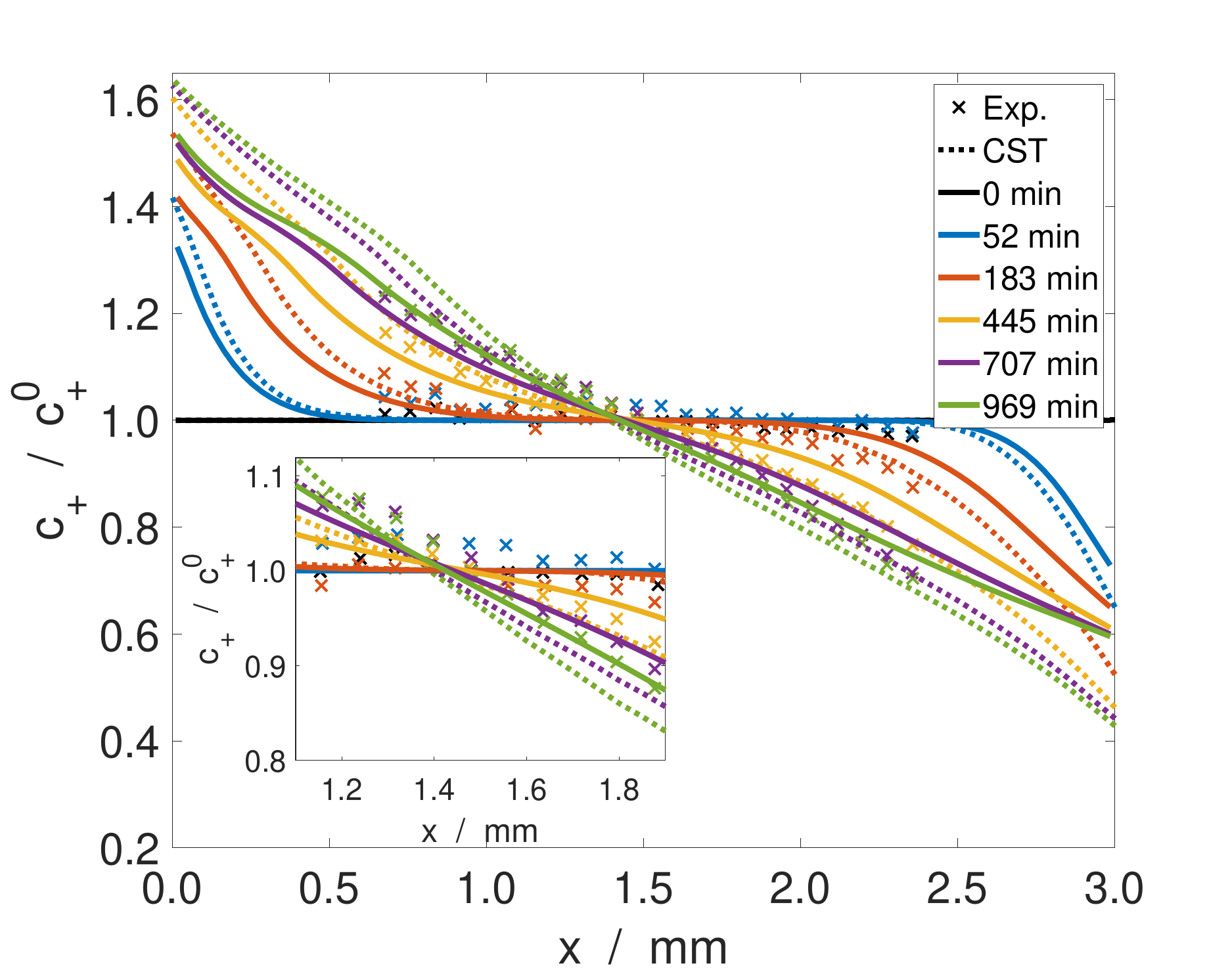}
    \caption{Comparison of our numerical results for the species concentrations (solid lines) with the experimental results (crosses) and the numerical results obtained from CST (dash line) both taken from Ref.~\citenum{Steinrueck2020}. The inset focuses on the middle of the electrolyte ($x=\SI{1.5}{\milli\meter}$).}
    \label{fig:peo_results_concentration}
\end{figure}

The slight gradient of $v_{\li}(t_3)$ is a result of the concentration gradient for $\cli$, as shown in \cref{fig:peo_results_concentration}.

The profiles shown in \cref{fig:peo_results_DOM_concentrations,fig:peo_results_DOM_velocities} are not perfectly linear and not symmetric (with respect to the position $x{=}\SI{1.5}{\milli\meter}$).
This is because the transport parameters depend on the species concentrations, $t_{\ce{Li}}^{\ce{p}}(\cli)$) and $\mathcal{D}^{\ce{p}}_{\ce{Li}}(\cli)$, which influences the delicate relation between migration and diffusion.

Overall, we observe concentration polarization for all three electrolyte species. Indeed, for ionic species, such a behaviour is typical for SPEs with two mobile ions and a small transference number $t_{\li} < 1$, and is well-described in the literature.\cite{Stolz2022,STOLZ20219}

However, the role of the neutral solvent-like polymer species in such systems has not yet been intensely discussed.

Our results show that the effect of concentration polarization of the ions is accompanied by a deformation of the polymer.
This deformation originates from the volume-preserving fluxes of the electrolyte species.
The accumulation of TFSI- and Li-ions near the positive electrode implies a volume-flux of  $\ctfsi\upnu_{\ce{TFSI^-}}$ and $\cli\upnu_{\li}$.
This is compensated by a volume-flux $c_{\p}\upnu_{\p}$ of the polymer, towards the opposite direction,  {\it i.e.} a polymer-deformation,  which ensures the volumetric constraint \cref{eqn:polymer_deformation_concentration}.

\begin{figure}[!htb]
    \centering
    \includegraphics[width=\columnwidth]{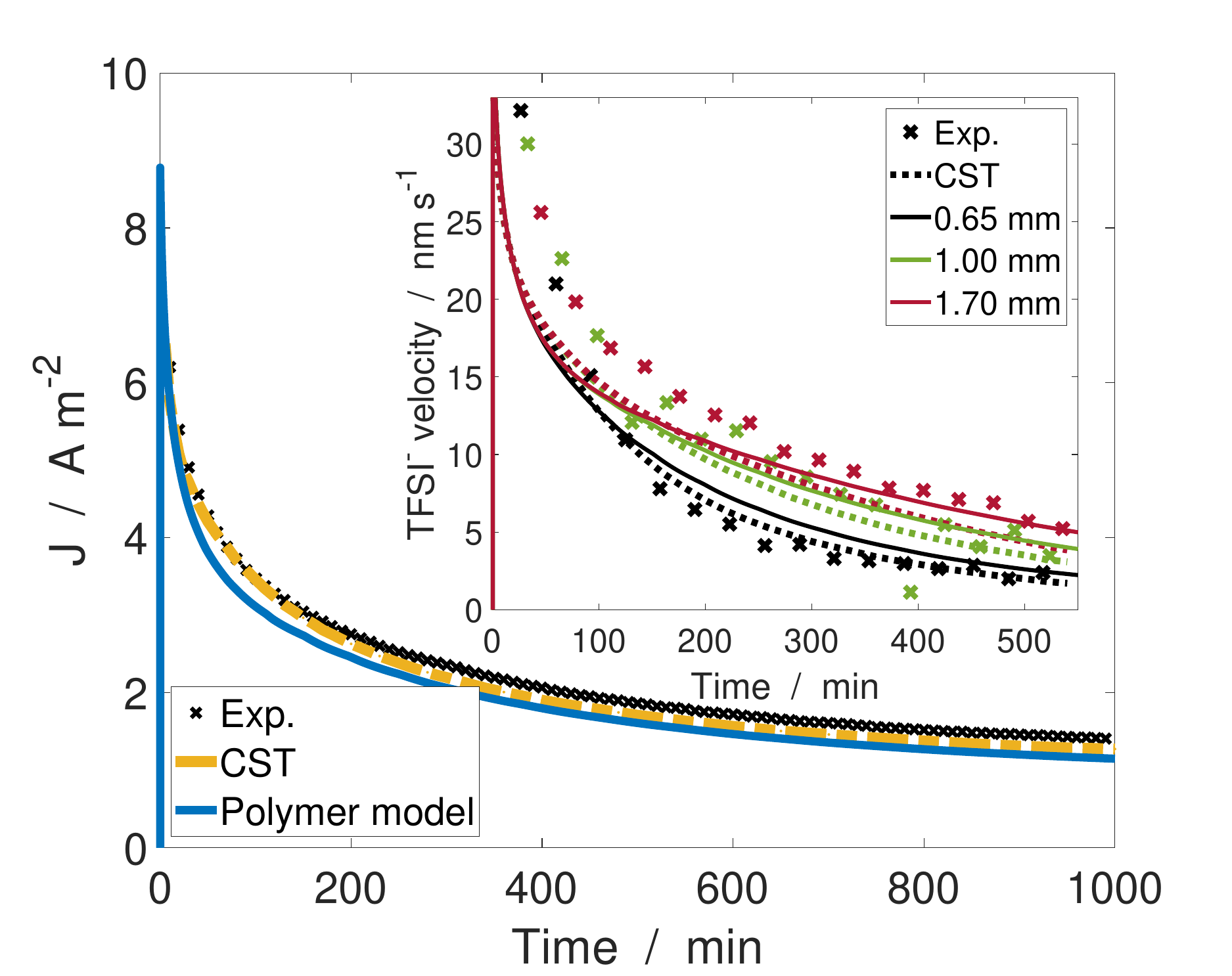}
    \caption{Comparison of our numerical results (blue line) for the current density over discharge time with experimental results (black crosses) and numerical results based on CST (yellow dashed line), both taken from Ref.~\citenum{Steinrueck2020}.  The inset shows the corresponding results for the species velocity $\vec{v}_{\tfsi}$ at three different locations in the electrolyte (colours). The numerical results obtained from our simulation are shown in solid lines, the crosses show the experimental results and the dashed lines show the CST results.}
    \label{fig:peo_results_current_velocity}
\end{figure}

\paragraph*{Validation.}
\label{sec:validation_detail}
In this section, we validate our description by comparing our numerical results with experimental and numerical results presented in Ref.~\citenum{Steinrueck2020}. 

In \cref{fig:peo_results_current_velocity}, we compare the results for the electric current density and species velocities.
The blue line illustrates our numerical results for the current density, whereas the yellow dashed line and the black crosses illustrate the numerical and the experimental results from Ref.~\citenum{Steinrueck2020}, respectively.
The inset in \cref{fig:peo_results_current_velocity} shows the velocity profiles of the TFSI-ions at different positions during discharge.

Our results for the current density show a decay over time which is very similar to the decay exhibited by the experimental results. However, the numerical results are slightly shifted to smaller values by a constant offset during the complete discharge time. In contrast, the deviation of the CST results from experiment varies over discharge time, where, initially, both results agree very well. Overall, our description reproduces the shape of the experimental results slightly better than CST.

 The inset in \cref{fig:peo_results_current_velocity} illustrates the results for the anion velocity $v_{\ce{TFSI^-}}(x_i,t)$, as function of time at three different positions $x_i$ (indicated by colors), 
as obtained from our numerical simulations (solid lines), from the experiments (crosses), and from the CST-simulations (dotted lines).
In the first \SI{100}{\minute}s, both CST and our transport model show a more rapid decrease of the $\ce{TFSI}^-$ velocity than the experiment. After that, CST reproduces the experimental values near the electrode very well, while our model reproduces the experimental values in the center of the electrolyte very well.

\Cref{fig:peo_results_concentration} shows a comparison of the Li-concentration profiles at different times (distinguished by the different colors) obtained from experiment (crosses), CST (dashed lines) and from our description (solid lines). 
The inset highlights the central part of the electrolyte for better clarity.
Up to $t = \SI{183}{\minute}$, CST underestimates the concentration polarization in comparison to the experimental measurements, while for larger times it overestimates it.
For nearly all times, our transport model shows a slower development of the concentration polarization than the experimental results and CST.
For the latest time of $\SI{969}{\minute}$ (green), the concentration profile from our transport model fits the experimental results better than CST, especially for the larger concentrations near the positive electrode.

Apparently, there exist some deviations between the numerical results obtained from our description, and from the CST description. The main difference between the two lies in the thermodynamic contribution to the species fluxes.
In CST, this contribution is constituted by the concentration-dependent thermodynamic factor comprised in $\partial \upmu / \partial c \cdot \grad\left( c \right)$, and was obtained by fitting to the experimental results.\cite{Pesko2018,Steinrueck2020}
This approach differs from our approach described in \cref{eqn:driving_forces_definition}. The driving forces $\partial \upmu / \partial c \cdot \grad\left( c \right)$ appearing in our transport model are derived from fundamental physical considerations.
In \cref*{sec:SI_thermodynamic_transport_contributions} we present a detailed discussion of the thermodynamic transport contributions.

To summarize, our results are in  good agreement with the experimental and numerical results presented in Ref.~\citenum{Steinrueck2020}, which validates our theoretical description.
\newline

\paragraph*{Free energy parameters.}
\label{subsubsec:freeenergyparameters}

In this section, we discuss the influence of the Flory-Huggins parameters and the elastic modulus of the polymer on the cell performance.
To address this goal, we focus our discussion on the current density and concentration profiles for the PEO/LiTFSI electrolyte.

First, we discuss the influence of the Flory-Huggins interaction parameters.
The interaction contributions to the free energy as derived by Flory are of the form $RT \chi_{\alpha\beta}c_\alpha\phi_\beta$.\cite{Flory1965}
Here, $\phi_\alpha{=}\upnu_\alpha c_\alpha$ is the volume fraction of species $\alpha$, and $\chi_{\alpha\beta}$ is the interaction energy between the species $\alpha$ and $\beta$.
Importantly, his lattice-based derivation assumes that all species occupy an equal amount of volume, and therefore the partial molar volumes are the same. 
This can be a bad approximation.
For example, the partial molar volumes of the species in the PEO-based electrolyte and the SIC electrolyte discussed in this work are very different (see \cref{subsec:SI_parameters}).
In contrast, our model explicitly accounts for the partial molar volumes of the electrolyte species via interaction contributions to the free energy of the form $RT \upxi_{\alpha \beta} \upnu_{\alpha}^0 c_{\alpha} \upnu_{\beta}^0 c_{\beta}$ for all species pairings (see \cref{eq:helmholtz_free_energy_total}).
By construction, the species volumes transfer the interaction parameters occurring in our theory and the Flory-Huggins parameters via $\upxi_{\alpha\beta}{=}\chi_{\alpha\beta}/\upnu_\alpha$.
While positive parameters $0{<}\chi_{\alpha\beta} {\lesssim} 0.5$ denote a possible mixing of two species $\alpha$ and $\beta$, the mixing occurs mainly due to an increase in entropy.\cite{Nedoma2008}
In contrast, a negative parameter $\chi_{\alpha\beta}{<}0$ implies that a configuration with higher volume fractions $\phi_\alpha$ and $\phi_\beta$ is energetically more favorable.
So, in addition to an increase in entropy, the decreasing energy amplifies mixing processes.\cite{Tomlin1992}
\begin{figure*}
    \centering
    \includegraphics[width=0.8\textwidth]{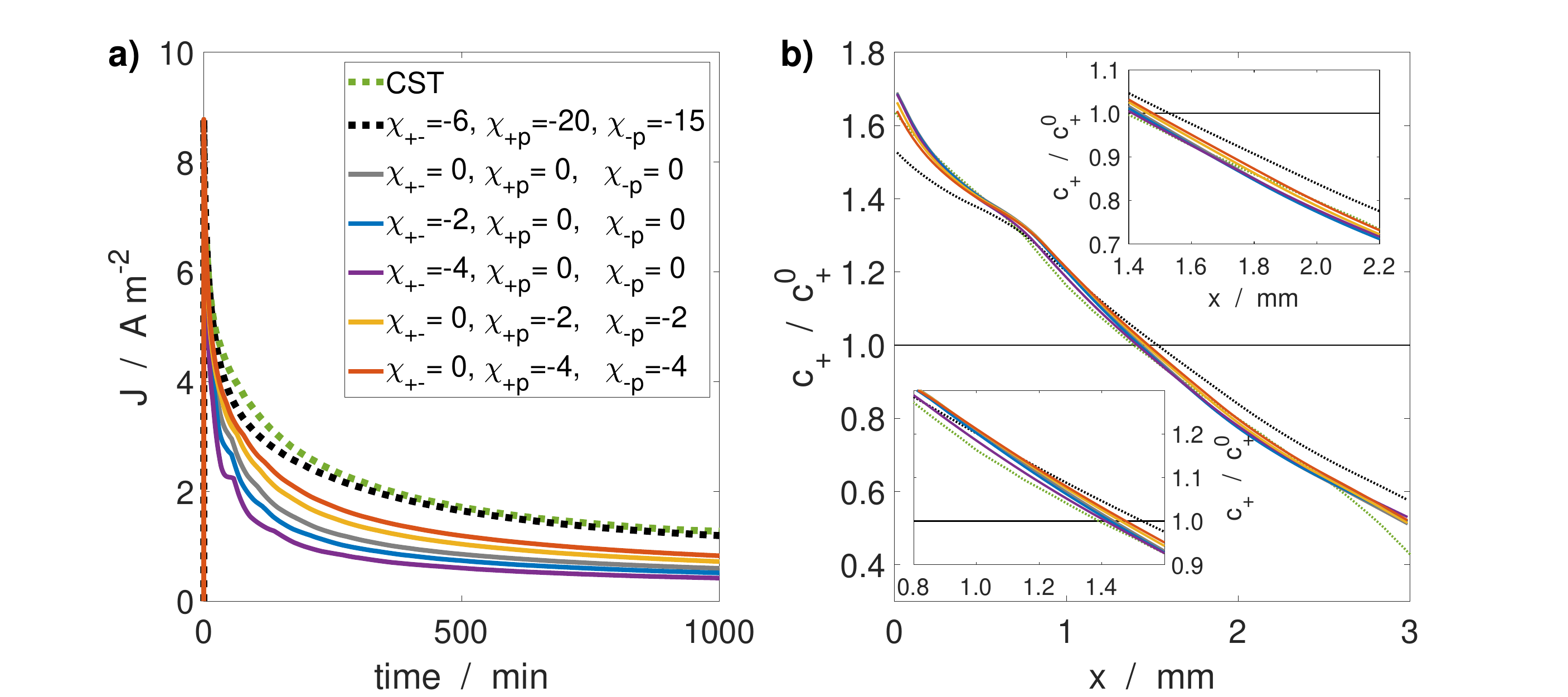}
    \caption{Comparison of the current density a) and concentration profiles b) for different Flory-Huggins parameters $\chi_{\alpha\beta}$ for the PEO/LiTFSI electrolyte. The dashed lines depict the results from CST (green) and for the set of parameters used in Numerical Results and Validation sections (black). The solid lines depict the results for different values for the Flory-Huggins parameters.}
    \label{fig:floryhuggins}
\end{figure*}
\Cref{fig:floryhuggins} a) shows the current density for different sets of Flory-Huggins parameters and the result obtained by Steinrück {\textit et al}. from CST.
For orientation, the results from CST and from the chosen set of parameters in the sections One Dimensional Model Equations and Validation are depicted by the dotted green and black curves, respectively.
The solid lines denote the current densities obtained for different sets of parameters to show their influence.
The solid gray line depicts the result obtained for vanishing Flory-Huggins parameters, {\textit i.e.} $\chi_{\li\tfsi}{=}\chi_{\li\p}{=}\chi_{\tfsi\p}{=}0$, for reference.
The solid blue and violet lines show the resulting current densities for negative values for the interaction parameter between the cation and anion, {\textit i.e.} $\chi_{\li\tfsi}{=}-2,-4$, with the other parameters zero.
The solid yellow and orange lines depict the current densities for negative values of the interaction parameters between the two ions and the polymer, {\textit i.e.} $\chi_{\li\p}{=}\chi_{\tfsi\p}{=}-2,-4$.
A negative interaction parameter $\chi_{\li\tfsi}$ results in a faster decrease of the current density with time than the reference current density (solid gray line), while negative interaction parameters $\chi_{\li\p}$ and $\chi_{\tfsi\p}$ of the two ions with the polymer result in a slower decrease.
This can be explained by the free energy of the system.
In our model, the respective contributions to the free energy are given by $RT\upxi_{\alpha\beta}\upnu_\alpha c_\alpha\upnu_\beta c_\beta=RT\chi_{\alpha\beta}c_\alpha\upnu_\beta c_\beta$.
We start discussing the first case of the negative ion-ion interaction $\chi_{\li\tfsi}{<}0$.
Using the electroneutral constraint $c_{\li}{=}c_{\tfsi}$, the energy contribution can be written as $RT\chi_{\li\tfsi}\upnu_{\tfsi} c_{\li}^2$.
Therefore, the energy decreases with increasing ion concentration.
During concentration polarization, the ion concentration in the PEO/LiTFSI electrolyte increases at the positive electrode and decreases the free energy for an amount $\Delta F_\text{H}^\text{pos}$.
Simultaneously, the decrease of ion concentration at the negative electrode increases the free energy for an amount $\Delta F_\text{H}^\text{neg}$.
However, due to the quadratic dependence on the ion concentration, the decrease of free energy outweighs $\Delta F_\text{H}^\text{pos}>\Delta F_\text{H}^\text{neg}$.
Therefore, for negative $\chi_{\li\tfsi}{<}0$, concentration polarization is energetically more favorable and develops more rapid, resulting in a faster decrease of the current density.
The second case of negative ion-polymer interactions $\chi_{\li\p},\chi_{\tfsi\p}{<}0$ can be explained analogously.
Using the electroneutrality constraint $c_{\li}{=}c_{\tfsi}$ and the Euler equation for volume $1{=}\sum_\alpha\upnu_\alpha c_\alpha$ the energy contribution becomes $RT\left(\chi_{\li\p}+\chi_{\tfsi\p}\right)c_{\li}\left(1-\upnu_{\litfsi}c_{\li}\right)$ with $\upnu_{\litfsi}{=}\upnu_{\li}{+}\upnu_{\tfsi}$.
Apparently, the energy contribution has a negative quadratic dependence on the ion concentration.
So, following the argument for the first case of $\chi_{\li\tfsi}{<}0$ but with opposite sign, negative ion-polymer interaction parameters lead to a reduced concentration polarization and therefore to a slower decrease of the current density.

\Cref{fig:floryhuggins} b) depicts the concentration profiles for the different sets of Flory-Huggins interaction parameters at the simulation time of $t{=}\SI{1000}{\minute}$.
As in \cref{fig:floryhuggins} a), the dotted lines depict the results for CST and the parameter set used in the One Dimensional Model Equations and Validation sections for orientation.
The solid lines depict the results for different parameter sets: gray for vanishing parameters, blue and violet for negative $\chi_{\li\tfsi}$, and yellow and orange for negative $\chi_{\li\p}$ and $\chi_{\tfsi\p}$.
The differences between the different parameter sets (solid) are small, so the two insets show smaller sections of the electrolyte.
For negative interactions between the ions, {\textit i.e.} $\chi_{\li\tfsi}{<}0$, the concentration is very slightly larger than for negative ion-polymer interactions ($\chi_{\li/\tfsi\p}{<}0$).
This behaviour is similiar to the behavior of the current densities, and a negative $\chi_{\li\tfsi}$ energetically favors a volumetric separation of ions from the polymer, while negative $\chi_{\li\p}$/$\chi_{\tfsi\p}$ favor a mixing of the ions with the polymer.

Second, we discuss the elastic parameters of the polymer.
In general, the elasticity of a body is described using two parameters, in our case the bulk modulus $K_{\p}$ and the shear modulus $G_{\p}$. However, for our discussion of the influence of these parameters on the simulation results, we cast them into one single parameter given by the elastic modulus $E_{\p}$.
We obtain the bulk and shear moduli by assuming a constant Poisson's ratio of $\nu^{\text{Poi}}{=}0.33$ and use the conversion formulae $K_{\p}{=}E_{\p}/3(1-2\nu^{\text{Poi}})$ and $G_{\p}{=}E_{\p}/2(1+\nu^{\text{Poi}})$.
The elastic modulus describes the resistance of the polymer to deformations with regards to a reference configuration.

\begin{figure*}
    \centering
    \includegraphics[width=0.8\textwidth]{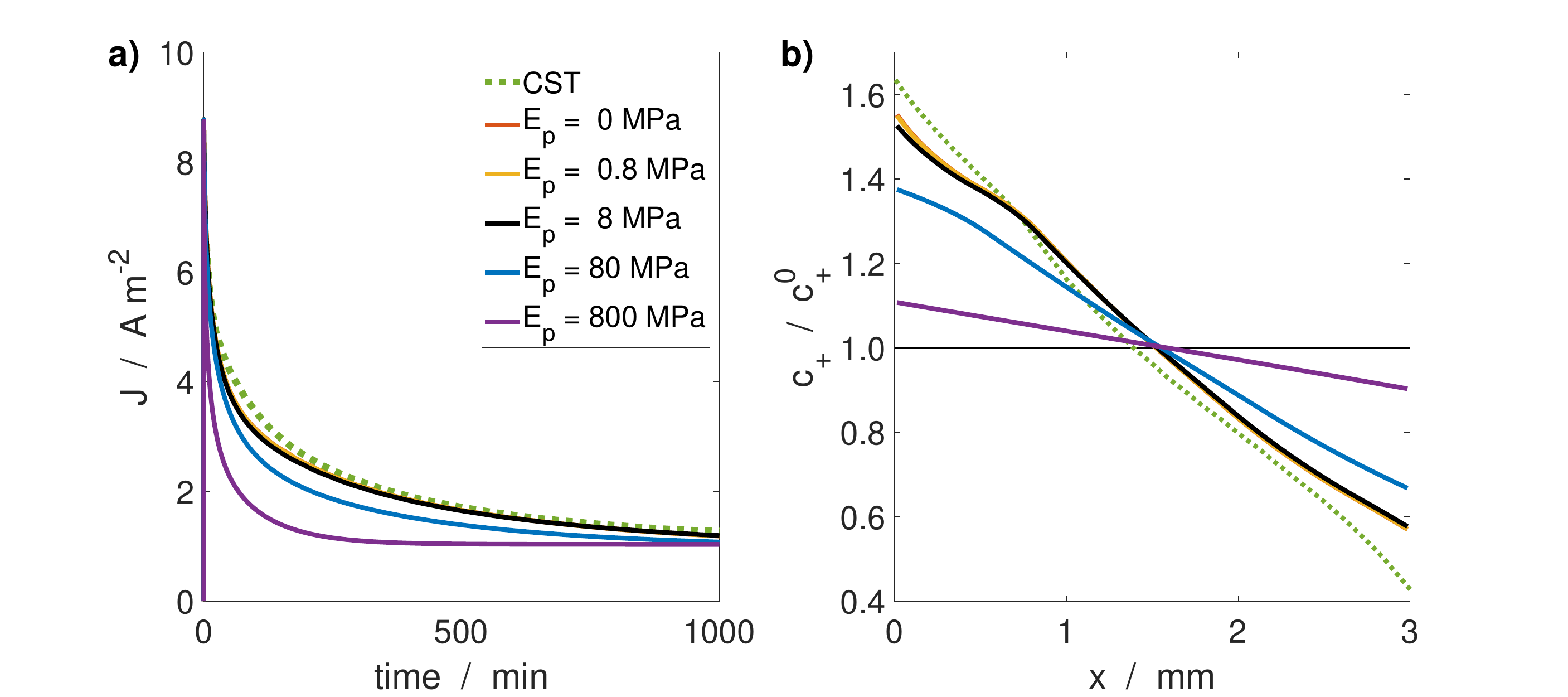}
    \caption{Comparison of the current density a) and concentration profiles b) for different values of the elastic modulus $E_{\p}$ for the PEO/LiTFSI electrolyte. The green dashed line depicts the results from CST. The solid lines depict the results of our model for different $E_{\p}$. In the sections Numerical Results and Validation, the elastic modulus of $E_{\p}{=}8\,\text{MPa}$ (black) was chosen.}
    \label{fig:mechanics}
\end{figure*}
\Cref{fig:mechanics} a) shows the current densities for different elastic moduli of the polymer (solid lines) and the current density from CST (dashed green line).
The elastic modulus increases from $E_{\p}{=}\SI{0}{\mega\pascal}$ (orange), which would correspond to a liquid electrolyte without elasticity, and $E_{\p}{=}\SI{0.8}{\mega\pascal}$ (yellow) to the literature value for PEO of $E_{\p}{=}\SI{8}{\mega\pascal}$ (black).
The larger values of $E_{\p}{=}\SI{80}{\mega\pascal}$ (blue) and $E_{\p}{=}\SI{800}{\mega\pascal}$ (violet) would more closely resemble rigid solid electrolytes.
Apparently, a higher elastic modulus leads to a faster decrease of the current density than a smaller elastic modulus.
While for the small elastic moduli of $\SI{0}{\mega\pascal}$ (orange), $\SI{0.8}{\mega\pascal}$ (yellow) and $\SI{8}{\mega\pascal}$ (black) the differences are negligible, the larger elastic moduli of $\SI{80}{\mega\pascal}$ (blue) and $\SI{800}{\mega\pascal}$ (violet) lead to a more rapid approach of the current density to the steady state.
For the largest elastic modulus (violet), this steady state is reached after about $\SI{400}{\minute}$.

This can be explained by the elastic energy contribution.
Concentration polarization leads to a change in the polymer concentration due to the volumetric displacement by the ion species.
The polymer concentration is coupled to the polymer deformation with $J{=}\frac{1}{\cpeo\upnu_{\p}}$ (\cref{eq:const_eq_J}).
So, concentration polarization leads to a deformation of the polymer from the reference configuration, which increases the elastic energy.
The elastic modulus is a measure for this energy increase.
So, a high elastic modulus prevents large concentration gradients and leads to larger diffusion fluxes due to the coupling of the chemical potentials with the mechanics in the driving forces (see \cref{eqn:driving_forces_definition}).

\Cref{fig:mechanics} b) depicts the resulting concentration profiles for the different elastic moduli and CST at a time of $\SI{1000}{\minute}$.
Apparently, a higher elastic modulus prevents larger concentration gradients.
This suggests that large enough elastic moduli may result in negligible concentration polarization.
This might explain the lack of concentration polarization in solid electrolytes which typically possess a high elastic modulus ($E_{\p}>\SI{1}{\giga\pascal}$).\cite{Stolz2022}

\section*{Application: Single-Ion Conducting Block Copolymer}
\label{sec:application}

The occurrence of concentration polarization in dual-ion conducting polymer-based electrolytes constitutes transport limitations, which hinder the battery performance (see section Numerical Results).\cite{STOLZ20219} Recently, it has been reported that single-ion-conducting (SIC) polymers with immobilized anions can prevent concentration polarization.\cite{Stolz2022} However, the precise impact of the immobilization of the anions on Li-transport is not yet completely clear. Here, theoretical methods can improve the understanding of these electrolytes.

In this section,  we use our transport theory and investigate such a novel SIC electrolyte. We focus on a polymer-based electrolyte composed of a single-ion conducting multi-block copolymer solvated with ethylene carbonate (EC), which was recently described by Nguyen and coworkers.\cite{Nguyen2018} This electrolyte exhibits promising properties, among them a good electrochemical and thermal stability, an ionic conductivity close to commercially available liquid electrolytes, and a highly reversible cycling behaviour in symmetric Lithium/Lithium-cells.

We structure our investigation into three parts. First, in the section Model, we state the equations of motion for the SIC electrolyte.
Second, in the section Potentiostatic Discharge Simulation, we show that our theory predicts that concentration polarization is negligible in SIC electrolytes.
We validate this analytical finding and perform numerical discharge simulations of a symmetrical Li-metal cell (see \cref{fig:peo_setup}).
Finally, in the section Polymer Deformation, we present an analytical analysis of the mechanical deformation of the SIC polymer matrix.
\newline

\paragraph*{\textit{\textbf{Model.}}}
\label{sec:model_sic_polymer}

Similar to our discussion of the PEO electrolyte in the section Validation: Simulation of a Li Cell With Polymer Based Electrolyte, we assume complete dissociation into three electrolyte species. Because  the anions are chemically grafted onto the polymer backbone, we model the polymer as negative electrolyte species. The remaining two species are given by the uncharged solvent, \textit{i.e.} the  EC species, and by the positively charged Li-ions. We neglect microscopic details of the polymer in our continuum theory, and account for the complex structuring of the ionophobic and ionophilic blocks of the polymer chains using a volume averaged description (see \cref{eqn:SIC_continuityequation_volumeframe,eq:SIC_zeroflux,eq:SIC_vvol}). 

In contrast to the PEO based electrolyte discussed in Validation: Simulation of a Li Cell With Polymer Based Electrolyte, where the neutral polymer-solvent dominated the volume fraction of the electrolyte, the share of the volume fractions in the SIC is almost equally distributed among the neutral solvent (EC) and the charged polymer solvent (which is affected by migration). Hence, the polymer-based frame of reference becomes irrelevant for the SIC, and we choose the reference frame based on the volume-averaged  convection velocity $\vec{v}^{\textnormal{v}}=\sum_{\alpha=1}^{\ce{N}} c_\alpha \upnu_\alpha \vec{v}_\alpha$ for the SIC electrolyte.
As shown in Ref.~\citenum{Kilchert2022}, this has the advantage that, for the (nearly) incompressible electrolyte, the transport parameters depend hardly on the frame of reference, which facilitates the parametrization of the electrolyte and the convection is completely determined by the surface reactions,
\begin{equation}
    \label{eq:SIC_vvol}
    \div\left( \epsilon \vec{v}^\text{V} \right) = \upnu_{\li} r_{\li}.
\end{equation}
Here, we used porous electrode theory to account for the porous ionophilic polymer-phases (which carry the Li-motion), where $\epsilon$ denotes the porosity, and $\upbeta$ denotes the Bruggemann coefficient. The volume-based transport equations for porous electrodes read,\cite{Kilchert2022}
\begin{gather}
    \label{eqn:SIC_continuityequation_volumeframe}
\partial_t \left( \epsilon \cli \right) = - \div{\left( \epsilon^\upbeta \vec{N}_{\ce{Li}}^{\text{V}} \right)} - \div{\left( \epsilon \cli \vec{v}^\text{V} \right)}
 + r_{\ce{Li}},
	\\
 \label{eq:SIC_zeroflux}
	0 = - \div{\left( \epsilon^\upbeta \vec{\mathcal{J}}^{\text{V}} \right)} + Fr_{\ce{Li}}.
\end{gather}
We state the volume-based expressions for the current and flux densities in the SI, see \cref*{subsec:SI_volume_frame}. 

Similar to the PEO electrolyte, the three independent (volume-based) transport parameters are the electric conductvity  $\kappa^{\ce{V}}$, the transference number of the Li-ions $t_{\ce{Li}}^{\ce{V}}$, and one Li-diffusion coefficient $\mathcal{D}_{\ce{Li}}^{\ce{V}}$. 
For more details on the volume-frame, see \cref*{subsec:SI_volume_frame} or Ref.~\citenum{Kilchert2022}.

We set the elastic modulus of the SIC polymer to  $E_{\ce{p}} = \SI{249}{\mega\pascal}$, which corresponds to  $K_{\ce{p}} = \SI{277}{\mega\pascal}$ and $G_{\ce{p}} = \SI{92}{\mega\pascal}$.\cite{Sahu2009}
For the Flory-Huggins interaction parameters we choose $\chi_{\ce{LiEC}} {=}\num{ -5}$, $\chi_{\ce{Lip}} {=} \num{-20}$ and $\chi_{\ce{ECp}} {=}\num{ -10}$.
The values for the diffusion coefficient, $\mathcal{D}_{\ce{Li}}^{\ce{V}} {=} \SI{1.94e-10}{\square\meter\per\second}$,  for the conductivity $\kappa^{\ce{V}} {=} \SI{8.5e-2}{\siemens\per\meter}$, and for the transference number, $t_{\ce{Li}}^{\ce{V}} {=} 1$, are taken from Nguyen et al. (all at $T{=}\SI{90}{\celsius}$).\cite{Nguyen2018}.

The symmetric simulation setup of the Li-metal electrodes is identical to the one presented in \cref{fig:peo_setup} for the PEO/LiTFSI electrolyte, where the electrodes are separated by a $\SI{3}{\milli\metre}$ thick polymer electrolyte.
We discharge the battery by applying a constant voltage difference of $\SI{0.3}{\volt}$ for the whole simulation run of $\SI{1000}{\minute}$.
\newline

\paragraph*{\textit{\textbf{Potentiostatic discharge simulation.}}}
\label{subsec:SI_potentiostatic}

In this section, we discuss the dynamics of the electrolyte during discharging the cell. First, we show that our theoretical description predicts that concentration polarization is negligible in SIC-based systems. This prediction is confirmed by experimental results.\cite{Stolz2022} Second, we validate this finding by numerical simulations. 

First, we modify \cref{eqn:SIC_continuityequation_volumeframe}, such that 
\begin{equation}
\label{eq:cli_eom_modified}
\partial_t \left( \epsilon \cli \right) {=} \left(1{-}t_{\ce{Li}}^{\text{V}} {-}\cli \upnu_{\ce{Li}}\right) r_{\ce{Li}}
{+} \nabla \cdot (\epsilon^\beta \mathcal{D}_{\li}^{\text{v}}\tilde{\vec{\mathcal{F}}}_{\li}^{\text{v}})
{-} (\epsilon \vec{v}^{\text{v}}{\cdot}\nabla)\cli.
\end{equation}
 Because of the immobilization of the anions, the transference number of the Li-ions is equal to one, $t_{\ce{Li}}^{\text{V}}{=}\num{1}$.\cite{Nguyen2018} Note that the second term in brackets is a function of the Li-concentration, $\epsilon^\beta\mathcal{D}_{\li}^{\text{v}}\tilde{\vec{\mathcal{F}}}_{\li}^{\text{v}}{=}f(\cli,\nabla\cli)$. We make use of both properties, such that \cref{eq:cli_eom_modified} becomes
\begin{equation}
    \label{eq:SIC_tisone}
    \partial_t(\epsilon\cli)= -\cli \upnu_{\ce{Li}}r_{\ce{Li}} + (\hat{\mathbb{D}}_{\li}\cdot \nabla)\cli    .
\end{equation}
Usually, the volume fraction of the Li-ions is very small $\cli \upnu_{\ce{Li}}{\ll}\num{ 1}$ 
. Hence, in the stationary state, there arise no relevant concentration gradients, 
\begin{equation}
    \label{eq:stationary_state_SIC}
    \cli \upnu_{\ce{Li}}r_{\ce{Li}} = (\hat{\mathbb{D}}_{\li}\cdot \nabla)\cli    , \quad \textnormal{where} \quad \cli \upnu_{\ce{Li}}{\ll}\num{ 1}.
\end{equation}
This rationalizes recent experimental observations that concentration polarization is a negligible effect in SICs.\cite{Stolz2022,Nguyen2018} We emphasize that our analytical result depends crucially on the assumption that $t_{\li}^{\text{v}}{=}1$, and on the volume-based description. 

\begin{figure}
    \centering
    \includegraphics[width=\columnwidth]{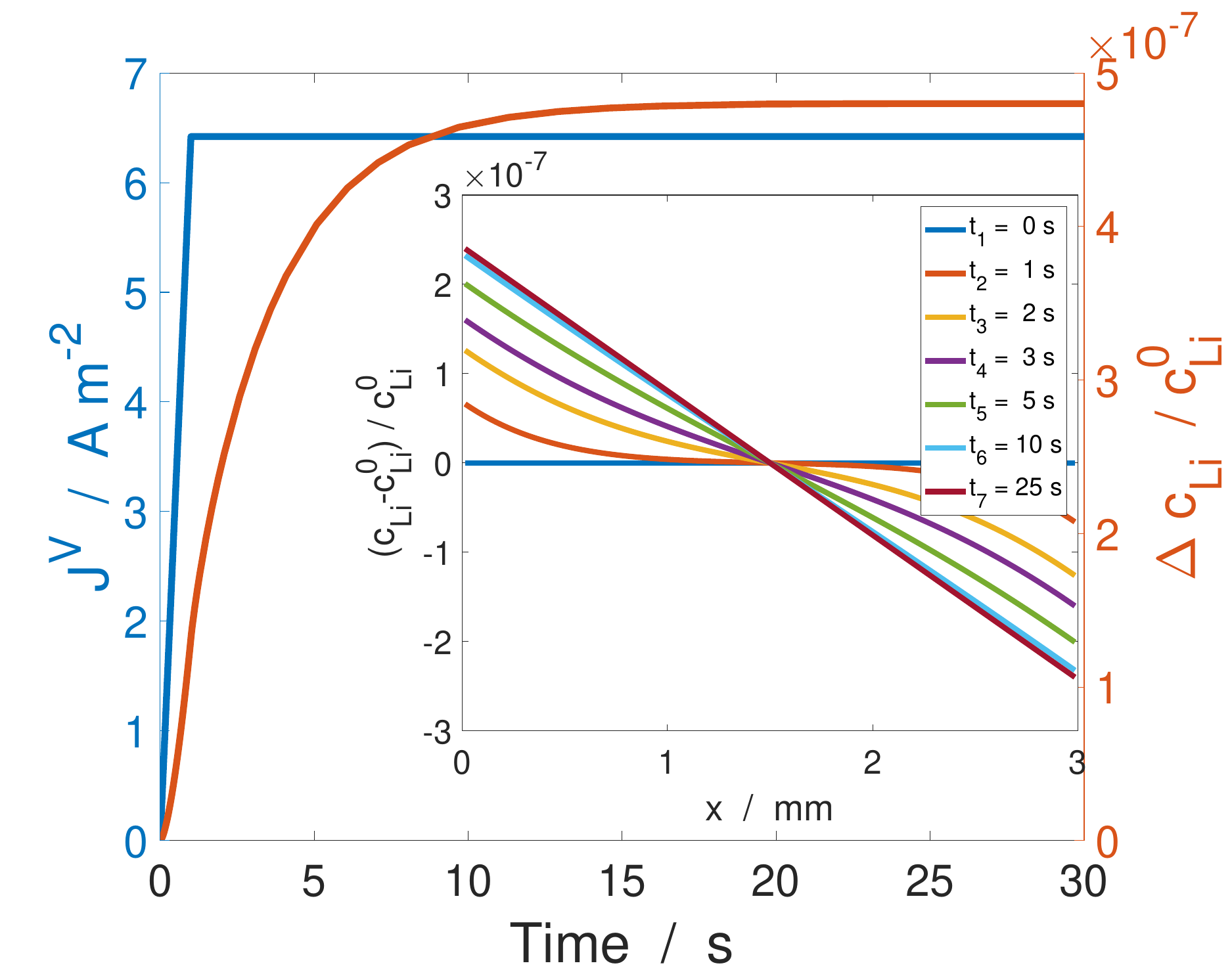}
    \caption{The current density (blue line, left axis) and the concentration difference $\Delta \cli$ (orange line, right axis) over the whole width of the electrolyte for the first $\SI{30}{\second}$ of the simulation run. The inset shows the concentration profiles over the whole width of the electrolyte at different times (colors).}
    \label{fig:sic_results_concentration}
\end{figure}

Next, we perform numerical discharge simulations of the as-described cell set-up (see \cref{fig:peo_setup} for an illustration).

\Cref{fig:sic_results_concentration} illustrates the temporal evolution of the electric current $\vec{\mathcal{J}}^{\text{v}}$ (blue line, left y-axis) and of the concentration differential of the Li-ions between the two electrodes $\Delta\cli(t)=|\cli(t,x{=}L)-\cli(t,x{=}\num{0})|$ (red line, right y-axis), normalized by the initial concentration $\cli^0$.
The electric current increases rapidly during the ramp up of the potential difference between the two electrodes (during the first second).
However, after this initial phase, it reaches a constant plateau at $\vec{\mathcal{J}}^{\text{v}}{\approx}\SI{6.5}{\ampere\per\meter\squared}$ and remains constant.
This corresponds to the system reaching a stationary state almost instantly.
The red line (right y-axis) illustrates the behaviour of the normalized concentration differential $\Delta\cli(t)/\cli^0$, which serves as indicator for the occurrence of concentration polarization ($\Delta\cli(t)/\cli^0{\ll}\num{1}$ for negligible concentration polarization).
The inset resolves the corresponding concentration profiles of the Li-ions over the cell length at different representative time steps $t_i$ during discharge.
During the complete discharge time, the difference between the Li-concentration at the two electrodes is of order $\mathcal{O}(\Delta\cli/\cli^0){=}\num{1e-7}$, \textit{i.e.} is negligible.
This behaviour is confirmed by the spatial profiles of the Li-concentrations shown in the inset.
A clear gradient evolves from the left side to the right side over discharge time.
In contrast to the PEO/LiTFSI electrolyte investigated in Validation: Simulation of a Li Cell With Polymer Based Electrolyte, the transport parameters do not depend on $\cli$.
This leads to symmetric concentration gradients.
Note that the spatial variation of the concentration profiles lies within the accuracy of our numerical simulation.

Altogether, we thus conclude from our numerical results that the occurrence of concentration polarization is negligible for the as-described system.
This confirms our analytical prediction and is in agreement with experimental results.\cite{Nguyen2018,Stolz2022}
\newline

\paragraph*{\textit{\textbf{Polymer deformation.}}}
\label{sec:polymer_deformation}

In this section, we supplement our analytical discussion from the section Potentiostatic Discharge Simulation, and focus on the deformation of the SIC polymer (see, also, \cref*{sec:SI_polymer_deformation} for more details).

Polymer deformation influences the electrolyte performance. Because the concentration of the Li-ions is the only independent species concentration, it determines the polymer deformation. Therefore, polymer deformation and polarization concentration are directly coupled.

\Cref{eq:stationary_state_SIC,eq:SIC_tisone,eq:cli_eom_modified} show the influence of the Li-transference number, the volume fraction of the Li-ions, and the reaction kinetics on the occurrence of concentration polarization.
We use this description and focus on the influence of these system parameters on the polymer-deformation as described by the deformation tensor $\ten{F}=\left(\begin{smallmatrix}J&0&0\\ 0&1&0\\0&0&1\end{smallmatrix}\right)$. Apparently, it suffices to focus on the volume ratio $J$, instead of the deformation $\vec{F}$.
In the SI (see \cref*{sec:SI_polymer_deformation}), we show that the volume ratio in the stationary state can be written as, 
\begin{equation}
    \label{eqn:polymerdeformation}
    \grad\left( J \right)
    = - \Omega \cdot j_{\ce{elde}}.
\end{equation}
where $\Omega$ depends on transport-/ and material-parameters and on the driving forces, and where $j_{\ce{elde}}$ is the current due to the interface reactions.

\begin{figure}
    \centering
    \includegraphics[width=\columnwidth]{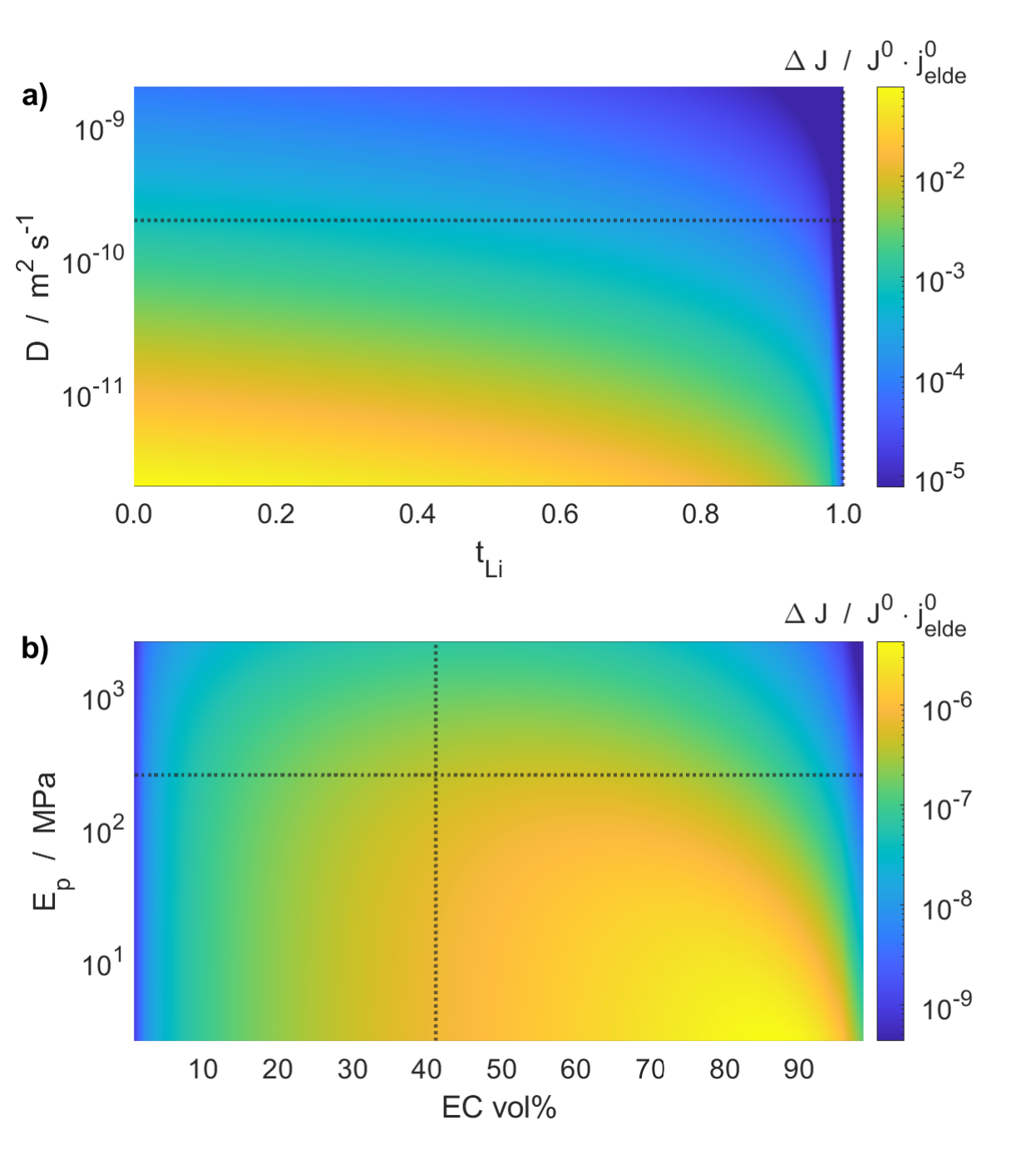}
    \caption{Polymer deformation as functions of transport parameters of the Li-ions a) and of material parameters b).
    The dotted lines denote the respective values used in the potentiostatic simulation in Application:Single-Ion Conducting Block Copolymer section.
    The colors indicate the polymer deformation as measured by the  difference $\Delta J$ of volume ratios across the SIC/EC electrolyte, normalized by the initial volume ratio $J^0$ and the non-dimensional current density $j_{\ce{elde}}^0$.}
    \label{fig:polymerdeformation_2d}
\end{figure}

\Cref{fig:polymerdeformation_2d} shows the influence of transport parameters and material parameters on the deformation of the polymer as predicted by \cref{eqn:polymerdeformation}.
Here, we use the normalized variation of $J$ from the left electrode to the right electrode, \textit{i.e.} $\Delta J = (J(x{=}L)-J(x{=}0)/J^0$, in units of the non-dimensional current density $j_{\ce{elde}}^0$ as measure for the polymer deformation.
The non-dimensionalized current density is the current density applied to the electrodes divided by the exchange current density $j_{\ce{elde}}^0 {=} j_{\ce{elde}}/j_0$, as given in the SI, \cref*{subsec:SI_parameters}.
The dotted lines denote the respective values used in the potentiostatic simulation in the section Potentiostatic Discharge Simulation.
\Cref{fig:polymerdeformation_2d} a) illustrates the influence of the transference number $t_{\li}^{\text{V}}$ and of the diffusion coefficient $\mathcal{D}_{\li}^{\text{V}}$ of the Li-ions on the polymer deformation.
The transference number $t_{\li}^{\text{V}}$ ranges from $0$ to $1$ and the lithium diffusion coefficient $\mathcal{D}_{\li}^{\text{V}}$ from $0.01\cdot D^0$ to $10\cdot D^0$, where $D^0 {=} \SI{1.94e-10}{\square\meter\per\second}$ is the diffusion coefficient measured by Nguyen et al.\cite{Nguyen2018}
Apparently, the polymer deformation decreases with increasing diffusion coefficient and with increasing lithium transference number $t_{\li}$.
Independently from the diffusion coefficient, the polymer deformation drops several orders of magnitudes to values smaller than $10^{-5}$ for $t_{\li}^{\text{V}} {\to} 1$.
This reproduces our finding of negligible concentration polarization for SICs.
In contrast, for $t_{\li}^{\text{V}} {\to} 0$, the flux of Li-ions in the stationary state is mainly driven by diffusion.
In particular, in this regime for $t_{\li}^{\text{V}}$, the deformation decreases with increasing diffusion coefficient.
Hence, for higher values of $\mathcal{D}_{\li}^{\text{V}}$, diffusion is fast enough to quickly equilibrate concentration gradients, yielding spatially homogeneous concentration profiles (small deformations).
This is in agreement with the property that the deformation increases with decreasing diffusion coefficients. 

\Cref{fig:polymerdeformation_2d} b) illustrates the influence of the ethylene carbonate volume ratio and of the elastic modulus of the polymer on the deformation.
Here, $\text{EC vol\%}$ ranges from $0\%$ to $99\%$, and the elastic modulus of the polymer $E_{\ce{p}}$ spans over a range of $0.01\cdot E_{\ce{p}}^0$ to $10\cdot E_{\ce{p}}^0$, with $E_{\ce{p}}^0 = \SI{249}{\mega\pascal}$.
Apparently, the polymer deformation is not very sensitive to variations of the elastic modulus for small EC volume ratios $< 30\%$.
However, for larger EC volume ratios $> 30\%$, the influence of the elastic modulus on the resulting polymer deformation increases noticably.
For the EC volume ratio however, the polymer deformation decreases at both ends of the range with a maximum in between, which results in an inverted U-shape.
Since the volume ratio of the Li-ions can be neglected ($\cli\upnu_{\li}{\ll}1$), small (large) EC volume ratios imply large (small) polymer volume ratios.
Hence, a more equal distribution of the volume fraction between the EC and the polymer favors polymer deformation.
This exact distribution for the largest polymer deformation also depends on the elastic modulus of the polymer.

Furthermore, it can be seen that the relative polymer deformations become much larger (up to nearly $10^{-1}$) under variations of the transport parameters (see  \cref{fig:polymerdeformation_2d}a), as compared with variations of the material parameters (relative polymer deformation ranges up to $10^{-5}$, see \cref{fig:polymerdeformation_2d}b). This suggests that the polymer deformation depends much stronger on the transport parameters than on the material parameters. For more details, see see \cref*{sec:SI_polymer_deformation}.

Finally, we compare the polymer deformation of the PEO system with the SIC results. In the section Potentiostatic Discharge Simulation it was shown that $\Delta J / \left( J^0 \cdot j_{\ce{elde}}^0 \right) {\approx} 1.8$ for the PEO/LiTFSI electrolyte. In comparison, for the SIC/EC electrolyte we found (with the parameters as denoted by the dotted lines in \cref{fig:polymerdeformation_2d}) $\Delta J / \left( J^0 \cdot j_{\ce{elde}}^0 \right) {\approx} 4.2 \cdot 10^{-7}$.
This huge difference between the two polymer electrolytes can be explained by different transport and material parameters.
The PEO/LiTFSI electrolyte exhibits a much smaller diffusion coefficient ($D^{\ce{PEO}}_{\li} {\approx} \SI{10e-11}{\square\meter\per\second}$), a smaller cation transference number ($t^{\ce{PEO}}_{\li} {\approx} 0.2$) and a smaller elastic modulus ($E^{\ce{PEO}}_{\ce{p}} {=} \SI{8}{\mega\pascal}$) as compared to the SIC/EC electrolyte (where $D_{\li}^{\ce{SIC}} {\approx} \SI{1.94e-10}{\square\meter\per\second}$, $t^{\ce{SIC}}_{\li} {\approx} 1$ and $E_{\ce{p}}^{\ce{SIC}} {=} \SI{249}{\mega\pascal}$).
In the SI, we supplement this discussion by a more detailed analysis (see \cref*{sec:SI_polymer_deformation}). 

Altogether, we conclude that as consequence of the immobilization of the anions (yielding a high Li-transference number),  a high diffusion coefficient and a high elastic modulus, the deformation of the SIC electrolyte is much smaller than that of other polymer-based electrolytes. This mechanical behaviour is beneficial for the performance of the electrolyte.

\section*{Discussion}
\label{sec:discussion}
\label{sec:comp_with_SE_LE}

In this section we discuss the derived continuum-model for polymer electrolytes and its positioning relative to previously developed transport models.

\begin{figure}[!htb]
    \centering
    \includegraphics[width=\columnwidth]{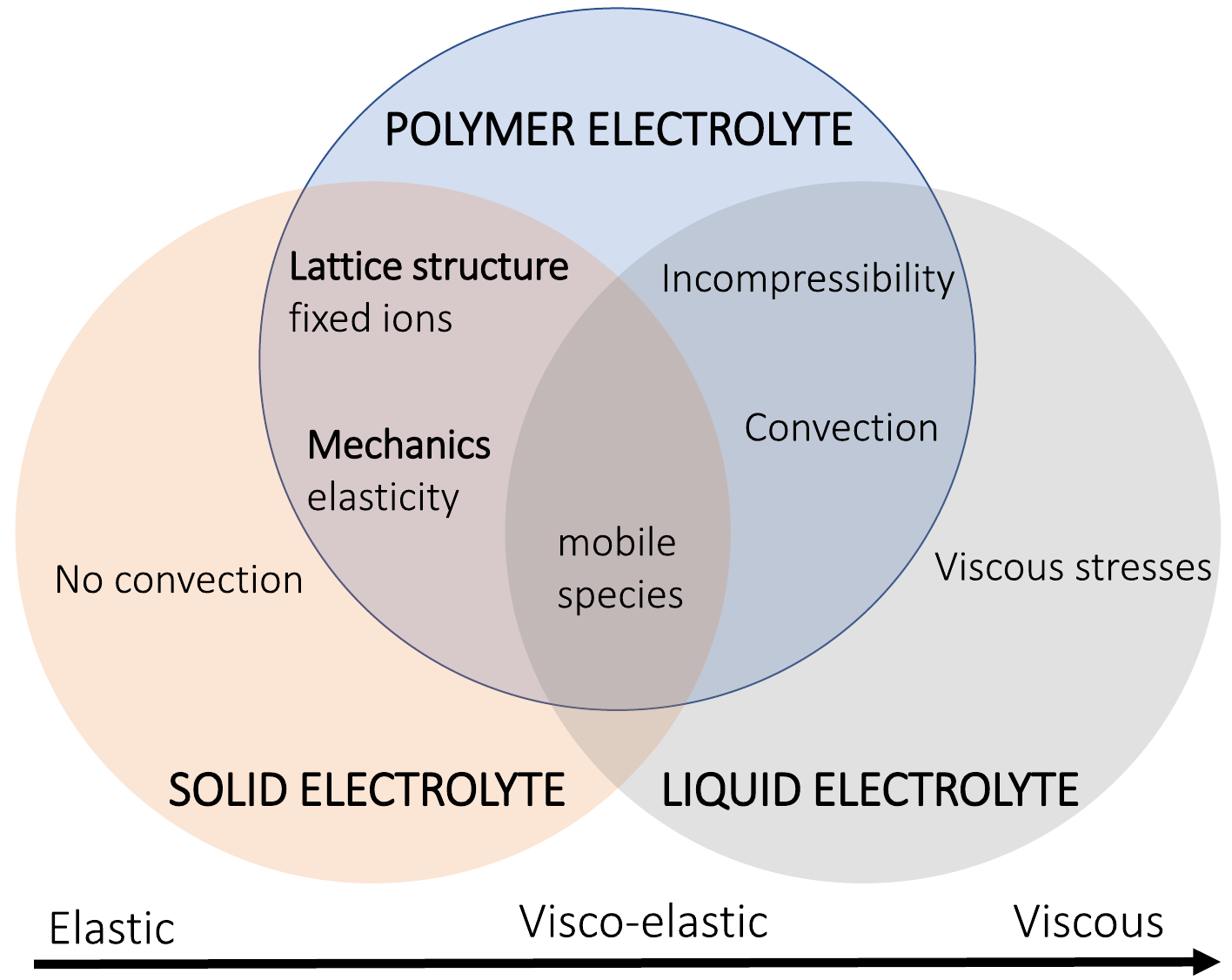}
    \caption{Venn diagram illustrating our polymer model in relation to models for viscous liquid electrolytes and solid electrolytes. }
    \label{fig:comp_models}
\end{figure}

As discussed in the Introduction, polymer electrolytes comprise a wide range of materials, and exhibit solid-like properties, as well as liquid-like properties.
For example, while gel polymer electrolytes behave more similar to liquid electrolytes, dry solid or composite polymer electrolytes have more similarities with solid electrolytes (like the SIC polymer electrolyte discussed in Application: Single-Ion Conducting Block Copolymer section).
The Venn diagram shown in \cref{fig:comp_models} illustrates the specific properties of solid electrolytes, polymer electrolytes and liquid electrolytes, and highlights their commonalities.

Recently, our working group has developed transport theories for highly concentrated liquid electrolytes,\cite{Schammer2021,doi:10.1021/acs.jpcb.2c00215,Kilchert2022} and for inorganic solid electrolytes.\cite{Braun2015, KBSarxiv} 
However, these theories are limited to either viscous materials, or elastic materials, and thus constitute mutually exclusive descriptions which renders the description of viscoelastic materials, \textit{e.g.} polymer electrolytes, insufficient.
Our polymer theory bridges this gap and provides a description for such materials. 

All three transport theories are based on the framework of rational thermodynamics (RT).
Thus, they are derived from the same universal assumptions and share a common rationale.
This facilitates the comparison of the three transport theories.
In RT, the focal quantity is the Helmholtz free energy, which comprises material-specific properties.
As consequence, the three transport theories can be differentiated via their model free energy. 

From a mechanical perspective, solid electrolytes can be described as elastic materials.
This implies a functional relation between the exertion of stress and the deformation (strain), which is taken into account in the model free energy.
Our model describes anions as stationary lattice structure, while cations and vacancies are mobile.
The immobile anions do not contribute to the mixing free energy.
However, the sum of cations and vacancies has to be constant, which is a constraint not generally applicable to liquids electrolytes.

In contrast to solids, liquid electrolytes exhibit a relation between the exertion of stress and the rate of deformation (rate of strain).
This accounts for viscous stresses and momentum dissipation.
Our transport theory for liquid electrolytes satisfies a kinematic constraint on the volume fractions, and gives a prediction for the convection velocity in incompressible electrolytes.
Convective effects due to local volume fluxes and surface reactions are important in multicomponent liquid electrolytes with high amount of salts.
Here, all electrolyte species are mobilized and thus susceptible to convection and diffusion, and, if they carry charge, migration. 

Finally, we put our model for polymer electrolytes into perspective. Our derived polymer electrolyte model comprises liquid-like properties. For example, our  polymer description does include the motion of the polymer matrix as convection, since this is an important consideration when obtaining transport parameters from measurements. \cite{Rosenwinkel2019,Shao2022,Kilchert2022} Furthermore, 
because transport processes on the molecular scale depend on the segmental mobility of the polymer chains, \cite{Diddens2010} we assume the polymer electrolyte to be incompressible.
However, our polymer model does not include viscous forces, which are typical for liquid electrolytes.
In addition, our model comprises also solid-like properties. For example, it allows for the description of mobile ion species and of immobilized ion-species, where the cation moves relative to a stationary and negatively charged background (as is the case for SIC/EC electrolyte in Application: Single-Ion Conducting Block Copolymer section). This is typical for inorganic SEs.
Our polymer model also includes mechanical aspects (isotropic elasticity of the polymer matrix), which are typical for solid electrolytes. 

Altogether, our polymer electrolyte model constitutes a  middle ground, where deformations of the polymer play a role for transport, but are negligible for materials with a high elastic modulus ({\it cf.} the section Validation: Simulation of a Li Cell With Polymer Based Electrolyte).

\section*{Conclusions}

In this work, we have derived a continuum transport model for polymer electrolytes using the thermodynamically consistent modelling approach already introduced for liquid and solid electrolytes as well as for ionic liquids \cite{Latz2011,Latz2015,KBSarxiv,Schammer2021}.
With this approach, we were able to derive a transport model that couples electro-chemical with mechanical processes while also including convection.
The inclusion of thermal and viscous processes is straightforward.
The formulation with respect to the polymer reference-frame made it simple to include the necessary transport parameters from other sources.

We validated our model with results from experiment and the "standard" concentration solution theory for the "benchmark" polymer electrolyte PEO/LiTFSI.
We could show that our approach is able to reproduce the thermodynamic behaviour of the electrolyte system without the need for an empirical thermodynamic factor or activities.
We also investigated a novel single-ion conducting polymer electrolyte.
We showed that changes in single transport or material parameters have only negligible influence on the resulting concentration polarization for nearly single-ion conducting materials. Furthermore, we rationalized the occurrence of concentration polarization in polymer-based electrolytes, and derived an analytical description thereof. 

The transport model presented in this work captures the behaviour of polymers with very different properties. It serves as a valid framework to model the behaviour of the vast range of polymer electrolytes and can fill the gap of materials treated by the already developed models for highly concentrated liquid electrolytes and ionic liquids, and inorganic solid electrolytes.
Additional material processes and properties can be included by choice of a suitable free energy model.
This supports the rational design of polymer electrolytes for novel high-performance batteries.

\section*{Acknowledgments}
\label{sec:ackn-luzibmbf}

This work was supported by the German Min\-istry of Education and
Research (BMBF) (project LUZI, BMBF: 03SF0499E) and by the European
Union's Horizon 2020 research and innovation programme via the
``Si-DRIVE'' project (grant agreement No 814464).

The authors acknowledge support by the state of Baden-W\"urttemberg
through bwHPC and the German Research Foundation (DFG) through grant
no INST 40/575-1 FUGG (JUSTUS 2 cluster).

\bibliographystyle{apsrev4-2}
\bibliography{Bibliography}

\begin{thebibliography}{82}%
\makeatletter
\providecommand \@ifxundefined [1]{%
 \@ifx{#1\undefined}
}%
\providecommand \@ifnum [1]{%
 \ifnum #1\expandafter \@firstoftwo
 \else \expandafter \@secondoftwo
 \fi
}%
\providecommand \@ifx [1]{%
 \ifx #1\expandafter \@firstoftwo
 \else \expandafter \@secondoftwo
 \fi
}%
\providecommand \natexlab [1]{#1}%
\providecommand \enquote  [1]{``#1''}%
\providecommand \bibnamefont  [1]{#1}%
\providecommand \bibfnamefont [1]{#1}%
\providecommand \citenamefont [1]{#1}%
\providecommand \href@noop [0]{\@secondoftwo}%
\providecommand \href [0]{\begingroup \@sanitize@url \@href}%
\providecommand \@href[1]{\@@startlink{#1}\@@href}%
\providecommand \@@href[1]{\endgroup#1\@@endlink}%
\providecommand \@sanitize@url [0]{\catcode `\\12\catcode `\$12\catcode
  `\&12\catcode `\#12\catcode `\^12\catcode `\_12\catcode `\%12\relax}%
\providecommand \@@startlink[1]{}%
\providecommand \@@endlink[0]{}%
\providecommand \url  [0]{\begingroup\@sanitize@url \@url }%
\providecommand \@url [1]{\endgroup\@href {#1}{\urlprefix }}%
\providecommand \urlprefix  [0]{URL }%
\providecommand \Eprint [0]{\href }%
\providecommand \doibase [0]{http://dx.doi.org/}%
\providecommand \selectlanguage [0]{\@gobble}%
\providecommand \bibinfo  [0]{\@secondoftwo}%
\providecommand \bibfield  [0]{\@secondoftwo}%
\providecommand \translation [1]{[#1]}%
\providecommand \BibitemOpen [0]{}%
\providecommand \bibitemStop [0]{}%
\providecommand \bibitemNoStop [0]{.\EOS\space}%
\providecommand \EOS [0]{\spacefactor3000\relax}%
\providecommand \BibitemShut  [1]{\csname bibitem#1\endcsname}%
\let\auto@bib@innerbib\@empty
\bibitem [{\citenamefont {Chu}\ and\ \citenamefont {Majumdar}(2012)}]{Chu2012}%
  \BibitemOpen
  \bibfield  {author} {\bibinfo {author} {\bibfnamefont {S.}~\bibnamefont
  {Chu}}\ and\ \bibinfo {author} {\bibfnamefont {A.}~\bibnamefont {Majumdar}},\
  }\href {\doibase 10.1038/nature11475} {\bibfield  {journal} {\bibinfo
  {journal} {Nature}\ }\textbf {\bibinfo {volume} {488}},\ \bibinfo {pages}
  {294} (\bibinfo {year} {2012})}\BibitemShut {NoStop}%
\bibitem [{\citenamefont {Bresser}\ \emph {et~al.}(2018)\citenamefont
  {Bresser}, \citenamefont {Hosoi}, \citenamefont {Howell}, \citenamefont {Li},
  \citenamefont {Zeisel}, \citenamefont {Amine},\ and\ \citenamefont
  {Passerini}}]{Bresser2018}%
  \BibitemOpen
  \bibfield  {author} {\bibinfo {author} {\bibfnamefont {D.}~\bibnamefont
  {Bresser}}, \bibinfo {author} {\bibfnamefont {K.}~\bibnamefont {Hosoi}},
  \bibinfo {author} {\bibfnamefont {D.}~\bibnamefont {Howell}}, \bibinfo
  {author} {\bibfnamefont {H.}~\bibnamefont {Li}}, \bibinfo {author}
  {\bibfnamefont {H.}~\bibnamefont {Zeisel}}, \bibinfo {author} {\bibfnamefont
  {K.}~\bibnamefont {Amine}}, \ and\ \bibinfo {author} {\bibfnamefont
  {S.}~\bibnamefont {Passerini}},\ }\href
  {https://www.sciencedirect.com/science/article/abs/pii/S0378775318301599}
  {\bibfield  {journal} {\bibinfo  {journal} {Journal of Power Sources}\
  }\textbf {\bibinfo {volume} {382}},\ \bibinfo {pages} {176} (\bibinfo {year}
  {2018})}\BibitemShut {NoStop}%
\bibitem [{\citenamefont {Armand}\ and\ \citenamefont
  {Taracson}(2008)}]{Armand2008}%
  \BibitemOpen
  \bibfield  {author} {\bibinfo {author} {\bibfnamefont {M.}~\bibnamefont
  {Armand}}\ and\ \bibinfo {author} {\bibfnamefont {J.-M.}\ \bibnamefont
  {Taracson}},\ }\href {https://doi.org/10.1038/451652a} {\bibfield  {journal}
  {\bibinfo  {journal} {Nature}\ }\textbf {\bibinfo {volume} {451}},\ \bibinfo
  {pages} {652} (\bibinfo {year} {2008})}\BibitemShut {NoStop}%
\bibitem [{\citenamefont {Whittingham}(2004)}]{Whittingham2004}%
  \BibitemOpen
  \bibfield  {author} {\bibinfo {author} {\bibfnamefont {M.~S.}\ \bibnamefont
  {Whittingham}},\ }\href {\doibase 10.1021/cr020731c} {\bibfield  {journal}
  {\bibinfo  {journal} {Chemical Reviews}\ }\textbf {\bibinfo {volume} {104}},\
  \bibinfo {pages} {4271} (\bibinfo {year} {2004})}\BibitemShut {NoStop}%
\bibitem [{\citenamefont {Kim}\ \emph {et~al.}(2015)\citenamefont {Kim},
  \citenamefont {Son}, \citenamefont {Mukherjee}, \citenamefont {Schuppert},
  \citenamefont {Bates}, \citenamefont {Kwon}, \citenamefont {Choi},
  \citenamefont {Chung},\ and\ \citenamefont {Park}}]{Kim2015}%
  \BibitemOpen
  \bibfield  {author} {\bibinfo {author} {\bibfnamefont {J.~G.}\ \bibnamefont
  {Kim}}, \bibinfo {author} {\bibfnamefont {B.}~\bibnamefont {Son}}, \bibinfo
  {author} {\bibfnamefont {S.}~\bibnamefont {Mukherjee}}, \bibinfo {author}
  {\bibfnamefont {N.}~\bibnamefont {Schuppert}}, \bibinfo {author}
  {\bibfnamefont {A.}~\bibnamefont {Bates}}, \bibinfo {author} {\bibfnamefont
  {O.}~\bibnamefont {Kwon}}, \bibinfo {author} {\bibfnamefont {M.~J.}\
  \bibnamefont {Choi}}, \bibinfo {author} {\bibfnamefont {H.~Y.}\ \bibnamefont
  {Chung}}, \ and\ \bibinfo {author} {\bibfnamefont {S.}~\bibnamefont {Park}},\
  }\href {\doibase 10.1016/j.jpowsour.2015.02.054} {\bibfield  {journal}
  {\bibinfo  {journal} {Journal of Power Sources}\ }\textbf {\bibinfo {volume}
  {282}},\ \bibinfo {pages} {299} (\bibinfo {year} {2015})}\BibitemShut
  {NoStop}%
\bibitem [{\citenamefont {Weiss}\ \emph {et~al.}(2021)\citenamefont {Weiss},
  \citenamefont {Ruess}, \citenamefont {Kasnatscheew}, \citenamefont
  {Levartovsky}, \citenamefont {Levy}, \citenamefont {Minnmann}, \citenamefont
  {Stolz}, \citenamefont {Waldmann}, \citenamefont {Wohlfahrt-Mehrens},
  \citenamefont {Aurbach}, \citenamefont {Winter}, \citenamefont {Ein-Eli},\
  and\ \citenamefont {Janek}}]{Weiss2021}%
  \BibitemOpen
  \bibfield  {author} {\bibinfo {author} {\bibfnamefont {M.}~\bibnamefont
  {Weiss}}, \bibinfo {author} {\bibfnamefont {R.}~\bibnamefont {Ruess}},
  \bibinfo {author} {\bibfnamefont {J.}~\bibnamefont {Kasnatscheew}}, \bibinfo
  {author} {\bibfnamefont {Y.}~\bibnamefont {Levartovsky}}, \bibinfo {author}
  {\bibfnamefont {N.~R.}\ \bibnamefont {Levy}}, \bibinfo {author}
  {\bibfnamefont {P.}~\bibnamefont {Minnmann}}, \bibinfo {author}
  {\bibfnamefont {L.}~\bibnamefont {Stolz}}, \bibinfo {author} {\bibfnamefont
  {T.}~\bibnamefont {Waldmann}}, \bibinfo {author} {\bibfnamefont
  {M.}~\bibnamefont {Wohlfahrt-Mehrens}}, \bibinfo {author} {\bibfnamefont
  {D.}~\bibnamefont {Aurbach}}, \bibinfo {author} {\bibfnamefont
  {M.}~\bibnamefont {Winter}}, \bibinfo {author} {\bibfnamefont
  {Y.}~\bibnamefont {Ein-Eli}}, \ and\ \bibinfo {author} {\bibfnamefont
  {J.}~\bibnamefont {Janek}},\ }\href {\doibase 10.1002/aenm.202101126}
  {\bibfield  {journal} {\bibinfo  {journal} {Advanced Energy Materials}\
  }\textbf {\bibinfo {volume} {11}},\ \bibinfo {pages} {2101126} (\bibinfo
  {year} {2021})}\BibitemShut {NoStop}%
\bibitem [{\citenamefont {Meng}\ \emph {et~al.}(2021)\citenamefont {Meng},
  \citenamefont {Zhu},\ and\ \citenamefont {Lian}}]{Meng2021}%
  \BibitemOpen
  \bibfield  {author} {\bibinfo {author} {\bibfnamefont {N.}~\bibnamefont
  {Meng}}, \bibinfo {author} {\bibfnamefont {X.}~\bibnamefont {Zhu}}, \ and\
  \bibinfo {author} {\bibfnamefont {F.}~\bibnamefont {Lian}},\ }\href {\doibase
  10.1016/j.partic.2021.04.002} {\bibfield  {journal} {\bibinfo  {journal}
  {Particuology}\ }\textbf {\bibinfo {volume} {60}},\ \bibinfo {pages} {14}
  (\bibinfo {year} {2021})}\BibitemShut {NoStop}%
\bibitem [{\citenamefont {Janek}\ and\ \citenamefont
  {Zeier}(2016)}]{Janek2016}%
  \BibitemOpen
  \bibfield  {author} {\bibinfo {author} {\bibfnamefont {J.}~\bibnamefont
  {Janek}}\ and\ \bibinfo {author} {\bibfnamefont {W.~G.}\ \bibnamefont
  {Zeier}},\ }\href {\doibase https://doi.org/10.1038/nenergy.2016.141}
  {\bibfield  {journal} {\bibinfo  {journal} {Nature Energy}\ }\textbf
  {\bibinfo {volume} {1}},\ \bibinfo {pages} {16141} (\bibinfo {year}
  {2016})}\BibitemShut {NoStop}%
\bibitem [{\citenamefont {Krauskopf}\ \emph {et~al.}(2020)\citenamefont
  {Krauskopf}, \citenamefont {Richter}, \citenamefont {Zeier},\ and\
  \citenamefont {Janek}}]{Krauskopf2020}%
  \BibitemOpen
  \bibfield  {author} {\bibinfo {author} {\bibfnamefont {T.}~\bibnamefont
  {Krauskopf}}, \bibinfo {author} {\bibfnamefont {F.~H.}\ \bibnamefont
  {Richter}}, \bibinfo {author} {\bibfnamefont {W.~G.}\ \bibnamefont {Zeier}},
  \ and\ \bibinfo {author} {\bibfnamefont {J.}~\bibnamefont {Janek}},\ }\href
  {\doibase 10.1021/acs.chemrev.0c00431} {\bibfield  {journal} {\bibinfo
  {journal} {Chemical Reviews}\ }\textbf {\bibinfo {volume} {120}},\ \bibinfo
  {pages} {7745} (\bibinfo {year} {2020})}\BibitemShut {NoStop}%
\bibitem [{\citenamefont {Li}\ \emph {et~al.}(2014)\citenamefont {Li},
  \citenamefont {Huang}, \citenamefont {{Yann Liaw}}, \citenamefont {Metzler},\
  and\ \citenamefont {Zhang}}]{Li2014}%
  \BibitemOpen
  \bibfield  {author} {\bibinfo {author} {\bibfnamefont {Z.}~\bibnamefont
  {Li}}, \bibinfo {author} {\bibfnamefont {J.}~\bibnamefont {Huang}}, \bibinfo
  {author} {\bibfnamefont {B.}~\bibnamefont {{Yann Liaw}}}, \bibinfo {author}
  {\bibfnamefont {V.}~\bibnamefont {Metzler}}, \ and\ \bibinfo {author}
  {\bibfnamefont {J.}~\bibnamefont {Zhang}},\ }\href {\doibase
  10.1016/j.jpowsour.2013.12.099} {\bibfield  {journal} {\bibinfo  {journal}
  {Journal of Power Sources}\ }\textbf {\bibinfo {volume} {254}},\ \bibinfo
  {pages} {168} (\bibinfo {year} {2014})}\BibitemShut {NoStop}%
\bibitem [{\citenamefont {Cheng}\ \emph {et~al.}(2015)\citenamefont {Cheng},
  \citenamefont {Zhang}, \citenamefont {Zhao}, \citenamefont {Wei},
  \citenamefont {Zhang},\ and\ \citenamefont {Zhang}}]{Cheng2015}%
  \BibitemOpen
  \bibfield  {author} {\bibinfo {author} {\bibfnamefont {X.~B.}\ \bibnamefont
  {Cheng}}, \bibinfo {author} {\bibfnamefont {R.}~\bibnamefont {Zhang}},
  \bibinfo {author} {\bibfnamefont {C.~Z.}\ \bibnamefont {Zhao}}, \bibinfo
  {author} {\bibfnamefont {F.}~\bibnamefont {Wei}}, \bibinfo {author}
  {\bibfnamefont {J.~G.}\ \bibnamefont {Zhang}}, \ and\ \bibinfo {author}
  {\bibfnamefont {Q.}~\bibnamefont {Zhang}},\ }\href {\doibase
  10.1002/advs.201500213} {\bibfield  {journal} {\bibinfo  {journal} {Advanced
  Science}\ }\textbf {\bibinfo {volume} {3}},\ \bibinfo {pages} {1} (\bibinfo
  {year} {2015})}\BibitemShut {NoStop}%
\bibitem [{\citenamefont {Zheng}\ \emph {et~al.}(2018)\citenamefont {Zheng},
  \citenamefont {Kotobuki}, \citenamefont {Song}, \citenamefont {Lai},\ and\
  \citenamefont {Lu}}]{Zheng2018}%
  \BibitemOpen
  \bibfield  {author} {\bibinfo {author} {\bibfnamefont {F.}~\bibnamefont
  {Zheng}}, \bibinfo {author} {\bibfnamefont {M.}~\bibnamefont {Kotobuki}},
  \bibinfo {author} {\bibfnamefont {S.}~\bibnamefont {Song}}, \bibinfo {author}
  {\bibfnamefont {M.~O.}\ \bibnamefont {Lai}}, \ and\ \bibinfo {author}
  {\bibfnamefont {L.}~\bibnamefont {Lu}},\ }\href {\doibase
  10.1016/j.jpowsour.2018.04.022} {\bibfield  {journal} {\bibinfo  {journal}
  {Journal of Power Sources}\ }\textbf {\bibinfo {volume} {389}},\ \bibinfo
  {pages} {198} (\bibinfo {year} {2018})}\BibitemShut {NoStop}%
\bibitem [{\citenamefont {Mindemark}\ \emph {et~al.}(2018)\citenamefont
  {Mindemark}, \citenamefont {Lacey}, \citenamefont {Bowden},\ and\
  \citenamefont {Brandell}}]{Mindemark2018}%
  \BibitemOpen
  \bibfield  {author} {\bibinfo {author} {\bibfnamefont {J.}~\bibnamefont
  {Mindemark}}, \bibinfo {author} {\bibfnamefont {M.~J.}\ \bibnamefont
  {Lacey}}, \bibinfo {author} {\bibfnamefont {T.}~\bibnamefont {Bowden}}, \
  and\ \bibinfo {author} {\bibfnamefont {D.}~\bibnamefont {Brandell}},\ }\href
  {\doibase 10.1016/j.progpolymsci.2017.12.004} {\bibfield  {journal} {\bibinfo
   {journal} {Progress in Polymer Science}\ }\textbf {\bibinfo {volume} {81}},\
  \bibinfo {pages} {114} (\bibinfo {year} {2018})}\BibitemShut {NoStop}%
\bibitem [{\citenamefont {Huo}\ \emph {et~al.}(2019)\citenamefont {Huo},
  \citenamefont {Chen}, \citenamefont {Luo}, \citenamefont {Yang},
  \citenamefont {Guo},\ and\ \citenamefont {Sun}}]{Huo2019}%
  \BibitemOpen
  \bibfield  {author} {\bibinfo {author} {\bibfnamefont {H.}~\bibnamefont
  {Huo}}, \bibinfo {author} {\bibfnamefont {Y.}~\bibnamefont {Chen}}, \bibinfo
  {author} {\bibfnamefont {J.}~\bibnamefont {Luo}}, \bibinfo {author}
  {\bibfnamefont {X.}~\bibnamefont {Yang}}, \bibinfo {author} {\bibfnamefont
  {X.}~\bibnamefont {Guo}}, \ and\ \bibinfo {author} {\bibfnamefont
  {X.}~\bibnamefont {Sun}},\ }\href {\doibase 10.1002/aenm.201804004}
  {\bibfield  {journal} {\bibinfo  {journal} {Advanced Energy Materials}\
  }\textbf {\bibinfo {volume} {9}},\ \bibinfo {pages} {1804004} (\bibinfo
  {year} {2019})}\BibitemShut {NoStop}%
\bibitem [{\citenamefont {Weiss}\ \emph {et~al.}(2020)\citenamefont {Weiss},
  \citenamefont {Simon}, \citenamefont {Busche}, \citenamefont {Nakamura},
  \citenamefont {Schr{\"{o}}der}, \citenamefont {Richter},\ and\ \citenamefont
  {Janek}}]{Weiss2020}%
  \BibitemOpen
  \bibfield  {author} {\bibinfo {author} {\bibfnamefont {M.}~\bibnamefont
  {Weiss}}, \bibinfo {author} {\bibfnamefont {F.~J.}\ \bibnamefont {Simon}},
  \bibinfo {author} {\bibfnamefont {M.~R.}\ \bibnamefont {Busche}}, \bibinfo
  {author} {\bibfnamefont {T.}~\bibnamefont {Nakamura}}, \bibinfo {author}
  {\bibfnamefont {D.}~\bibnamefont {Schr{\"{o}}der}}, \bibinfo {author}
  {\bibfnamefont {F.~H.}\ \bibnamefont {Richter}}, \ and\ \bibinfo {author}
  {\bibfnamefont {J.}~\bibnamefont {Janek}},\ }\href {\doibase
  10.1007/s41918-020-00062-7} {\bibfield  {journal} {\bibinfo  {journal}
  {Electrochemical Energy Reviews}\ }\textbf {\bibinfo {volume} {3}},\ \bibinfo
  {pages} {221} (\bibinfo {year} {2020})}\BibitemShut {NoStop}%
\bibitem [{\citenamefont {Neumann}\ \emph {et~al.}(2021)\citenamefont
  {Neumann}, \citenamefont {Hamann}, \citenamefont {Danner}, \citenamefont
  {Hein}, \citenamefont {Becker-Steinberger}, \citenamefont {Wachsman},\ and\
  \citenamefont {Latz}}]{Neumann2021}%
  \BibitemOpen
  \bibfield  {author} {\bibinfo {author} {\bibfnamefont {A.}~\bibnamefont
  {Neumann}}, \bibinfo {author} {\bibfnamefont {T.~R.}\ \bibnamefont {Hamann}},
  \bibinfo {author} {\bibfnamefont {T.}~\bibnamefont {Danner}}, \bibinfo
  {author} {\bibfnamefont {S.}~\bibnamefont {Hein}}, \bibinfo {author}
  {\bibfnamefont {K.}~\bibnamefont {Becker-Steinberger}}, \bibinfo {author}
  {\bibfnamefont {E.}~\bibnamefont {Wachsman}}, \ and\ \bibinfo {author}
  {\bibfnamefont {A.}~\bibnamefont {Latz}},\ }\href {\doibase
  https://doi.org/10.1021/acsaem.1c00362} {\bibfield  {journal} {\bibinfo
  {journal} {ACS Applied Energy Materials}\ ,\ \bibinfo {pages} {4786–4804}}
  (\bibinfo {year} {2021})}\BibitemShut {NoStop}%
\bibitem [{\citenamefont {Jetybayeva}\ \emph {et~al.}(2021)\citenamefont
  {Jetybayeva}, \citenamefont {Uzakbaiuly}, \citenamefont {Mukanova},
  \citenamefont {Myung},\ and\ \citenamefont {Bakenov}}]{Jetybayeva2021}%
  \BibitemOpen
  \bibfield  {author} {\bibinfo {author} {\bibfnamefont {A.}~\bibnamefont
  {Jetybayeva}}, \bibinfo {author} {\bibfnamefont {B.}~\bibnamefont
  {Uzakbaiuly}}, \bibinfo {author} {\bibfnamefont {A.}~\bibnamefont
  {Mukanova}}, \bibinfo {author} {\bibfnamefont {S.~T.}\ \bibnamefont {Myung}},
  \ and\ \bibinfo {author} {\bibfnamefont {Z.}~\bibnamefont {Bakenov}},\ }\href
  {\doibase 10.1039/d1ta02652f} {\bibfield  {journal} {\bibinfo  {journal}
  {Journal of Materials Chemistry A}\ }\textbf {\bibinfo {volume} {9}},\
  \bibinfo {pages} {15140} (\bibinfo {year} {2021})}\BibitemShut {NoStop}%
\bibitem [{\citenamefont {Liu}\ \emph {et~al.}(2020)\citenamefont {Liu},
  \citenamefont {Cheng}, \citenamefont {Huang}, \citenamefont {Yuan},
  \citenamefont {Lu}, \citenamefont {Yan}, \citenamefont {Zhu}, \citenamefont
  {Xu}, \citenamefont {Zhao}, \citenamefont {Hou}, \citenamefont {He},
  \citenamefont {Kaskel},\ and\ \citenamefont {Zhang}}]{Liu2020}%
  \BibitemOpen
  \bibfield  {author} {\bibinfo {author} {\bibfnamefont {H.}~\bibnamefont
  {Liu}}, \bibinfo {author} {\bibfnamefont {X.~B.}\ \bibnamefont {Cheng}},
  \bibinfo {author} {\bibfnamefont {J.~Q.}\ \bibnamefont {Huang}}, \bibinfo
  {author} {\bibfnamefont {H.}~\bibnamefont {Yuan}}, \bibinfo {author}
  {\bibfnamefont {Y.}~\bibnamefont {Lu}}, \bibinfo {author} {\bibfnamefont
  {C.}~\bibnamefont {Yan}}, \bibinfo {author} {\bibfnamefont {G.~L.}\
  \bibnamefont {Zhu}}, \bibinfo {author} {\bibfnamefont {R.}~\bibnamefont
  {Xu}}, \bibinfo {author} {\bibfnamefont {C.~Z.}\ \bibnamefont {Zhao}},
  \bibinfo {author} {\bibfnamefont {L.~P.}\ \bibnamefont {Hou}}, \bibinfo
  {author} {\bibfnamefont {C.}~\bibnamefont {He}}, \bibinfo {author}
  {\bibfnamefont {S.}~\bibnamefont {Kaskel}}, \ and\ \bibinfo {author}
  {\bibfnamefont {Q.}~\bibnamefont {Zhang}},\ }\href {\doibase
  10.1021/acsenergylett.9b02660} {\bibfield  {journal} {\bibinfo  {journal}
  {ACS Energy Letters}\ }\textbf {\bibinfo {volume} {5}},\ \bibinfo {pages}
  {833} (\bibinfo {year} {2020})}\BibitemShut {NoStop}%
\bibitem [{\citenamefont {Bocharova}\ and\ \citenamefont
  {Sokolov}(2020)}]{doi:10.1021/acs.macromol.9b02742}%
  \BibitemOpen
  \bibfield  {author} {\bibinfo {author} {\bibfnamefont {V.}~\bibnamefont
  {Bocharova}}\ and\ \bibinfo {author} {\bibfnamefont {A.~P.}\ \bibnamefont
  {Sokolov}},\ }\href {https://doi.org/10.1021/acs.macromol.9b02742} {\bibfield
   {journal} {\bibinfo  {journal} {Macromolecules}\ }\textbf {\bibinfo {volume}
  {53}},\ \bibinfo {pages} {4141} (\bibinfo {year} {2020})}\BibitemShut
  {NoStop}%
\bibitem [{\citenamefont {Frenck}\ \emph {et~al.}(2019)\citenamefont {Frenck},
  \citenamefont {Sethi}, \citenamefont {Maslyn},\ and\ \citenamefont
  {Balsara}}]{Frenck2019}%
  \BibitemOpen
  \bibfield  {author} {\bibinfo {author} {\bibfnamefont {L.}~\bibnamefont
  {Frenck}}, \bibinfo {author} {\bibfnamefont {G.~K.}\ \bibnamefont {Sethi}},
  \bibinfo {author} {\bibfnamefont {J.~A.}\ \bibnamefont {Maslyn}}, \ and\
  \bibinfo {author} {\bibfnamefont {N.~P.}\ \bibnamefont {Balsara}},\ }\href
  {\doibase 10.3389/fenrg.2019.00115} {\bibfield  {journal} {\bibinfo
  {journal} {Frontiers in Energy Research}\ }\textbf {\bibinfo {volume} {7}},\
  \bibinfo {pages} {115} (\bibinfo {year} {2019})}\BibitemShut {NoStop}%
\bibitem [{\citenamefont {Wright}(1975)}]{Wright1975}%
  \BibitemOpen
  \bibfield  {author} {\bibinfo {author} {\bibfnamefont {P.~V.}\ \bibnamefont
  {Wright}},\ }\href {\doibase 10.1002/pi.4980070505} {\bibfield  {journal}
  {\bibinfo  {journal} {British Polymer Journal}\ }\textbf {\bibinfo {volume}
  {7}},\ \bibinfo {pages} {319} (\bibinfo {year} {1975})}\BibitemShut {NoStop}%
\bibitem [{\citenamefont {Bresser}\ \emph {et~al.}(2019)\citenamefont
  {Bresser}, \citenamefont {Lyonnard}, \citenamefont {Iojoiu}, \citenamefont
  {Picard},\ and\ \citenamefont {Passerini}}]{Bresser2019}%
  \BibitemOpen
  \bibfield  {author} {\bibinfo {author} {\bibfnamefont {D.}~\bibnamefont
  {Bresser}}, \bibinfo {author} {\bibfnamefont {S.}~\bibnamefont {Lyonnard}},
  \bibinfo {author} {\bibfnamefont {C.}~\bibnamefont {Iojoiu}}, \bibinfo
  {author} {\bibfnamefont {L.}~\bibnamefont {Picard}}, \ and\ \bibinfo {author}
  {\bibfnamefont {S.}~\bibnamefont {Passerini}},\ }\href {\doibase
  10.1039/c9me00038k} {\bibfield  {journal} {\bibinfo  {journal} {Molecular
  Systems Design and Engineering}\ }\textbf {\bibinfo {volume} {4}},\ \bibinfo
  {pages} {779} (\bibinfo {year} {2019})}\BibitemShut {NoStop}%
\bibitem [{\citenamefont {Nguyen}\ \emph {et~al.}(2018)\citenamefont {Nguyen},
  \citenamefont {Kim}, \citenamefont {Shi}, \citenamefont {Paillard},
  \citenamefont {Judeinstein}, \citenamefont {Lyonnard}, \citenamefont
  {Bresser},\ and\ \citenamefont {Iojoiu}}]{Nguyen2018}%
  \BibitemOpen
  \bibfield  {author} {\bibinfo {author} {\bibfnamefont {H.~D.}\ \bibnamefont
  {Nguyen}}, \bibinfo {author} {\bibfnamefont {G.~T.}\ \bibnamefont {Kim}},
  \bibinfo {author} {\bibfnamefont {J.}~\bibnamefont {Shi}}, \bibinfo {author}
  {\bibfnamefont {E.}~\bibnamefont {Paillard}}, \bibinfo {author}
  {\bibfnamefont {P.}~\bibnamefont {Judeinstein}}, \bibinfo {author}
  {\bibfnamefont {S.}~\bibnamefont {Lyonnard}}, \bibinfo {author}
  {\bibfnamefont {D.}~\bibnamefont {Bresser}}, \ and\ \bibinfo {author}
  {\bibfnamefont {C.}~\bibnamefont {Iojoiu}},\ }\href {\doibase
  10.1039/c8ee02093k} {\bibfield  {journal} {\bibinfo  {journal} {Energy and
  Environmental Science}\ }\textbf {\bibinfo {volume} {11}},\ \bibinfo {pages}
  {3298} (\bibinfo {year} {2018})}\BibitemShut {NoStop}%
\bibitem [{\citenamefont {Nematdoust}\ \emph {et~al.}(2020)\citenamefont
  {Nematdoust}, \citenamefont {Najjar}, \citenamefont {Bresser},\ and\
  \citenamefont {Passerini}}]{Nematdoust2020}%
  \BibitemOpen
  \bibfield  {author} {\bibinfo {author} {\bibfnamefont {S.}~\bibnamefont
  {Nematdoust}}, \bibinfo {author} {\bibfnamefont {R.}~\bibnamefont {Najjar}},
  \bibinfo {author} {\bibfnamefont {D.}~\bibnamefont {Bresser}}, \ and\
  \bibinfo {author} {\bibfnamefont {S.}~\bibnamefont {Passerini}},\ }\href
  {\doibase 10.1021/acs.jpcc.0c08749} {\bibfield  {journal} {\bibinfo
  {journal} {Journal of Physical Chemistry C}\ }\textbf {\bibinfo {volume}
  {124}},\ \bibinfo {pages} {27907} (\bibinfo {year} {2020})}\BibitemShut
  {NoStop}%
\bibitem [{\citenamefont {Butzelaar}\ \emph {et~al.}(2021)\citenamefont
  {Butzelaar}, \citenamefont {Liu}, \citenamefont {R{\"{o}}ring}, \citenamefont
  {Brunklaus}, \citenamefont {Winter},\ and\ \citenamefont
  {Theato}}]{Butzelaar2021}%
  \BibitemOpen
  \bibfield  {author} {\bibinfo {author} {\bibfnamefont {A.~J.}\ \bibnamefont
  {Butzelaar}}, \bibinfo {author} {\bibfnamefont {K.~L.}\ \bibnamefont {Liu}},
  \bibinfo {author} {\bibfnamefont {P.}~\bibnamefont {R{\"{o}}ring}}, \bibinfo
  {author} {\bibfnamefont {G.}~\bibnamefont {Brunklaus}}, \bibinfo {author}
  {\bibfnamefont {M.}~\bibnamefont {Winter}}, \ and\ \bibinfo {author}
  {\bibfnamefont {P.}~\bibnamefont {Theato}},\ }\href {\doibase
  10.1021/acsapm.0c01398} {\bibfield  {journal} {\bibinfo  {journal} {ACS
  Applied Polymer Materials}\ }\textbf {\bibinfo {volume} {3}},\ \bibinfo
  {pages} {1573} (\bibinfo {year} {2021})}\BibitemShut {NoStop}%
\bibitem [{\citenamefont {Angell}\ \emph {et~al.}(1993)\citenamefont {Angell},
  \citenamefont {Liu},\ and\ \citenamefont {Sanchez}}]{Angell1993}%
  \BibitemOpen
  \bibfield  {author} {\bibinfo {author} {\bibfnamefont {C.~A.}\ \bibnamefont
  {Angell}}, \bibinfo {author} {\bibfnamefont {C.}~\bibnamefont {Liu}}, \ and\
  \bibinfo {author} {\bibfnamefont {E.}~\bibnamefont {Sanchez}},\ }\href
  {\doibase 10.1038/362137a0} {\bibfield  {journal} {\bibinfo  {journal}
  {Nature}\ }\textbf {\bibinfo {volume} {362}},\ \bibinfo {pages} {137}
  (\bibinfo {year} {1993})}\BibitemShut {NoStop}%
\bibitem [{\citenamefont {Zhou}\ \emph {et~al.}(2016)\citenamefont {Zhou},
  \citenamefont {Wang}, \citenamefont {Li}, \citenamefont {Xin}, \citenamefont
  {Manthiram},\ and\ \citenamefont {Goodenough}}]{Zhou2016}%
  \BibitemOpen
  \bibfield  {author} {\bibinfo {author} {\bibfnamefont {W.}~\bibnamefont
  {Zhou}}, \bibinfo {author} {\bibfnamefont {S.}~\bibnamefont {Wang}}, \bibinfo
  {author} {\bibfnamefont {Y.}~\bibnamefont {Li}}, \bibinfo {author}
  {\bibfnamefont {S.}~\bibnamefont {Xin}}, \bibinfo {author} {\bibfnamefont
  {A.}~\bibnamefont {Manthiram}}, \ and\ \bibinfo {author} {\bibfnamefont
  {J.~B.}\ \bibnamefont {Goodenough}},\ }\href {\doibase 10.1021/jacs.6b05341}
  {\bibfield  {journal} {\bibinfo  {journal} {Journal of the American Chemical
  Society}\ }\textbf {\bibinfo {volume} {138}},\ \bibinfo {pages} {9385}
  (\bibinfo {year} {2016})}\BibitemShut {NoStop}%
\bibitem [{\citenamefont {Maitra}\ and\ \citenamefont
  {Heuer}(2007)}]{Maitra2007}%
  \BibitemOpen
  \bibfield  {author} {\bibinfo {author} {\bibfnamefont {A.}~\bibnamefont
  {Maitra}}\ and\ \bibinfo {author} {\bibfnamefont {A.}~\bibnamefont {Heuer}},\
  }\href {\doibase 10.1103/PhysRevLett.98.227802} {\bibfield  {journal}
  {\bibinfo  {journal} {Physical Review Letters}\ }\textbf {\bibinfo {volume}
  {98}},\ \bibinfo {pages} {1} (\bibinfo {year} {2007})}\BibitemShut {NoStop}%
\bibitem [{\citenamefont {Diddens}\ \emph {et~al.}(2010)\citenamefont
  {Diddens}, \citenamefont {Heuer},\ and\ \citenamefont
  {Borodin}}]{Diddens2010}%
  \BibitemOpen
  \bibfield  {author} {\bibinfo {author} {\bibfnamefont {D.}~\bibnamefont
  {Diddens}}, \bibinfo {author} {\bibfnamefont {A.}~\bibnamefont {Heuer}}, \
  and\ \bibinfo {author} {\bibfnamefont {O.}~\bibnamefont {Borodin}},\ }\href
  {\doibase 10.1021/ma901893h} {\bibfield  {journal} {\bibinfo  {journal}
  {Macromolecules}\ }\textbf {\bibinfo {volume} {43}},\ \bibinfo {pages} {2028}
  (\bibinfo {year} {2010})}\BibitemShut {NoStop}%
\bibitem [{\citenamefont {Mackanic}\ \emph {et~al.}(2018)\citenamefont
  {Mackanic}, \citenamefont {Michaels}, \citenamefont {Lee}, \citenamefont
  {Feng}, \citenamefont {Lopez}, \citenamefont {Qin}, \citenamefont {Cui},\
  and\ \citenamefont {Bao}}]{Mackanic2018}%
  \BibitemOpen
  \bibfield  {author} {\bibinfo {author} {\bibfnamefont {D.~G.}\ \bibnamefont
  {Mackanic}}, \bibinfo {author} {\bibfnamefont {W.}~\bibnamefont {Michaels}},
  \bibinfo {author} {\bibfnamefont {M.}~\bibnamefont {Lee}}, \bibinfo {author}
  {\bibfnamefont {D.}~\bibnamefont {Feng}}, \bibinfo {author} {\bibfnamefont
  {J.}~\bibnamefont {Lopez}}, \bibinfo {author} {\bibfnamefont
  {J.}~\bibnamefont {Qin}}, \bibinfo {author} {\bibfnamefont {Y.}~\bibnamefont
  {Cui}}, \ and\ \bibinfo {author} {\bibfnamefont {Z.}~\bibnamefont {Bao}},\
  }\href {\doibase 10.1002/aenm.201800703} {\bibfield  {journal} {\bibinfo
  {journal} {Advanced Energy Materials}\ }\textbf {\bibinfo {volume} {8}},\
  \bibinfo {pages} {1} (\bibinfo {year} {2018})}\BibitemShut {NoStop}%
\bibitem [{\citenamefont {Liivat}(2011)}]{Liivat2011}%
  \BibitemOpen
  \bibfield  {author} {\bibinfo {author} {\bibfnamefont {A.}~\bibnamefont
  {Liivat}},\ }\href {\doibase 10.1016/j.electacta.2011.06.097} {\bibfield
  {journal} {\bibinfo  {journal} {Electrochimica Acta}\ }\textbf {\bibinfo
  {volume} {57}},\ \bibinfo {pages} {244} (\bibinfo {year} {2011})}\BibitemShut
  {NoStop}%
\bibitem [{\citenamefont {Ebadi}\ \emph {et~al.}(2017)\citenamefont {Ebadi},
  \citenamefont {Costa}, \citenamefont {Araujo},\ and\ \citenamefont
  {Brandell}}]{Ebadi2017}%
  \BibitemOpen
  \bibfield  {author} {\bibinfo {author} {\bibfnamefont {M.}~\bibnamefont
  {Ebadi}}, \bibinfo {author} {\bibfnamefont {L.~T.}\ \bibnamefont {Costa}},
  \bibinfo {author} {\bibfnamefont {C.~M.}\ \bibnamefont {Araujo}}, \ and\
  \bibinfo {author} {\bibfnamefont {D.}~\bibnamefont {Brandell}},\ }\href
  {\doibase 10.1016/j.electacta.2017.03.030} {\bibfield  {journal} {\bibinfo
  {journal} {Electrochimica Acta}\ }\textbf {\bibinfo {volume} {234}},\
  \bibinfo {pages} {43} (\bibinfo {year} {2017})}\BibitemShut {NoStop}%
\bibitem [{\citenamefont {Thum}\ \emph {et~al.}(2021)\citenamefont {Thum},
  \citenamefont {Diddens},\ and\ \citenamefont {Heuer}}]{Thum2021}%
  \BibitemOpen
  \bibfield  {author} {\bibinfo {author} {\bibfnamefont {A.}~\bibnamefont
  {Thum}}, \bibinfo {author} {\bibfnamefont {D.}~\bibnamefont {Diddens}}, \
  and\ \bibinfo {author} {\bibfnamefont {A.}~\bibnamefont {Heuer}},\ }\href
  {\doibase 10.1021/acs.jpcc.1c07751} {\bibfield  {journal} {\bibinfo
  {journal} {Journal of Physical Chemistry C}\ }\textbf {\bibinfo {volume}
  {125}},\ \bibinfo {pages} {25392} (\bibinfo {year} {2021})}\BibitemShut
  {NoStop}%
\bibitem [{\citenamefont {Johansson}(2015)}]{Johansson2015}%
  \BibitemOpen
  \bibfield  {author} {\bibinfo {author} {\bibfnamefont {P.}~\bibnamefont
  {Johansson}},\ }\href {\doibase 10.1016/j.electacta.2015.03.116} {\bibfield
  {journal} {\bibinfo  {journal} {Electrochimica Acta}\ }\textbf {\bibinfo
  {volume} {175}},\ \bibinfo {pages} {42} (\bibinfo {year} {2015})}\BibitemShut
  {NoStop}%
\bibitem [{\citenamefont {Siekierski}\ \emph {et~al.}(2007)\citenamefont
  {Siekierski}, \citenamefont {Wieczorek},\ and\ \citenamefont
  {Nadara}}]{Siekierski2007}%
  \BibitemOpen
  \bibfield  {author} {\bibinfo {author} {\bibfnamefont {M.}~\bibnamefont
  {Siekierski}}, \bibinfo {author} {\bibfnamefont {W.}~\bibnamefont
  {Wieczorek}}, \ and\ \bibinfo {author} {\bibfnamefont {K.}~\bibnamefont
  {Nadara}},\ }\href {\doibase 10.1016/j.electacta.2007.04.033} {\bibfield
  {journal} {\bibinfo  {journal} {Electrochimica Acta}\ }\textbf {\bibinfo
  {volume} {53}},\ \bibinfo {pages} {1556} (\bibinfo {year}
  {2007})}\BibitemShut {NoStop}%
\bibitem [{\citenamefont {Katzenmeier}\ \emph {et~al.}(2022)\citenamefont
  {Katzenmeier}, \citenamefont {Manuel}, \citenamefont {Gagliardi},\ and\
  \citenamefont {Bandarenka}}]{Katzenmeier2022}%
  \BibitemOpen
  \bibfield  {author} {\bibinfo {author} {\bibfnamefont {L.}~\bibnamefont
  {Katzenmeier}}, \bibinfo {author} {\bibfnamefont {G.}~\bibnamefont {Manuel}},
  \bibinfo {author} {\bibfnamefont {A.}~\bibnamefont {Gagliardi}}, \ and\
  \bibinfo {author} {\bibfnamefont {A.~S.}\ \bibnamefont {Bandarenka}},\ }\href
  {\doibase 10.1021/acs.jpcc.2c02481} {\bibfield  {journal} {\bibinfo
  {journal} {Journal of Physical Chemistry C}\ }\textbf {\bibinfo {volume}
  {26}},\ \bibinfo {pages} {10900} (\bibinfo {year} {2022})}\BibitemShut
  {NoStop}%
\bibitem [{\citenamefont {Natsiavas}\ \emph {et~al.}(2016)\citenamefont
  {Natsiavas}, \citenamefont {Weinberg}, \citenamefont {Rosato},\ and\
  \citenamefont {Ortiz}}]{Natsiavas2016}%
  \BibitemOpen
  \bibfield  {author} {\bibinfo {author} {\bibfnamefont {P.~P.}\ \bibnamefont
  {Natsiavas}}, \bibinfo {author} {\bibfnamefont {K.}~\bibnamefont {Weinberg}},
  \bibinfo {author} {\bibfnamefont {D.}~\bibnamefont {Rosato}}, \ and\ \bibinfo
  {author} {\bibfnamefont {M.}~\bibnamefont {Ortiz}},\ }\href {\doibase
  10.1016/j.jmps.2016.05.007} {\bibfield  {journal} {\bibinfo  {journal}
  {Journal of the Mechanics and Physics of Solids}\ }\textbf {\bibinfo {volume}
  {95}},\ \bibinfo {pages} {92} (\bibinfo {year} {2016})}\BibitemShut {NoStop}%
\bibitem [{\citenamefont {Bucci}\ \emph {et~al.}(2016)\citenamefont {Bucci},
  \citenamefont {Chiang},\ and\ \citenamefont {Carter}}]{Bucci2016}%
  \BibitemOpen
  \bibfield  {author} {\bibinfo {author} {\bibfnamefont {G.}~\bibnamefont
  {Bucci}}, \bibinfo {author} {\bibfnamefont {Y.~M.}\ \bibnamefont {Chiang}}, \
  and\ \bibinfo {author} {\bibfnamefont {W.~C.}\ \bibnamefont {Carter}},\
  }\href {\doibase 10.1016/j.actamat.2015.11.030} {\bibfield  {journal}
  {\bibinfo  {journal} {Acta Materialia}\ }\textbf {\bibinfo {volume} {104}},\
  \bibinfo {pages} {33} (\bibinfo {year} {2016})}\BibitemShut {NoStop}%
\bibitem [{\citenamefont {Grazioli}\ \emph {et~al.}(2019)\citenamefont
  {Grazioli}, \citenamefont {Verners}, \citenamefont {Zadin}, \citenamefont
  {Brandell},\ and\ \citenamefont {Simone}}]{Grazioli2019}%
  \BibitemOpen
  \bibfield  {author} {\bibinfo {author} {\bibfnamefont {D.}~\bibnamefont
  {Grazioli}}, \bibinfo {author} {\bibfnamefont {O.}~\bibnamefont {Verners}},
  \bibinfo {author} {\bibfnamefont {V.}~\bibnamefont {Zadin}}, \bibinfo
  {author} {\bibfnamefont {D.}~\bibnamefont {Brandell}}, \ and\ \bibinfo
  {author} {\bibfnamefont {A.}~\bibnamefont {Simone}},\ }\href {\doibase
  10.1016/j.electacta.2018.07.234} {\bibfield  {journal} {\bibinfo  {journal}
  {Electrochimica Acta}\ }\textbf {\bibinfo {volume} {296}},\ \bibinfo {pages}
  {1122} (\bibinfo {year} {2019})}\BibitemShut {NoStop}%
\bibitem [{\citenamefont {Dickinson}\ and\ \citenamefont
  {Smith}(2020)}]{Dickinson2020}%
  \BibitemOpen
  \bibfield  {author} {\bibinfo {author} {\bibfnamefont {E.~J.}\ \bibnamefont
  {Dickinson}}\ and\ \bibinfo {author} {\bibfnamefont {G.}~\bibnamefont
  {Smith}},\ }\href {\doibase 10.3390/membranes10110310} {\bibfield  {journal}
  {\bibinfo  {journal} {Membranes}\ }\textbf {\bibinfo {volume} {10}},\
  \bibinfo {pages} {1} (\bibinfo {year} {2020})}\BibitemShut {NoStop}%
\bibitem [{\citenamefont {Narayan}\ and\ \citenamefont
  {Anand}(2021)}]{Narayan2021}%
  \BibitemOpen
  \bibfield  {author} {\bibinfo {author} {\bibfnamefont {S.}~\bibnamefont
  {Narayan}}\ and\ \bibinfo {author} {\bibfnamefont {L.}~\bibnamefont
  {Anand}},\ }\href {\doibase 10.1016/j.jmps.2021.104734} {\bibfield  {journal}
  {\bibinfo  {journal} {Journal of the Mechanics and Physics of Solids}\
  }\textbf {\bibinfo {volume} {159}},\ \bibinfo {pages} {104734} (\bibinfo
  {year} {2021})}\BibitemShut {NoStop}%
\bibitem [{\citenamefont {Latz}\ and\ \citenamefont {Zausch}(2011)}]{Latz2011}%
  \BibitemOpen
  \bibfield  {author} {\bibinfo {author} {\bibfnamefont {A.}~\bibnamefont
  {Latz}}\ and\ \bibinfo {author} {\bibfnamefont {J.}~\bibnamefont {Zausch}},\
  }\href {\doibase 10.1016/j.jpowsour.2010.11.088} {\bibfield  {journal}
  {\bibinfo  {journal} {Journal of Power Sources}\ }\textbf {\bibinfo {volume}
  {196}},\ \bibinfo {pages} {3296} (\bibinfo {year} {2011})}\BibitemShut
  {NoStop}%
\bibitem [{\citenamefont {Latz}\ and\ \citenamefont {Zausch}(2015)}]{Latz2015}%
  \BibitemOpen
  \bibfield  {author} {\bibinfo {author} {\bibfnamefont {A.}~\bibnamefont
  {Latz}}\ and\ \bibinfo {author} {\bibfnamefont {J.}~\bibnamefont {Zausch}},\
  }\href {\doibase 10.3762/bjnano.6.102} {\bibfield  {journal} {\bibinfo
  {journal} {Beilstein Journal of Nanotechnology}\ }\textbf {\bibinfo {volume}
  {6}},\ \bibinfo {pages} {987} (\bibinfo {year} {2015})}\BibitemShut {NoStop}%
\bibitem [{\citenamefont {Braun}\ \emph {et~al.}(2015)\citenamefont {Braun},
  \citenamefont {Yada},\ and\ \citenamefont {Latz}}]{Braun2015}%
  \BibitemOpen
  \bibfield  {author} {\bibinfo {author} {\bibfnamefont {S.}~\bibnamefont
  {Braun}}, \bibinfo {author} {\bibfnamefont {C.}~\bibnamefont {Yada}}, \ and\
  \bibinfo {author} {\bibfnamefont {A.}~\bibnamefont {Latz}},\ }\href {\doibase
  10.1021/acs.jpcc.5b02679} {\bibfield  {journal} {\bibinfo  {journal} {Journal
  of Physical Chemistry C}\ }\textbf {\bibinfo {volume} {119}},\ \bibinfo
  {pages} {22281} (\bibinfo {year} {2015})}\BibitemShut {NoStop}%
\bibitem [{\citenamefont {Schammer}\ \emph {et~al.}(2021)\citenamefont
  {Schammer}, \citenamefont {Horstmann},\ and\ \citenamefont
  {Latz}}]{Schammer2021}%
  \BibitemOpen
  \bibfield  {author} {\bibinfo {author} {\bibfnamefont {M.}~\bibnamefont
  {Schammer}}, \bibinfo {author} {\bibfnamefont {B.}~\bibnamefont {Horstmann}},
  \ and\ \bibinfo {author} {\bibfnamefont {A.}~\bibnamefont {Latz}},\ }\href
  {\doibase 10.1149/1945-7111/abdddf} {\bibfield  {journal} {\bibinfo
  {journal} {Journal of the Electrochemical Society}\ }\textbf {\bibinfo
  {volume} {168}},\ \bibinfo {pages} {026511} (\bibinfo {year}
  {2021})}\BibitemShut {NoStop}%
\bibitem [{\citenamefont {von Kolzenberg}\ \emph {et~al.}(2021)\citenamefont
  {von Kolzenberg}, \citenamefont {Latz},\ and\ \citenamefont
  {Horstmann}}]{VonKolzenberg2021}%
  \BibitemOpen
  \bibfield  {author} {\bibinfo {author} {\bibfnamefont {L.}~\bibnamefont {von
  Kolzenberg}}, \bibinfo {author} {\bibfnamefont {A.}~\bibnamefont {Latz}}, \
  and\ \bibinfo {author} {\bibfnamefont {B.}~\bibnamefont {Horstmann}},\ }\href
  {\doibase 10.1002/batt.202100216} {\bibfield  {journal} {\bibinfo  {journal}
  {Batteries and Supercaps}\ }\textbf {\bibinfo {volume} {202100216}},\
  \bibinfo {pages} {1} (\bibinfo {year} {2021})}\BibitemShut {NoStop}%
\bibitem [{\citenamefont {Kilchert}\ \emph {et~al.}(2023)\citenamefont
  {Kilchert}, \citenamefont {Lorenz}, \citenamefont {Schammer}, \citenamefont
  {N{\"{u}}rnberg}, \citenamefont {Sch{\"{o}}nhoff}, \citenamefont {Latz},\
  and\ \citenamefont {Horstmann}}]{Kilchert2022}%
  \BibitemOpen
  \bibfield  {author} {\bibinfo {author} {\bibfnamefont {F.}~\bibnamefont
  {Kilchert}}, \bibinfo {author} {\bibfnamefont {M.}~\bibnamefont {Lorenz}},
  \bibinfo {author} {\bibfnamefont {M.}~\bibnamefont {Schammer}}, \bibinfo
  {author} {\bibfnamefont {P.}~\bibnamefont {N{\"{u}}rnberg}}, \bibinfo
  {author} {\bibfnamefont {M.}~\bibnamefont {Sch{\"{o}}nhoff}}, \bibinfo
  {author} {\bibfnamefont {A.}~\bibnamefont {Latz}}, \ and\ \bibinfo {author}
  {\bibfnamefont {B.}~\bibnamefont {Horstmann}},\ }\href {\doibase
  https://doi.org/10.1039/D2CP04423D} {\bibfield  {journal} {\bibinfo
  {journal} {Physical Chemistry Chemical Physics}\ }\textbf {\bibinfo {volume}
  {25}},\ \bibinfo {pages} {25965} (\bibinfo {year} {2023})}\BibitemShut
  {NoStop}%
\bibitem [{\citenamefont {Lorenz}\ \emph {et~al.}(2022)\citenamefont {Lorenz},
  \citenamefont {Kilchert}, \citenamefont {Nürnberg}, \citenamefont
  {Schammer}, \citenamefont {Latz}, \citenamefont {Horstmann},\ and\
  \citenamefont {Schönhoff}}]{doi:10.1021/acs.jpclett.2c02398}%
  \BibitemOpen
  \bibfield  {author} {\bibinfo {author} {\bibfnamefont {M.}~\bibnamefont
  {Lorenz}}, \bibinfo {author} {\bibfnamefont {F.}~\bibnamefont {Kilchert}},
  \bibinfo {author} {\bibfnamefont {P.}~\bibnamefont {Nürnberg}}, \bibinfo
  {author} {\bibfnamefont {M.}~\bibnamefont {Schammer}}, \bibinfo {author}
  {\bibfnamefont {A.}~\bibnamefont {Latz}}, \bibinfo {author} {\bibfnamefont
  {B.}~\bibnamefont {Horstmann}}, \ and\ \bibinfo {author} {\bibfnamefont
  {M.}~\bibnamefont {Schönhoff}},\ }\href {\doibase
  https://doi.org/10.1021/acs.jpclett.2c02398} {\bibfield  {journal} {\bibinfo
  {journal} {The Journal of Physical Chemistry Letters}\ }\textbf {\bibinfo
  {volume} {13}},\ \bibinfo {pages} {8761} (\bibinfo {year}
  {2022})}\BibitemShut {NoStop}%
\bibitem [{\citenamefont {Rudin}(1998)}]{rudin1998elements}%
  \BibitemOpen
  \bibfield  {author} {\bibinfo {author} {\bibfnamefont {A.}~\bibnamefont
  {Rudin}},\ }\href@noop {} {\emph {\bibinfo {title} {Elements of Polymer
  Science \& Engineering}}}\ (\bibinfo  {publisher} {Elsevier, Amsterdam},\
  \bibinfo {year} {1998})\BibitemShut {NoStop}%
\bibitem [{\citenamefont {Ligia}\ and\ \citenamefont
  {Deodato}(2009)}]{Ligia2009}%
  \BibitemOpen
  \bibfield  {author} {\bibinfo {author} {\bibfnamefont {G.}~\bibnamefont
  {Ligia}}\ and\ \bibinfo {author} {\bibfnamefont {R.}~\bibnamefont
  {Deodato}},\ }\href@noop {} {\emph {\bibinfo {title} {Physicochemical
  Behavior and Supramolecular Organization of Polymers}}}\ (\bibinfo
  {publisher} {Springer Netherlands},\ \bibinfo {year} {2009})\BibitemShut
  {NoStop}%
\bibitem [{\citenamefont {Schammer}\ \emph {et~al.}(2022)\citenamefont
  {Schammer}, \citenamefont {Latz},\ and\ \citenamefont
  {Horstmann}}]{doi:10.1021/acs.jpcb.2c00215}%
  \BibitemOpen
  \bibfield  {author} {\bibinfo {author} {\bibfnamefont {M.}~\bibnamefont
  {Schammer}}, \bibinfo {author} {\bibfnamefont {A.}~\bibnamefont {Latz}}, \
  and\ \bibinfo {author} {\bibfnamefont {B.}~\bibnamefont {Horstmann}},\ }\href
  {\doibase https://doi.org/10.1021/acs.jpcb.2c00215} {\bibfield  {journal}
  {\bibinfo  {journal} {The Journal of Physical Chemistry B}\ }\textbf
  {\bibinfo {volume} {126}},\ \bibinfo {pages} {2761} (\bibinfo {year}
  {2022})}\BibitemShut {NoStop}%
\bibitem [{\citenamefont {Becker-Steinberger}\ \emph
  {et~al.}(2021)\citenamefont {Becker-Steinberger}, \citenamefont {Schardt},
  \citenamefont {Horstmann},\ and\ \citenamefont {Latz}}]{KBSarxiv}%
  \BibitemOpen
  \bibfield  {author} {\bibinfo {author} {\bibfnamefont {K.}~\bibnamefont
  {Becker-Steinberger}}, \bibinfo {author} {\bibfnamefont {S.}~\bibnamefont
  {Schardt}}, \bibinfo {author} {\bibfnamefont {B.}~\bibnamefont {Horstmann}},
  \ and\ \bibinfo {author} {\bibfnamefont {A.}~\bibnamefont {Latz}},\
  }\href@noop {} {\bibfield  {journal} {\bibinfo  {journal} {arXiv}\ }
  (\bibinfo {year} {2021})},\ \Eprint
  {http://arxiv.org/abs/2101.10294v1}{arXiv:2101.10294v1}\BibitemShut {NoStop}%
\bibitem [{\citenamefont {M{\"{u}}ller}(2001)}]{Mueller2001}%
  \BibitemOpen
  \bibfield  {author} {\bibinfo {author} {\bibfnamefont {I.}~\bibnamefont
  {M{\"{u}}ller}},\ }\href@noop {} {\emph {\bibinfo {title} {{Grundz{\"{u}}ge
  der Thermodynamik}}}}\ (\bibinfo  {publisher} {Springer Berling,
  Heidelberg},\ \bibinfo {year} {2001})\BibitemShut {NoStop}%
\bibitem [{\citenamefont {Holzapfel}(2000)}]{Holzapfel2000}%
  \BibitemOpen
  \bibfield  {author} {\bibinfo {author} {\bibfnamefont {G.~A.}\ \bibnamefont
  {Holzapfel}},\ }\href@noop {} {\emph {\bibinfo {title} {{Nonlinear Solid
  Mechanics}}}}\ (\bibinfo  {publisher} {John Wiley \& Sons Ltd., Chichester},\
  \bibinfo {year} {2000})\BibitemShut {NoStop}%
\bibitem [{\citenamefont {Kovetz}(2000)}]{Kovetz2000}%
  \BibitemOpen
  \bibfield  {author} {\bibinfo {author} {\bibfnamefont {A.}~\bibnamefont
  {Kovetz}},\ }\href@noop {} {\emph {\bibinfo {title} {{Electromagnetic
  Theory}}}}\ (\bibinfo  {publisher} {Oxford University Press Inc., New York},\
  \bibinfo {year} {2000})\BibitemShut {NoStop}%
\bibitem [{\citenamefont {Medina}\ and\ \citenamefont
  {Stephany}(2014)}]{Medina2014}%
  \BibitemOpen
  \bibfield  {author} {\bibinfo {author} {\bibfnamefont {R.}~\bibnamefont
  {Medina}}\ and\ \bibinfo {author} {\bibfnamefont {J.}~\bibnamefont
  {Stephany}},\ }\href@noop {} {\bibfield  {journal} {\bibinfo  {journal}
  {arXiv}\ } (\bibinfo {year} {2014})},\ \Eprint
  {http://arxiv.org/abs/1404.5250}{arXiv:1404.5250}\BibitemShut {NoStop}%
\bibitem [{\citenamefont {de~Groot}\ and\ \citenamefont
  {Mazur}(1984)}]{DeGroot1984}%
  \BibitemOpen
  \bibfield  {author} {\bibinfo {author} {\bibfnamefont {S.~R.}\ \bibnamefont
  {de~Groot}}\ and\ \bibinfo {author} {\bibfnamefont {P.}~\bibnamefont
  {Mazur}},\ }\href@noop {} {\emph {\bibinfo {title} {{Non-Equilibrium
  Thermodynamics}}}}\ (\bibinfo  {publisher} {Dover Publications, Inc., New
  York},\ \bibinfo {year} {1984})\BibitemShut {NoStop}%
\bibitem [{\citenamefont {Henjes}(1993)}]{henjes1993pressure}%
  \BibitemOpen
  \bibfield  {author} {\bibinfo {author} {\bibfnamefont {K.}~\bibnamefont
  {Henjes}},\ }\href {\doibase 10.1006/aphy.1993.1035} {\bibfield  {journal}
  {\bibinfo  {journal} {Annals of Physics}\ }\textbf {\bibinfo {volume}
  {223}},\ \bibinfo {pages} {277} (\bibinfo {year} {1993})}\BibitemShut
  {NoStop}%
\bibitem [{\citenamefont {Flory}(1953)}]{Flory1965}%
  \BibitemOpen
  \bibfield  {author} {\bibinfo {author} {\bibfnamefont {P.~J.}\ \bibnamefont
  {Flory}},\ }\href@noop {} {\emph {\bibinfo {title} {Principles of Polymer
  Chemistry}}}\ (\bibinfo  {publisher} {Cornell University Press, Ithaca, New
  York},\ \bibinfo {year} {1953})\BibitemShut {NoStop}%
\bibitem [{\citenamefont {Ogden}\ and\ \citenamefont {Hill}(1972)}]{Ogden1972}%
  \BibitemOpen
  \bibfield  {author} {\bibinfo {author} {\bibfnamefont {R.~W.}\ \bibnamefont
  {Ogden}}\ and\ \bibinfo {author} {\bibfnamefont {R.}~\bibnamefont {Hill}},\
  }\href {\doibase 10.1098/rspa.1972.0096} {\bibfield  {journal} {\bibinfo
  {journal} {Proceedings of the Royal Society of London. A. Mathematical and
  Physical Sciences}\ }\textbf {\bibinfo {volume} {328}},\ \bibinfo {pages}
  {567} (\bibinfo {year} {1972})}\BibitemShut {NoStop}%
\bibitem [{\citenamefont {Steinr{\"{u}}ck}\ \emph {et~al.}(2020)\citenamefont
  {Steinr{\"{u}}ck}, \citenamefont {Takacs}, \citenamefont {Kim}, \citenamefont
  {MacKanic}, \citenamefont {Holladay}, \citenamefont {Cao}, \citenamefont
  {Narayanan}, \citenamefont {Dufresne}, \citenamefont {Chushkin},
  \citenamefont {Ruta}, \citenamefont {Zontone}, \citenamefont {Will},
  \citenamefont {Borodin}, \citenamefont {Sinha}, \citenamefont {Srinivasan},\
  and\ \citenamefont {Toney}}]{Steinrueck2020}%
  \BibitemOpen
  \bibfield  {author} {\bibinfo {author} {\bibfnamefont {H.~G.}\ \bibnamefont
  {Steinr{\"{u}}ck}}, \bibinfo {author} {\bibfnamefont {C.~J.}\ \bibnamefont
  {Takacs}}, \bibinfo {author} {\bibfnamefont {H.~K.}\ \bibnamefont {Kim}},
  \bibinfo {author} {\bibfnamefont {D.~G.}\ \bibnamefont {MacKanic}}, \bibinfo
  {author} {\bibfnamefont {B.}~\bibnamefont {Holladay}}, \bibinfo {author}
  {\bibfnamefont {C.}~\bibnamefont {Cao}}, \bibinfo {author} {\bibfnamefont
  {S.}~\bibnamefont {Narayanan}}, \bibinfo {author} {\bibfnamefont {E.~M.}\
  \bibnamefont {Dufresne}}, \bibinfo {author} {\bibfnamefont {Y.}~\bibnamefont
  {Chushkin}}, \bibinfo {author} {\bibfnamefont {B.}~\bibnamefont {Ruta}},
  \bibinfo {author} {\bibfnamefont {F.}~\bibnamefont {Zontone}}, \bibinfo
  {author} {\bibfnamefont {J.}~\bibnamefont {Will}}, \bibinfo {author}
  {\bibfnamefont {O.}~\bibnamefont {Borodin}}, \bibinfo {author} {\bibfnamefont
  {S.~K.}\ \bibnamefont {Sinha}}, \bibinfo {author} {\bibfnamefont
  {V.}~\bibnamefont {Srinivasan}}, \ and\ \bibinfo {author} {\bibfnamefont
  {M.~F.}\ \bibnamefont {Toney}},\ }\href {\doibase 10.1039/d0ee02193h}
  {\bibfield  {journal} {\bibinfo  {journal} {Energy and Environmental
  Science}\ }\textbf {\bibinfo {volume} {13}},\ \bibinfo {pages} {4312}
  (\bibinfo {year} {2020})}\BibitemShut {NoStop}%
\bibitem [{\citenamefont {Pesko}\ \emph {et~al.}(2018)\citenamefont {Pesko},
  \citenamefont {Feng}, \citenamefont {Sawhney}, \citenamefont {Newman},
  \citenamefont {Srinivasan},\ and\ \citenamefont {Balsara}}]{Pesko2018}%
  \BibitemOpen
  \bibfield  {author} {\bibinfo {author} {\bibfnamefont {D.~M.}\ \bibnamefont
  {Pesko}}, \bibinfo {author} {\bibfnamefont {Z.}~\bibnamefont {Feng}},
  \bibinfo {author} {\bibfnamefont {S.}~\bibnamefont {Sawhney}}, \bibinfo
  {author} {\bibfnamefont {J.}~\bibnamefont {Newman}}, \bibinfo {author}
  {\bibfnamefont {V.}~\bibnamefont {Srinivasan}}, \ and\ \bibinfo {author}
  {\bibfnamefont {N.~P.}\ \bibnamefont {Balsara}},\ }\href {\doibase
  10.1149/2.0921813jes} {\bibfield  {journal} {\bibinfo  {journal} {Journal of
  The Electrochemical Society}\ }\textbf {\bibinfo {volume} {165}},\ \bibinfo
  {pages} {A3186} (\bibinfo {year} {2018})}\BibitemShut {NoStop}%
\bibitem [{\citenamefont {Wen}\ \emph {et~al.}(2003)\citenamefont {Wen},
  \citenamefont {Itoh}, \citenamefont {Uno}, \citenamefont {Kubo},\ and\
  \citenamefont {Yamamoto}}]{Wen2003}%
  \BibitemOpen
  \bibfield  {author} {\bibinfo {author} {\bibfnamefont {Z.}~\bibnamefont
  {Wen}}, \bibinfo {author} {\bibfnamefont {T.}~\bibnamefont {Itoh}}, \bibinfo
  {author} {\bibfnamefont {T.}~\bibnamefont {Uno}}, \bibinfo {author}
  {\bibfnamefont {M.}~\bibnamefont {Kubo}}, \ and\ \bibinfo {author}
  {\bibfnamefont {O.}~\bibnamefont {Yamamoto}},\ }\href {\doibase
  10.1016/S0167-2738(03)00129-2} {\bibfield  {journal} {\bibinfo  {journal}
  {Solid State Ionics}\ }\textbf {\bibinfo {volume} {160}},\ \bibinfo {pages}
  {141} (\bibinfo {year} {2003})}\BibitemShut {NoStop}%
\bibitem [{\citenamefont {Newman}\ and\ \citenamefont
  {Thomas-Alyea}(2004)}]{Newman2004}%
  \BibitemOpen
  \bibfield  {author} {\bibinfo {author} {\bibfnamefont {J.}~\bibnamefont
  {Newman}}\ and\ \bibinfo {author} {\bibfnamefont {K.~E.}\ \bibnamefont
  {Thomas-Alyea}},\ }\href@noop {} {\emph {\bibinfo {title} {{Electrochemical
  Systems}}}}\ (\bibinfo  {publisher} {John Wiley {\&} Sons, Inc., Hoboken},\
  \bibinfo {year} {2004})\BibitemShut {NoStop}%
\bibitem [{\citenamefont {Pesko}\ \emph {et~al.}(2017)\citenamefont {Pesko},
  \citenamefont {Timachova}, \citenamefont {Bhattacharya}, \citenamefont
  {Smith}, \citenamefont {Villaluenga}, \citenamefont {Newman},\ and\
  \citenamefont {Balsara}}]{Pesko2017}%
  \BibitemOpen
  \bibfield  {author} {\bibinfo {author} {\bibfnamefont {D.~M.}\ \bibnamefont
  {Pesko}}, \bibinfo {author} {\bibfnamefont {K.}~\bibnamefont {Timachova}},
  \bibinfo {author} {\bibfnamefont {R.}~\bibnamefont {Bhattacharya}}, \bibinfo
  {author} {\bibfnamefont {M.~C.}\ \bibnamefont {Smith}}, \bibinfo {author}
  {\bibfnamefont {I.}~\bibnamefont {Villaluenga}}, \bibinfo {author}
  {\bibfnamefont {J.}~\bibnamefont {Newman}}, \ and\ \bibinfo {author}
  {\bibfnamefont {N.~P.}\ \bibnamefont {Balsara}},\ }\href {\doibase
  10.1149/2.0581711jes} {\bibfield  {journal} {\bibinfo  {journal} {Journal of
  The Electrochemical Society}\ }\textbf {\bibinfo {volume} {164}},\ \bibinfo
  {pages} {E3569} (\bibinfo {year} {2017})}\BibitemShut {NoStop}%
\bibitem [{\citenamefont {Zhang}\ \emph {et~al.}(2023)\citenamefont {Zhang},
  \citenamefont {Yang}, \citenamefont {Yang}, \citenamefont {Zhang},
  \citenamefont {Zhang}, \citenamefont {Duan},\ and\ \citenamefont
  {Jiang}}]{Zhang2023}%
  \BibitemOpen
  \bibfield  {author} {\bibinfo {author} {\bibfnamefont {A.}~\bibnamefont
  {Zhang}}, \bibinfo {author} {\bibfnamefont {X.}~\bibnamefont {Yang}},
  \bibinfo {author} {\bibfnamefont {F.}~\bibnamefont {Yang}}, \bibinfo {author}
  {\bibfnamefont {C.}~\bibnamefont {Zhang}}, \bibinfo {author} {\bibfnamefont
  {Q.}~\bibnamefont {Zhang}}, \bibinfo {author} {\bibfnamefont
  {G.}~\bibnamefont {Duan}}, \ and\ \bibinfo {author} {\bibfnamefont
  {S.}~\bibnamefont {Jiang}},\ }\href {\doibase 10.3390/molecules28052042}
  {\bibfield  {journal} {\bibinfo  {journal} {Molecules}\ }\textbf {\bibinfo
  {volume} {28}},\ \bibinfo {pages} {2042} (\bibinfo {year}
  {2023})}\BibitemShut {NoStop}%
\bibitem [{\citenamefont {Lazaridis}\ and\ \citenamefont
  {Paulaitis}(1993)}]{Lazaridis1993}%
  \BibitemOpen
  \bibfield  {author} {\bibinfo {author} {\bibfnamefont {T.}~\bibnamefont
  {Lazaridis}}\ and\ \bibinfo {author} {\bibfnamefont {M.~E.}\ \bibnamefont
  {Paulaitis}},\ }\href {\doibase 10.1002/aic.690390614} {\bibfield  {journal}
  {\bibinfo  {journal} {AIChE Journal}\ }\textbf {\bibinfo {volume} {39}},\
  \bibinfo {pages} {1051} (\bibinfo {year} {1993})}\BibitemShut {NoStop}%
\bibitem [{\citenamefont {Ingenmey}\ \emph {et~al.}(2019)\citenamefont
  {Ingenmey}, \citenamefont {Blasius}, \citenamefont {Marchelli}, \citenamefont
  {Riegel},\ and\ \citenamefont {Kirchner}}]{Ingenmey2019}%
  \BibitemOpen
  \bibfield  {author} {\bibinfo {author} {\bibfnamefont {J.}~\bibnamefont
  {Ingenmey}}, \bibinfo {author} {\bibfnamefont {J.}~\bibnamefont {Blasius}},
  \bibinfo {author} {\bibfnamefont {G.}~\bibnamefont {Marchelli}}, \bibinfo
  {author} {\bibfnamefont {A.}~\bibnamefont {Riegel}}, \ and\ \bibinfo {author}
  {\bibfnamefont {B.}~\bibnamefont {Kirchner}},\ }\href {\doibase
  10.1021/acs.jced.8b00779} {\bibfield  {journal} {\bibinfo  {journal} {Journal
  of Chemical and Engineering Data}\ }\textbf {\bibinfo {volume} {64}},\
  \bibinfo {pages} {255} (\bibinfo {year} {2019})}\BibitemShut {NoStop}%
\bibitem [{\citenamefont {Fang}\ \emph {et~al.}(2021)\citenamefont {Fang},
  \citenamefont {Loo},\ and\ \citenamefont {Wang}}]{Fang2021}%
  \BibitemOpen
  \bibfield  {author} {\bibinfo {author} {\bibfnamefont {C.}~\bibnamefont
  {Fang}}, \bibinfo {author} {\bibfnamefont {W.~S.}\ \bibnamefont {Loo}}, \
  and\ \bibinfo {author} {\bibfnamefont {R.}~\bibnamefont {Wang}},\ }\href
  {\doibase 10.1021/acs.macromol.0c01850} {\bibfield  {journal} {\bibinfo
  {journal} {Macromolecules}\ }\textbf {\bibinfo {volume} {54}},\ \bibinfo
  {pages} {2873} (\bibinfo {year} {2021})}\BibitemShut {NoStop}%
\bibitem [{\citenamefont {Jee}\ \emph {et~al.}(2013)\citenamefont {Jee},
  \citenamefont {Lee}, \citenamefont {Lee},\ and\ \citenamefont
  {Lee}}]{Jee2013}%
  \BibitemOpen
  \bibfield  {author} {\bibinfo {author} {\bibfnamefont {A.~Y.}\ \bibnamefont
  {Jee}}, \bibinfo {author} {\bibfnamefont {H.}~\bibnamefont {Lee}}, \bibinfo
  {author} {\bibfnamefont {Y.}~\bibnamefont {Lee}}, \ and\ \bibinfo {author}
  {\bibfnamefont {M.}~\bibnamefont {Lee}},\ }\href {\doibase
  10.1016/j.chemphys.2012.12.028} {\bibfield  {journal} {\bibinfo  {journal}
  {Chemical Physics}\ }\textbf {\bibinfo {volume} {422}},\ \bibinfo {pages}
  {246} (\bibinfo {year} {2013})}\BibitemShut {NoStop}%
\bibitem [{\citenamefont {Ushakova}\ \emph {et~al.}(2020)\citenamefont
  {Ushakova}, \citenamefont {Sergeev}, \citenamefont {Morzhukhin},
  \citenamefont {Napolskiy}, \citenamefont {Kristavchuk}, \citenamefont
  {Chertovich}, \citenamefont {Yashina},\ and\ \citenamefont
  {Itkis}}]{Ushakova2020}%
  \BibitemOpen
  \bibfield  {author} {\bibinfo {author} {\bibfnamefont {E.~E.}\ \bibnamefont
  {Ushakova}}, \bibinfo {author} {\bibfnamefont {A.~V.}\ \bibnamefont
  {Sergeev}}, \bibinfo {author} {\bibfnamefont {A.}~\bibnamefont {Morzhukhin}},
  \bibinfo {author} {\bibfnamefont {F.~S.}\ \bibnamefont {Napolskiy}}, \bibinfo
  {author} {\bibfnamefont {O.}~\bibnamefont {Kristavchuk}}, \bibinfo {author}
  {\bibfnamefont {A.~V.}\ \bibnamefont {Chertovich}}, \bibinfo {author}
  {\bibfnamefont {L.~V.}\ \bibnamefont {Yashina}}, \ and\ \bibinfo {author}
  {\bibfnamefont {D.~M.}\ \bibnamefont {Itkis}},\ }\href {\doibase
  10.1039/d0ra02325f} {\bibfield  {journal} {\bibinfo  {journal} {RSC
  Advances}\ }\textbf {\bibinfo {volume} {10}},\ \bibinfo {pages} {16118}
  (\bibinfo {year} {2020})}\BibitemShut {NoStop}%
\bibitem [{\citenamefont {Lee}\ \emph {et~al.}(2022)\citenamefont {Lee},
  \citenamefont {Rottmayer},\ and\ \citenamefont {Huang}}]{Lee2022}%
  \BibitemOpen
  \bibfield  {author} {\bibinfo {author} {\bibfnamefont {J.}~\bibnamefont
  {Lee}}, \bibinfo {author} {\bibfnamefont {M.}~\bibnamefont {Rottmayer}}, \
  and\ \bibinfo {author} {\bibfnamefont {H.}~\bibnamefont {Huang}},\ }\href
  {\doibase 10.3390/jcs6010012} {\bibfield  {journal} {\bibinfo  {journal}
  {Journal of Composites Science}\ }\textbf {\bibinfo {volume} {6}},\ \bibinfo
  {pages} {1} (\bibinfo {year} {2022})}\BibitemShut {NoStop}%
\bibitem [{\citenamefont {Nikoli{\'{c}}}\ \emph {et~al.}(2013)\citenamefont
  {Nikoli{\'{c}}}, \citenamefont {Moffat}, \citenamefont {Farrugia},
  \citenamefont {Kobryn}, \citenamefont {Gusarov}, \citenamefont {Wosnick},\
  and\ \citenamefont {Kovalenko}}]{Nikolic2013}%
  \BibitemOpen
  \bibfield  {author} {\bibinfo {author} {\bibfnamefont {D.}~\bibnamefont
  {Nikoli{\'{c}}}}, \bibinfo {author} {\bibfnamefont {K.~A.}\ \bibnamefont
  {Moffat}}, \bibinfo {author} {\bibfnamefont {V.~M.}\ \bibnamefont
  {Farrugia}}, \bibinfo {author} {\bibfnamefont {A.~E.}\ \bibnamefont
  {Kobryn}}, \bibinfo {author} {\bibfnamefont {S.}~\bibnamefont {Gusarov}},
  \bibinfo {author} {\bibfnamefont {J.~H.}\ \bibnamefont {Wosnick}}, \ and\
  \bibinfo {author} {\bibfnamefont {A.}~\bibnamefont {Kovalenko}},\ }\href
  {\doibase 10.1039/c3cp44285c} {\bibfield  {journal} {\bibinfo  {journal}
  {Physical Chemistry Chemical Physics}\ }\textbf {\bibinfo {volume} {15}},\
  \bibinfo {pages} {6128} (\bibinfo {year} {2013})}\BibitemShut {NoStop}%
\bibitem [{\citenamefont {Ding}\ \emph {et~al.}(2023)\citenamefont {Ding},
  \citenamefont {Wu}, \citenamefont {Lin}, \citenamefont {Lou}, \citenamefont
  {Tang}, \citenamefont {Guo}, \citenamefont {Guo}, \citenamefont {Wang},\ and\
  \citenamefont {Yu}}]{Ding2023}%
  \BibitemOpen
  \bibfield  {author} {\bibinfo {author} {\bibfnamefont {P.}~\bibnamefont
  {Ding}}, \bibinfo {author} {\bibfnamefont {L.}~\bibnamefont {Wu}}, \bibinfo
  {author} {\bibfnamefont {Z.}~\bibnamefont {Lin}}, \bibinfo {author}
  {\bibfnamefont {C.}~\bibnamefont {Lou}}, \bibinfo {author} {\bibfnamefont
  {M.}~\bibnamefont {Tang}}, \bibinfo {author} {\bibfnamefont {X.}~\bibnamefont
  {Guo}}, \bibinfo {author} {\bibfnamefont {H.}~\bibnamefont {Guo}}, \bibinfo
  {author} {\bibfnamefont {Y.}~\bibnamefont {Wang}}, \ and\ \bibinfo {author}
  {\bibfnamefont {H.}~\bibnamefont {Yu}},\ }\href {\doibase
  10.1021/jacs.2c06512} {\bibfield  {journal} {\bibinfo  {journal} {Journal of
  the American Chemical Society}\ }\textbf {\bibinfo {volume} {145}},\ \bibinfo
  {pages} {1548} (\bibinfo {year} {2023})}\BibitemShut {NoStop}%
\bibitem [{\citenamefont {Maitra}\ and\ \citenamefont
  {Heuer}(2008)}]{Maitra2008}%
  \BibitemOpen
  \bibfield  {author} {\bibinfo {author} {\bibfnamefont {A.}~\bibnamefont
  {Maitra}}\ and\ \bibinfo {author} {\bibfnamefont {A.}~\bibnamefont {Heuer}},\
  }\href {\doibase 10.1021/jp711563a} {\bibfield  {journal} {\bibinfo
  {journal} {Journal of Physical Chemistry B}\ }\textbf {\bibinfo {volume}
  {112}},\ \bibinfo {pages} {9641} (\bibinfo {year} {2008})},\ \Eprint
  {http://arxiv.org/abs/0804.2212}{0804.2212}\BibitemShut {NoStop}%
\bibitem [{\citenamefont {Stolz}\ \emph {et~al.}(2022)\citenamefont {Stolz},
  \citenamefont {Hochst{\"{a}}dt}, \citenamefont {R{\"{o}}ser}, \citenamefont
  {Hansen}, \citenamefont {Winter},\ and\ \citenamefont
  {Kasnatscheew}}]{Stolz2022}%
  \BibitemOpen
  \bibfield  {author} {\bibinfo {author} {\bibfnamefont {L.}~\bibnamefont
  {Stolz}}, \bibinfo {author} {\bibfnamefont {S.}~\bibnamefont
  {Hochst{\"{a}}dt}}, \bibinfo {author} {\bibfnamefont {S.}~\bibnamefont
  {R{\"{o}}ser}}, \bibinfo {author} {\bibfnamefont {M.~R.}\ \bibnamefont
  {Hansen}}, \bibinfo {author} {\bibfnamefont {M.}~\bibnamefont {Winter}}, \
  and\ \bibinfo {author} {\bibfnamefont {J.}~\bibnamefont {Kasnatscheew}},\
  }\href {\doibase 10.1021/acsami.2c00084} {\bibfield  {journal} {\bibinfo
  {journal} {ACS Applied Materials and Interfaces}\ }\textbf {\bibinfo {volume}
  {14}},\ \bibinfo {pages} {11559} (\bibinfo {year} {2022})}\BibitemShut
  {NoStop}%
\bibitem [{\citenamefont {Stolz}\ \emph {et~al.}(2021)\citenamefont {Stolz},
  \citenamefont {Homann}, \citenamefont {Winter},\ and\ \citenamefont
  {Kasnatscheew}}]{STOLZ20219}%
  \BibitemOpen
  \bibfield  {author} {\bibinfo {author} {\bibfnamefont {L.}~\bibnamefont
  {Stolz}}, \bibinfo {author} {\bibfnamefont {G.}~\bibnamefont {Homann}},
  \bibinfo {author} {\bibfnamefont {M.}~\bibnamefont {Winter}}, \ and\ \bibinfo
  {author} {\bibfnamefont {J.}~\bibnamefont {Kasnatscheew}},\ }\href {\doibase
  https://doi.org/10.1016/j.mattod.2020.11.025} {\bibfield  {journal} {\bibinfo
   {journal} {Materials Today}\ }\textbf {\bibinfo {volume} {44}},\ \bibinfo
  {pages} {9} (\bibinfo {year} {2021})}\BibitemShut {NoStop}%
\bibitem [{\citenamefont {Nedoma}\ \emph {et~al.}(2008)\citenamefont {Nedoma},
  \citenamefont {Robertson}, \citenamefont {Wanakule},\ and\ \citenamefont
  {Balsara}}]{Nedoma2008}%
  \BibitemOpen
  \bibfield  {author} {\bibinfo {author} {\bibfnamefont {A.~J.}\ \bibnamefont
  {Nedoma}}, \bibinfo {author} {\bibfnamefont {M.~L.}\ \bibnamefont
  {Robertson}}, \bibinfo {author} {\bibfnamefont {N.~S.}\ \bibnamefont
  {Wanakule}}, \ and\ \bibinfo {author} {\bibfnamefont {N.~P.}\ \bibnamefont
  {Balsara}},\ }\href {\doibase 10.1021/ma800698r} {\bibfield  {journal}
  {\bibinfo  {journal} {Macromolecules}\ }\textbf {\bibinfo {volume} {41}},\
  \bibinfo {pages} {5773} (\bibinfo {year} {2008})}\BibitemShut {NoStop}%
\bibitem [{\citenamefont {Tomlin}\ and\ \citenamefont
  {Roland}(1992)}]{Tomlin1992}%
  \BibitemOpen
  \bibfield  {author} {\bibinfo {author} {\bibfnamefont {D.~W.}\ \bibnamefont
  {Tomlin}}\ and\ \bibinfo {author} {\bibfnamefont {C.~M.}\ \bibnamefont
  {Roland}},\ }\href {\doibase 10.1021/ma00037a033} {\bibfield  {journal}
  {\bibinfo  {journal} {Macromolecules}\ }\textbf {\bibinfo {volume} {25}},\
  \bibinfo {pages} {2994} (\bibinfo {year} {1992})}\BibitemShut {NoStop}%
\bibitem [{\citenamefont {Sahu}\ \emph {et~al.}(2009)\citenamefont {Sahu},
  \citenamefont {Pitchumani}, \citenamefont {Sridhar},\ and\ \citenamefont
  {Shukla}}]{Sahu2009}%
  \BibitemOpen
  \bibfield  {author} {\bibinfo {author} {\bibfnamefont {A.~K.}\ \bibnamefont
  {Sahu}}, \bibinfo {author} {\bibfnamefont {S.}~\bibnamefont {Pitchumani}},
  \bibinfo {author} {\bibfnamefont {P.}~\bibnamefont {Sridhar}}, \ and\
  \bibinfo {author} {\bibfnamefont {A.~K.}\ \bibnamefont {Shukla}},\ }\href
  {\doibase 10.1007/s12034-009-0042-8} {\bibfield  {journal} {\bibinfo
  {journal} {Bulletin of Materials Science}\ }\textbf {\bibinfo {volume}
  {32}},\ \bibinfo {pages} {285} (\bibinfo {year} {2009})}\BibitemShut
  {NoStop}%
\bibitem [{\citenamefont {Rosenwinkel}\ and\ \citenamefont
  {Sch{\"{o}}nhoff}(2019)}]{Rosenwinkel2019}%
  \BibitemOpen
  \bibfield  {author} {\bibinfo {author} {\bibfnamefont {M.~P.}\ \bibnamefont
  {Rosenwinkel}}\ and\ \bibinfo {author} {\bibfnamefont {M.}~\bibnamefont
  {Sch{\"{o}}nhoff}},\ }\href {\doibase 10.1149/2.0831910jes} {\bibfield
  {journal} {\bibinfo  {journal} {Journal of The Electrochemical Society}\
  }\textbf {\bibinfo {volume} {166}},\ \bibinfo {pages} {A1977} (\bibinfo
  {year} {2019})}\BibitemShut {NoStop}%
\bibitem [{\citenamefont {Shao}\ \emph {et~al.}(2022)\citenamefont {Shao},
  \citenamefont {Gudla}, \citenamefont {Brandell},\ and\ \citenamefont
  {Zhang}}]{Shao2022}%
  \BibitemOpen
  \bibfield  {author} {\bibinfo {author} {\bibfnamefont {Y.}~\bibnamefont
  {Shao}}, \bibinfo {author} {\bibfnamefont {H.}~\bibnamefont {Gudla}},
  \bibinfo {author} {\bibfnamefont {D.}~\bibnamefont {Brandell}}, \ and\
  \bibinfo {author} {\bibfnamefont {C.}~\bibnamefont {Zhang}},\ }\href
  {\doibase 10.1021/jacs.2c02389} {\bibfield  {journal} {\bibinfo  {journal}
  {Journal of the American Chemical Society}\ }\textbf {\bibinfo {volume}
  {144}},\ \bibinfo {pages} {7583} (\bibinfo {year} {2022})}\BibitemShut
  {NoStop}%
\end{thebibliography}%

\end{document}